\documentclass[journal]{IEEEtran}
\ifCLASSINFOpdf
\else
\fi
\usepackage{syntax}
\usepackage{listings}
\usepackage{resizegather}
\usepackage[T1]{fontenc}
\usepackage{lmodern}
\usepackage{standalone}
\usepackage{tikz,pgfplots}
\usetikzlibrary{positioning}
\usetikzlibrary{arrows.meta}
\usepackage{subcaption}
\pgfplotsset{compat=1.11}
\usepackage{amsmath}
\usepackage{csquotes}
\usepackage{tabularx}
\usepackage{rotating}
\usepackage{multirow} 
\usepackage{color} 
\usepgfplotslibrary{fillbetween}
\usepgfplotslibrary{statistics}
\usepackage{xcolor,colortbl}
\usepackage[listings,skins,breakable]{tcolorbox}
\usetikzlibrary{backgrounds}
\usetikzlibrary{fadings}
\usepackage{enumitem}
\usepackage{tcolorbox}
\usepackage{lmodern}
\usepackage{amsthm}
\usepackage[numbers]{natbib}
\usepackage{amsfonts}
\usepackage{newtxmath}
\usepackage{booktabs}
\usepackage{url}
\usepackage{threeparttable}
\usepackage[tikz]{bclogo}
\theoremstyle{plain}

\newtheorem*{def*}{}

\usetikzlibrary{pgfplots.groupplots}
\pgfplotsset{
        /pgfplots/ybar legend/.style={
        /pgfplots/legend image code/.code={%
        \draw[##1,/tikz/.cd,bar width=3pt,yshift=-0.2em,bar shift=0pt]
                plot coordinates {(0cm,0.8em)};},
},
}

\newcolumntype{P}[1]{>{\centering\arraybackslash}m{#1}}
\newcolumntype{Y}{>{\centering\arraybackslash}X}

\begin{document}
%
\title{Synergizing Domain Expertise with Self-Awareness in Software Systems: A Patternized Architecture Guideline}
%
%
%

\author{Tao~Chen,~\IEEEmembership{Member,~IEEE,}
        Rami~Bahsoon,~\IEEEmembership{Member,~IEEE,}
        and~Xin~Yao,~\IEEEmembership{Fellow,~IEEE}
\thanks{Tao Chen (co-corresponding author) is with the Department of Computer Science, Loughborough University, Epinal Way, Loughborough, LE11 3TU, UK (e-mail: t.t.chen@lboro.ac.uk)}
\thanks{Rami Bahsoon is with the School of Computer Science, University of Birmingham, Birmingham, B15 2TT, UK (e-mail: r.bahsoon@cs.bham.ac.uk)}
\thanks{Xin Yao (co-corresponding author) is with the Guangdong Provincial Key Laboratory of Brain-inspired Intelligent Computation, Department of Computer Science and Engineering, Southern University of Science and Technology, Shenzhen 518055, Guangdong, China (e-mail: xiny@sustech.edu.cn)}
}

%
%

\markboth{Accepted by the Proceedings of the IEEE}{Author Copy}
%



\maketitle

\begin{abstract}

To promote engineering self-aware and self-adaptive software systems in a reusable manner, architectural patterns and the related methodology provide an unified solution to handle the recurring problems in the engineering process. However, in existing patterns and methods, domain knowledge and engineers' expertise that is built over time are not explicitly linked to the self-aware processes. This linkage is important, as the knowledge is a valuable asset for the related problems and its absence would cause unnecessary overhead, possibly misleading results and unwise waste of the tremendous benefit that could have been brought by the domain expertise. This paper highlights the importance of synergizing domain expertise and the self-awareness to enable better self-adaptation in software systems, relying on well-defined expertise representation, algorithms and techniques. In particular, we present a holistic framework of notions, enriched patterns and methodology, dubbed DBASES, that offers a principled guideline for the engineers to perform difficulty and benefit analysis on possible synergies, in an attempt to keep ``engineers-in-the-loop". Through three tutorial case studies, we demonstrate how DBASES can be applied in different domains, within which a carefully selected set of candidates with different synergies can be used for quantitative investigation, providing more informed decisions of the design choices.


\end{abstract}


%
\IEEEpeerreviewmaketitle

\section{Introduction}
%
%
%
%
\IEEEPARstart{E}{ngineering} software systems has been becoming increasingly complex, and labor-intensive due to the continuous changes in requirements, the underlying environments and the relevant data. Such complexity is prevalent when engineering self-aware and self-adaptive software systems\textemdash a category of systems that is capable of obtaining and maintaining knowledge on themselves and the environment, reasoning about this knowledge, and eventually adapting their operations to better cope with the changes. In this respect, engineers need some sets of high-level guideline that provides a clear overview of the software system to be built, based on which they are able to make better-informed decisions during the engineering process. Such a high-level guideline for engineering software systems can be represented in the form of architectural patterns and their methodologies. In essence, architectural patterns are particular solutions for common and recurring domain specific problems, culminating best practices, and described at high-level~\cite{Zdun:2005}. A variety of architectural patterns and methodologies exist, each of which aims at a different context, e.g., distributed systems~\cite{schmidt2003patterns,schmidt2013pattern}, service-oriented systems~\cite{1605179,endrei2004patterns}, self-adaptive systems~\cite{computing2006architectural,1350726}, and more recently, self-aware and self-adaptive software systems~\cite{2014epicshandbook,Giese2017,7185305,Chen2016:book,6827105} (a.k.a. self-awareness architectural patterns).

Unlike the other patterns, self-awareness architectural patterns particularly document the common primitives and different capabilities of self-awareness for obtaining and maintaining knowledge about different aspects, such as time, goals, or interactions between different nodes of software systems. While these patterns are abstract, they can be instantiated to meet particular needs for engineering a self-aware and self-adaptive software system, thereby providing more concrete guideline on how to align the capabilities of self-awareness with the requirements. Over the last few years, those patterns and the related methodology have proved to be promising when engineering self-aware and self-adaptive software systems, as evident by the fact that they have been referenced and used in various autonomic domains, such as cloud resource and configuration management~\cite{Chen:2015:tse,tsc:chen:2015,7274426}, multi-processors systems scheduling~\cite{7372548}, sensor network control~\cite{8016147} and multi-camera coordination~\cite{7163249}. 

Traditionally, engineering self-awareness in software systems have been primarily supported by various Artificial Intelligence (AI) algorithms, which serve as cheap and ``black boxes" that can be directly applied with little specialization~\cite{DBLP:journals/ese/BinkleyLM18,DBLP:journals/infsof/FuMS16}. However, as reviewed by Menzies~\cite{DBLP:journals/software/Menzies20}, an emerging question in the application of AI algorithms to various engineering problems is the validity of the assumptions that underlie the creation of those algorithms. Therefore, the practice of applying standard AI algorithms as ``black boxes", where researchers do not tinker with internal design choices of these algorithms with respect to their expertise on the problem is not ideal~\cite{DBLP:journals/software/Menzies20}. Indeed, self-aware and self-adaptive software systems have never been created by non-experts. This means that software and systems engineers often accumulate domain expertise that is built over time. Such expertise, if captured and exploited, would provide an important added value to consolidate the self-awareness capability of the software system. Utilizing domain expertise to guide the processes of underlying AI algorithms, and thus the self-awareness, can bear additional benefit. In this way, the software system would be more controllable, which helps to monitor and avoid some abnormalities in behaviors, providing a foundation for keeping ``engineers-in-the-loop". There existing many researches~\cite{DBLP:conf/icse/PanichellaDOPPL13,femosaa,chen-icpe,seeding,DBLP:journals/infsof/ChenLY19} that show superior results can be obtained by specializing AI algorithm to the particulars of engineering problem with domain expertise. 


With this in mind, despite the successful applications of the existing architectural patterns and methodology for engineering self-awareness, the consideration of engineers' expertise, particularly on how they can be `synergized' into the self-awareness capabilities, is weak, ad-hoc and left implicit. By the term \textbf{synergy}, we refer to the process of incorporating domain expertise, which involves the knowledge of the problem that is not naturally initiative (on contrary to, e.g., the range and type of parameters, various equality and inequality constraints) but can be extracted following engineering principles, into the underlying algorithms/techniques that realize self-awareness. Indeed, the lack of a holistic framework of patterns and methodology would inevitably create barrier for the domain information/knowledge to be maintained, reused and exploited to steer the design process, especially given a large variety of existing expertise representations and AI algorithms. This absence can eventually result in some strong domain expertise being overlooked, causing unncessary overhead and possibly misleading results~\cite{Chen:2018:STS,elhabbash2019self}. 


To overcome such a gap, in this paper, we formalize a holistic framework that provides a principled guideline to perform \textbf{\underline{d}}ifficulty and \textbf{\underline{b}}enefit \textbf{\underline{a}}nalysis for \textbf{\underline{s}}ynergizing domain \textbf{\underline{e}}xpertise and \textbf{\underline{s}}elf-awareness (hence dubbed as DBASES). Our aim is to elaborate and showcase how DBASES can support the ``synergy" and reveal its importance, taking into account the self-awareness in software systems based on well-defined and widely used expertise representations, algorithms and techniques. It is indeed an ambitious plan, therefore we intend to be introductory rather than comprehensive. However, we hope that this work can spark a dialog about the diverse and representative research on combining domain expertise with self-awareness, and that some level of consensus on the design of such synergy will be achieved.




Specifically, our key contributions of the DBASES framework in the paper are:

\begin{itemize}

\item We introduce general notions that captures the domain expertise of the engineers and their synergies with the concepts of self-awareness, providing intuitive, extracted and readily available information to enrich the self-awareness architectural patterns. Specifically, we contribute to the followings:

\begin{itemize}
\item We present the notions of expertise representation with concrete examples, based on which we form a classification and the related rules that helps to capture the expertise knowledge.

\item Drawing on the expertise representation, we codify a taxonomy that describes their nature in terms of structurability and tangibility.

\item We then discuss their possible synergies with different capabilities of self-awareness, and present rules that classify different levels of synergies and the relative difficulty\footnote{Difficulty is related to complexity, which can impact both the implementation and maintenance of a software system.}, with respect to the structurability and tangibility of expertise representation.
\end{itemize}

\item We illustrate, by means of examples, how the proposed notions can be used to enrich the well-defined self-awareness architectural patterns from the literature~\cite{2014epicshandbook}, and in what ways they can be instantiated to cope with different styles of synergies.

\item Supporting by the proposed notions and the enriched patterns, we present a practical, intuitive and step-by-step methodology that assists the engineers to analyze the difficulty and benefit for alternative synergies of domain expertise with self-awareness, revealing their importance. This would help the engineers to elicit the most preferred candidate(s) for further investigation, while ruling out some of the options that are clearly undesired, thus saving great effort in the development.


\item We demonstrate three recent tutorial case studies~\cite{femosaa,chen-icpe,DBLP:journals/infsof/ChenLY19}, which relied on DBASES, that seek to build self-aware and self-adaptive software systems. Through quantitative results, we show how DBASES can be applied to the engineering process for analyzing the difficulty and benefit of different synergy candidates, leading to a set of more promising ones for further investigation.


\end{itemize}

The remaining paper is organized as follows: we motivate the needs and discuss the related work in Section~\ref{sec:motivation}, following by a brief overview of the capabilities of self-awareness and the existing self-awareness architectural patterns in Section~\ref{sec:self-aware} and~\ref{sec:patterns}. After such, in Section~\ref{sec:expertise}, we present the notions and theoretical foundation that underpins DBASES. In Section~\ref{sec:transposition}, we illustrate how the existing self-awareness patterns can be enriched with DBASES. In Section~\ref{sec:method}, we present, as part of DBASES, a practical step-by-step methodology that assists the engineers in selecting the possible ways of synergies. Three tutorial case studies from different domains are drawn in Section~\ref{sec:case-study} to demonstrate how DBASES can be practically applied. Finally, Section~\ref{sec:conclusion} concludes the paper with discussion on future work.

\section{Preliminaries}
\label{sec:motivation}

\subsection{Problem Nature and Domain Expertise}
\label{sec:domain-expertise}

As mentioned, self-aware and self-adaptive software systems have been increasingly relying on AI algorithms and techniques. Indeed, given the significant growth of the AI community, it is not uncommon to see that successful engineering of self-awareness is underpinned by several AI algorithms~\cite{DBLP:conf/icse/Chen19b,DBLP:journals/corr/abs-2002-09040,icse2020-empirical,ai-and-se,femosaa,wada2012e3,DBLP:journals/infsof/ChenLY19,DBLP:conf/icpads/KumarBCLB18,DBLP:conf/icse/ChenB14,DBLP:journals/corr/abs-2001-08236}, which conducts learning, reasoning and problem solving. 

In general, the application of standard AI algorithms and techniques may need to be combined with sufficient domain information, and thus they can better serve the purpose. Yet, it is important to note that domain information can be distinguished between the problem nature and the domain expertise. In fact, the former refers to the nature information of the problem, which is the basic elements required to appropriately apply the algorithm/techniques (e.g., the range of parameters)~\cite{Chen:2018:STS,elhabbash2019self}; and the latter is the engineers' domain expertise, which is specifically related to the engineering problem to be addressed and is often deemed as optional, but desirable. In essence, what make the additional engineers' domain expertise differs from the basic problem nature is that, 
\begin{itemize}
\item \textbf{Problem nature} refers to commonly known properties and characteristics of the problem domain, such that the AI algorithms have to comply with in order to be used appropriately. This may, for example, include the range of parameters, sparsity of the values, various equality and inequality constraints, etc. Directly applying standard AI algorithms is often considered as exploiting only the problem's nature, due primarily to the generality of existing AI algorithms~\cite{DBLP:journals/software/Menzies20}.
\item In contrast, the \textbf{domain expertise} is represented as or produced by typical software and system engineering methods, practices and models. Most commonly, the knowledge of domain expertise is not naturally intuitive form the problem context, but can be extracted through engineering practices, skills and tools, e.g., design models, formatted documents or even concepts. 
\end{itemize}

\subsection{Lessons from Applying Standard AI Algorithms in Engineering}
\label{sec:related-ai}


In the software and system engineering community, there is an increasing recognition on the limitation of applying standard AI algorithms to various engineering problems. A very recent study, conducted by Agrawal and Menzies~\cite{ai-and-se}, on a wide range of software engineering problems have revealed the following fact:

\begin{tcolorbox}[breakable,left=5pt,right=5pt,top=5pt,bottom=5pt] 
\textit{Our conclusion is that the algorithms which we call ``general AI tools" may not be ``general" at all. Rather, they are tools which are powerful in their home domain but need to be used with care if applied to some new domains like software engineering. Hence, we argue that it is not good enough to just take AI tools developed elsewhere, then apply them verbatim to software engineering problems. Software engineers need to develop AI tools that are better suited to the particulars of software engineering problems.} 
\end{tcolorbox}

The conclusion delivers a very clear message that the standard AI algorithms combined with the necessary information of problem nature, including those that realize self-awareness~\cite{Chen:2018:STS,elhabbash2019self}, cannot fully meet the complex requirements of engineering software systems. It therefore calls for better specialization of these AI algorithms based on the domain expertise of engineers.

From the literature, it is not uncommon to see that greater benefits can be obtained by synergizing domain expertise. For example, there is a thread of research that seeks to synergize Feature Model, which represents domain expertise on requirement analysis, with evolutionary search to reason about behaviors of the self-aware and self-adaptive software systems~\cite{DBLP:journals/jss/PascualLPFE15}. This is motivated by the fact that domain expertise on the requirement cannot be easily captured by simply applying the AI algorithms. Another example of ``software/system engineering needs different AI algorithms" comes from the work of Hindle et al.~\cite{DBLP:journals/cacm/HindleBGS16}, in which they stated that, unlike the common areas where AI was most originally applied, domain expertise in software engineering may suggest some important terms in the code which is used exponentially less frequent. This can provide useful information when model the software system with AI, enabling more accurate self-awareness of the faults.

The domain expertise of engineers can often serve as useful information to engineer self-aware and self-adaptive software systems, thus they should not be simply ignored. To this end, a better synergy between domain expertise and AI algorithms is required. Although in this case the AI algorithms may be made less general and pose extra difficulty, they are expected to work better under the given problem where the domain expertise lies, and more importantly, rendering the self-awareness more controllable. It is in fact part of our contributions in DBASES to provide better analysis on the trade-off between difficulty and benefit when designing different ways of synergy.


\subsection{The Problems}




As summarized in the recent surveys~\cite{Chen:2018:STS,elhabbash2019self}, majority of the work incorporates merely the necessary information of the problem nature with self-awareness in an ad-hoc manner, which is the direct application of the standard AI algorithms. This is because the problem nature can be easily obtained and the existing AI algorithms are designed to be as general as possible, such that they will cope with the basic properties of different domains. However, it is clearly difficult to perform the same for synergizing domain expertise without omission. The key issue is that there is a lack of general guideline that assists the engineers when engineering self-awareness into software systems with explicit consideration of the domain expertise. For example, it is not uncommon that engineers would have certain domain expertise represented as models, documentations, or even artifacts of a software systems, but how they may be related to self-awareness is unknown. Below, we illustrate some common, but difficult decisions and problem to deal with during the synergy process, together with what contributions in DBASES can help on each: 

\begin{itemize}
\item Which available domain expertise can be synergized into which aspect of self-awareness? 
\begin{itemize}
\item[---] Answering such would require understanding on both the available domain expertise and which aspect of the self-awareness is required, e.g., time, goal or interaction~\cite{DBLP:series/ncs/2016LP}. Clearly, there will be constraint that prevents certain synergies, e.g., a feature model cannot usually help in terms of interaction, as its notation does not embed any knowledge of it. In essence, the feasible synergies form the possible candidates for the engineers to make design decision. Yet, it is challenging to build the set of candidates for synergy in the absence of systematic guideline, especially when multiple forms of domain expertise and aspects of self-awareness exist.
\end{itemize}

\begin{bclogo}[couleur=gray!10,arrondi=0.1,epBord=1.5,couleurBord=black!70,logo=\bctrombone]{{\normalsize What parts in DBASES can address this?}}Self-awareness architecture patterns with explicit synergy between domain expertise and self-awareness, as discussed in Section~\ref{sec:patterns} and~\ref{sec:transposition}. 
\end{bclogo}

\item To what extent can a synergy be completed and what are the difficulties?  
\begin{itemize}
\item[---] Synergies can often be done in different levels, e.g., whether the domain expertise can be directly incorporated into the algorithm/techniques or certain internal components need to be specialized~\cite{DBLP:series/ncs/ChenFB16,2014epicshandbook}. This is a crucial design decision to make and it should not be conducted without knowing the relative difficulty, which directly related to the cost of the engineering and maintenance process. However, without guideline, it would be difficult for the engineers to obtain a full picture of the possible extents of synergy and their difficulties.
\end{itemize}

\begin{bclogo}[couleur=gray!10,arrondi=0.1,epBord=1.5,couleurBord=black!70,logo=\bctrombone]{{\normalsize What parts in DBASES can help this?}}Generic notions and categories of different synergy levels and their relative difficulties, as shown in Section~\ref{sec:expertise}.  
\end{bclogo}

\item How to make decision taking into consideration the difficulty of synergy and the expected benefit? 
\begin{itemize}
\item[---] The different candidates of alternative design options would inevitably lead to a decision space~\cite{paul2002evaluating,DBLP:conf/icse/KazmanAK01}. As a result, it would be challenging to enable well-informed decision making without the support of quantifiable and intuitive metrics. In particular, given the potentially large number of alternatives, it would be nicer to intuitively understand which can be ruled out and what needs to be investigated further.
\end{itemize}

\begin{bclogo}[couleur=gray!10,arrondi=0.1,epBord=1.5,couleurBord=black!70,logo=\bctrombone]{{\normalsize What parts in DBASES can improve this?}}Methodological guideline and quantification metrics for visualizing possible candidates of synergies with respect to their difficulties and benefits, as elaborated in Section~\ref{sec:method}. 
\end{bclogo}

\end{itemize}

As a result, the lack of general guideline on how to exploit engineers' expertise when engineering self-awareness into software systems would hinder the benefits of domain expertise synergy, causing barrier to create more advanced and controllable self-awareness driven by the expertise of engineers.

This is what we seek to achieve in the paper with DBASES for engineering self-aware and self-adaptive software systems, in which we conduct the first attempt to propose a general, yet holistic framework to assist the engineers in making decisions of synergy, or at least a more concise set of options that are subject to further investigation.


\subsection{Related Work on Architectural Pattern and Methods}
\label{sec:related-ar}

Software and system architecture, as the highest level of abstraction for all software systems, serves as the framework for satisfying requirements; as the managerial basis for cost estimation and process management; and as an effective basis for reuse and dependency analysis~\cite{Perry:1992}. From the community of software and system engineering, architectural patterns and the related methods seek to abstract common features of architecture instances in a specific domain, which is known to serve as a useful guide to the engineers when designing software systems~\cite{Perry:1992}. Among others, Cost-Benefit Analysis Method (CBAM)~\cite{DBLP:conf/icse/KazmanAK01} and Architecture Trade-off Analysis Method (ATAM)~\cite{paul2002evaluating} are two most widely used methodologies that help to reason about different design options on architectures and their patterns. However, they were designed to deal with general software system and thus are irrelevant to the concept of self-awareness.

Over the past two decades of research for architecting self-aware and self-adaptive software systems, several architecture patterns and their methodologies have emerged. Among others, feedback loop based architecture pattern~\cite{Hellerstein:2004:FCC:975344}, whether as single loop or multiple loops, have been the most widely adopted approach. Such pattern merely assume that the software system can be monitored, and that it can be influenced after certain process is completed based upon the collected data. The reason behind its popularity is due to its simplicity and flexibility, such that there is no constraint on how and what should be architected in the patterns. Yet, as the software system becomes more complex, such simplicity turns into a barrier, as the software and system engineers require more specific guideline when designing the architecture which has been missing from the feedback loop based architecture pattern.

In light of these, the MAPE-K architecture pattern~\cite{computing2006architectural} and its design guidelines are proposed to provide more specific codification about what should be achieved within a feedback loop when engineering self-aware and self-adaptive software systems. In MAPE-K, the (K)nowledge component is shared by the (M)onitor, (A)nalyser, (P)lanner and (E)xecutor components, which provides primitives for expressing domain knowledge in K. This knowledge is used to reason about run-time adaptation. Two other patternized methods, which are sub-classes of MAPE-K but generic enough to be classified as representative styles with distinct qualities. The first is proposed by Oreizy et al.~\cite{769885} such that the pattern consists of an adaptation layer and an evolution layer. Particularly, the adaptation layer is responsible for monitoring and adapting, while the evolution layer caters to ensuring changes in the running system are performed in such a way that the operation of the system is not disrupted. The second patternized method is Rainbow~\cite{1350726}, which is explicitly designed for engineering rule-based adaptation in software systems. Since the above patterns assume a centralized scenario where there is only one instance of software system to be adapted, the MAPE-K is then further extended by Weyns et al.~\cite{weyns2013patterns} into a decentralized version, such that they are specialized into contexts with different degree of decentralization that the software system encounters, with some guidelines.

Inspired by Gat's three layered architecture in the robotics domain~\cite{gat1998three}, Kramer and Magee presented a conceptual three layered architecture patterns and methods~\cite{Kramer:2007} for self-adaptive software system. The three layer, namely goal layer, change layer and component control layer, work in a hierarchical way such that the goal layer provides change plans, which are then further translated into change actions by the change layer, and eventually those actions are run by the component control layer. The opposite of the direction would occur when data needs to be collected.

Alternative to the MAPE-K and the three-layer pattern, SEEC~\cite{hoffmann2011seec} is another set of architecture patterns and methods that claim self-aware capabilities. In a nutshell, SEEC relies on the basic (O)bserve-(D)ecide-(A)ct (ODA) loop~\cite{hoffmann2011seec}. Here, the O and A components in ODA are equivalent to the M and E components in MAPE-K respectively, while analysis and planning tasks are subsumed in the Decide component. Another more recent effort, namely LRA-M loop~\cite{Giese2017}, aims to capture the knowledge of self-awareness in terms of universal models, which can then be exploited by the reasoning.

However, all the above patterns and methods have focused on providing guideline about how to exploit the obtained knowledge to inform adaptation, but limited in modeling the knowledge at a coarse grain, without explicit distinction between knowledge concerns for different levels, e.g., at goals, time, or interaction. In 2014, we proposed a set of self-awareness patterns and methodolgoy~\cite{2014epicshandbook,7185305,Chen2016:book,6827105} derived from the general concept of self-awareness~\cite{epics} for engineering self-aware and self-adaptive software systems. Unlike others, we explicitly encode the pattern based on the fine-grained capability of self-awareness with respect to stimulus, goal, time, interaction and meta-self, considering their distinctions and interplays (we elaborate the patterns in Section~\ref{sec:self-aware}). Those patterns have been followed by a considerable amount of work from other research groups and have attracted a wide range of attentions. However, our experience with industrial partners when using those patterns and methods (together with the other state-of-the-art) is that they fail to capture how domain expertise, and more importantly how they can be combined with the AI algorithms that underpin self-awareness, which is now become the major barrier for them to follow.


\section{The Capabilities of Self-Awareness in Software Systems}
\label{sec:self-aware}

Self-awareness is certainly not new in the other disciplines, but it is challenging to model such a concept in the context of software systems. In this work, we use the term \textbf{node} to refer to a software system that can either work alone, or as one individual in a networked group of different systems. Drawing on Neisser's notions on the self-awareness from the psychology domain, different capabilities of computational self-awareness have been codified~\cite{epics}, which are what the DBASES framework based upon. As illustrated below, each \textbf{self-aware capability} captures distinct knowledge that a software system would need in order to perform self-adaptation and self-expression at certain degree:

\begin{itemize}
\item \textbf{Stimulus-awareness:} A software system is stimulus-aware, if it has knowledge of stimuli. The software system is not able to distinguish between the sources of stimuli. It does not have knowledge of past/future stimuli, but enables the ability in a software system to respond to events. It is a prerequisite for all other capabilities of self-awareness.

\item \textbf{Time-awareness:} A software system is time-aware if it has knowledge of historical and/or likely future phenomena. Implementing time-awareness may involve the software system possessing an explicit memory, capabilities of time series modeling and/or anticipation. 

\item \textbf{Interaction-awareness:} A software system is interaction-aware if it has knowledge that stimuli and its own actions form part of interactions with other systems and the environment. It has knowledge via feedback loops that its actions can provoke, generate or cause specific reactions from the environment. It enables a software system to distinguish between other nodes of software systems and environments. Simple interaction-awareness may just enable a software system to reason about individual interactions. More advanced interaction-awareness may involve the possessing knowledge of social structures such as communities or network topology. In this work, from the pattern's perspective, we strictly treat interaction awareness with respect to the different nodes of software systems and/or the environment, and thus the internal information about interactions between different elements within a single software system is not considered as knowledge of interaction.

\item \textbf{Goal-awareness:} A software system is goal-aware if it has knowledge of current goals, objectives, preferences and constraints. It is important to note that there is a difference between a goal existing implicitly in the design of a software system, and it having knowledge of that goal in such a way that it can reason about it. The former does not describe goal-awareness; the latter does. Example implementations of such knowledge in a software system include state based goals (i.e. knowing what is a goal state and what is not) and utility based goals (i.e. having a utility or objective function). 

\item \textbf{Meta-self-awareness:} A software system is meta-self-aware if it has knowledge of its own capability(ies) of awareness and the degree of complexity with which the capabilities(ies) are exercised. Such awareness permits a software system to reason about the benefits and costs of maintaining a certain capability of awareness (and degree of complexity with which it exercises this capability). 

\end{itemize}

\section{Self-Awareness Architectural Patterns}

\label{sec:patterns}

While the notions of self-awareness can be well conceptualized with respect to a software system, the presence of various requirements would still need more concrete guideline on how those concepts can be modeled within the needs. This urges a formal documentation of the self-awareness as architectural patterns when engineering self-aware and self-adaptive software systems. An architectural pattern refers to an architectural problem-solution pair for a given domain, which in the context of self-aware software systems, means that they are linked to the capabilities of self-awareness. Our previously proposed self-awareness architectural patterns~\cite{2014epicshandbook,7185305,Chen2016:book,6827105} have been showing great potential in engineering self-aware and self-adaptive software systems. 

In such context, different capabilities of computational self-awareness enable capability of the systems to obtain and react upon certain knowledge, which could be either about its own states or about the environment. The patterns provide a formal way to ensure that only relevant capabilities of self-awareness are included, and their inclusion justified by identified benefits. There is no need for a system to become unnecessarily complex, learning and maintaining self-awareness capabilities which do nothing to advance the outcomes for that system, generating only overhead. Each of the self-awareness architectural patterns is decentralized by design. That is, structurally they resemble a peer-to-peer network of interconnecting self-aware nodes, varying only in the number of the capabilities and the type of interconnection between them. Even with the decentralized expression, a centralized software system can be easily modelled by considering only one node. In this section, we provide an overview of these well-defined patterns with selected examples.

\begin{figure}[!t]
  \centering 
   \includegraphics[width=0.7\columnwidth]{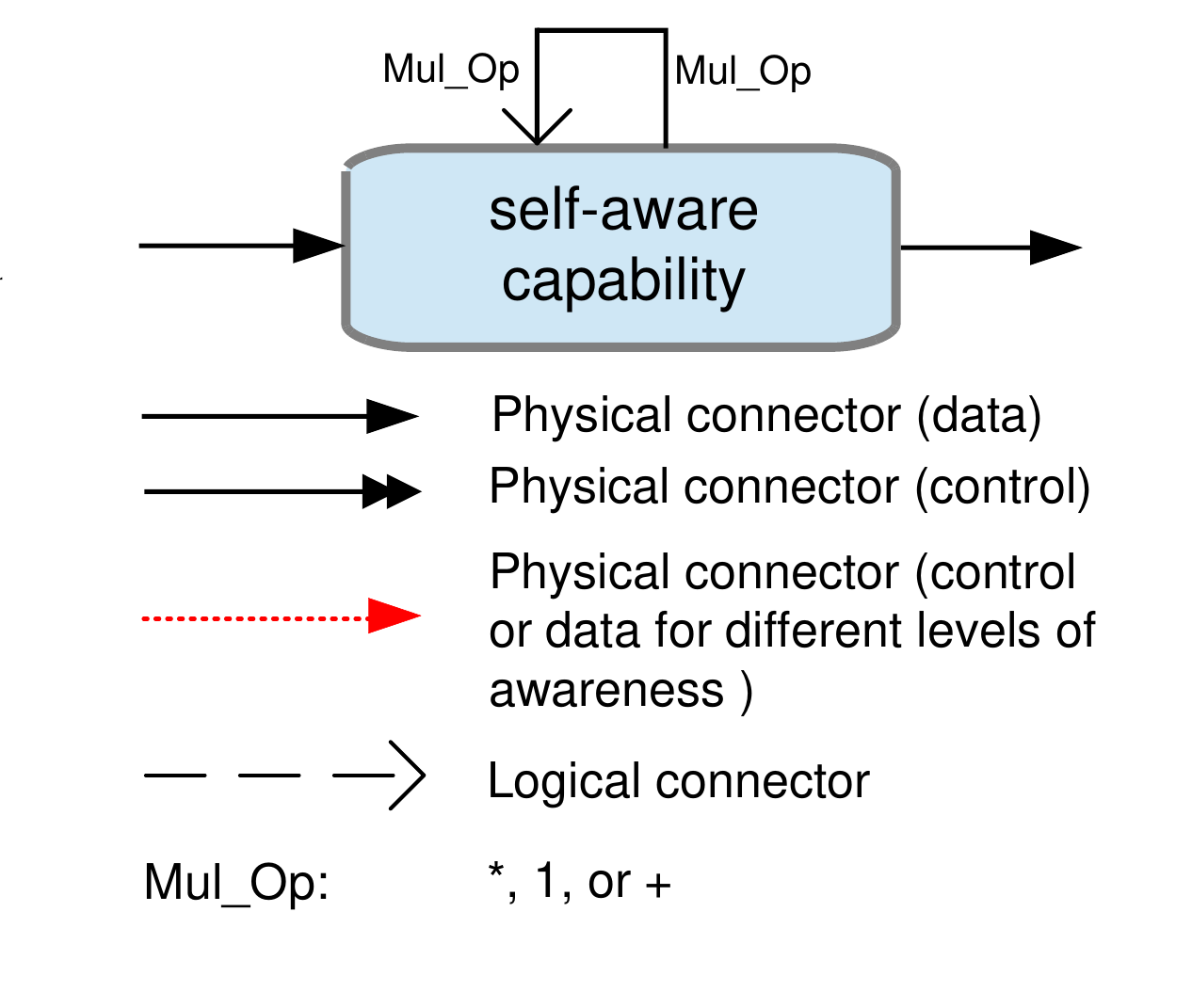}
    \caption{The basic notations for self-awareness architectural patterns.}
 \label{fig:notation}
  \end{figure}

\subsection{Notations}

In general, an architecture of software system consists of two fundamental elements, the component and connector, which are described as below~\cite{2014epicshandbook}:

\begin{itemize}

\item \textbf{Component:} A smaller and more manageable part of a software system, which is often divided based on requirements, functionality and purpose.

\item \textbf{Connector:} A bridge that represents the possible interaction between components and the multiplicity involved.

\end{itemize}

The uniqueness of the self-awareness architectural patterns is that, the component is replaced by the notion of capability of self-awareness, sensoring and actuating, in which case they are not necessarily to be a one-to-one mapping. In other words, depending on the context, two or more capabilities may be combined and realized in one component; or one capability can be implemented in separate components. The basic notations used to describe the patterns are depicted in Figure~\ref{fig:notation}.

In particular, the connectors are used to express the physical and logical interactions, which have different notations: 

\begin{itemize}

\item \textbf{Physical connector:} This means there is a direct interaction between the components, and each component is required to directly interact with the others. Notably, the  physical connectors are further divided into two types. The first type, expressed red arrow, particularly refers to the interactions for different capabilities of the self-awareness (e.g., goal and time-awareness); in contrast, the second type, denoted by black solid arrows, represents the interactions for the self-awareness of the same capability (e.g., the interaction-awareness from different interacting software systems)

\item \textbf{Logical connector:} This does not require direct interaction, but rather the data or control in the interaction is sent/received through the sensors and actuators, which have the physical connector. For instance, self-expression and self-adaptation might be logically required to reach consensus amongst different nodes, but such interaction is physically realized through \emph{Sensors} and \emph{Actuators}.

\end{itemize}

Note that the \emph{Sensors} and \emph{Actuators} can be either external or internal, where the former refers to the case that information/control is aimed for external nodes; while the latter means such data/control exchange only happen internally at the current node. The benefit of additionally introducing the logical connector is that, for example, when designing a capability of self-awareness where the communication protocol is not needed, the pattern can still illustrate that the software system needs to interact with the others. Thus, this provides the engineers with a more precise view on the architecture.

The multiplicity operators are used to represent how many concrete components (which may realize one or more capabilities of self-awareness), including those from different nodes of software systems, are involved in the interaction. In the self-awareness architectural patterns, there are three types of multiplicity operators (denoted as \emph{Mul\_Op}):

\begin{itemize}
\item \textbf{+} expresses that the number of components that realize the same capability in the interaction is restricted to at least one.
\item \textbf{1} indicates that one and only one component that realizes the same capability is permitted.
\item \textbf{*} indicates that zero, one or many components that realize the capability specified is permitted in the interaction
\end{itemize}

\subsection{The Patterns}

Drawing on the feasible combinations of the self-aware capabilities, we have previously documented eight well-defined patterns for engineering self-aware software systems~\cite{2014epicshandbook}. In a nutshell, these patterns are summarized in Table~\ref{table:patterns}. Noteworthily, the meta-self-awareness is considered as an optional capability, and thereby it is not explicitly coded into a particular given pattern. Each pattern was documented using standard pattern template~\cite{Buschmann:2007:POS:1215359} as follows.

\begin{table*}[t!]
\centering
   \caption{Self-Awareness Architectural Patterns.}
\label{table:patterns}

\begin{tabularx}{\textwidth}{p{4.5cm}p{5.5cm}X}
\toprule 
\textbf{Pattern}&\textbf{Self-Aware Capabilities}&\textbf{Characteristics}\\ \midrule
Basic Pattern&stimulus-awareness&For cases where some actions need to be triggered in order to cope with emergent events and stimuli \\
Basic Information Sharing Pattern&stimulus- and interaction-awareness&For cases where more
nodes may be required with loosely shared data to meet the scalability requirement of the system \\
Coordinated Decision-making Pattern&stimulus- and interaction-awareness (with additional interactions to external nodes)&For cases requiring consistent global decision making in a cooperative setting \\
Temporal Knowledge Sharing Pattern&stimulus-, interaction- and time-awareness&For cases where  timing of actions and availability of historical knowledge have an impact on the integrity of information sharing in the software system \\
Temporal Knowledge Aware Pattern&stimulus- and time-awareness&For cases where timing of actions and availability of historical knowledge is required only at the local level \\
Goal Sharing Pattern&stimulus-, interaction- and goal-awareness&For cases where goal reasoning and optimization is required with strong consensus \\
Temporal Goal Aware Pattern&stimulus-, time- and goal-awareness&For cases where timing of actions and availability of historical knowledge are required for local optimization and reasoning of goal \\
Fully Self-Aware Pattern&stimulus-, interaction-, time- and goal-awareness&For cases where timing and historical knowledge is required for performing goal reasoning with strong consensus \\

\bottomrule

\end{tabularx}
\end{table*}

%
%
%
\begin{itemize}
\item \textbf{Problem/Motivation:} A scenario where the pattern is applicable
\item \textbf{Solution:} A representation of the said pattern in a graphical form
\item \textbf{Consequences:} A narration of the outcome of applying the pattern
\item \textbf{Example:} Instance of the pattern in real applications or systems
\end{itemize}

We designed the patterns following the principles of architectural patterns:

\begin{tcolorbox}[breakable,left=5pt,right=5pt,top=5pt,bottom=5pt] 
\textit{An architectural pattern is a named collection of architectural design decisions that are applicable to a recurring design problem parameterized to account for different software development contexts in which that problem appears~\cite{paakki2000software}.}
\end{tcolorbox}

In other words, we provide a collection of architecture design decisions to realize the self-aware capability, parametrized by the level of knowledge available(which is behavioral in essence).  The provided description is generic, providing template solution to a recurring problems. Level of knowledge can range from stimuli,  time, goal, interaction and so the parameterization of the design decisions that invoke the self-aware capability. 

In fact, codifying different structures has been commonly used as the way to patternize software architecture when engineering self-aware and self-adaptive systems~\cite{weyns2013patterns}. Our ways of formulating and describing the pattens were inspired by Weyns et al.~\cite{weyns2013patterns}, who propose patterns for self-adaptive systems with special focus on their interactions in a decentralized manner. It is worth noting that the pattern can be instantiated, such that a capability may be decomposed into more than one actual components.

Indeed, the key differences of the patterns are what combination of self-aware capabilities is involved, but they also exhibits different forms of interactions and multiplicity. This is important, as the combination of capabilities cannot be done arbitrarily. For example, all the patterns would need stimulus-awareness; stimulus- and goal-awareness cannot be the only capabilities to form a pattern, as merely obtaining information about the stimulus does little help to reason about goals. There are also examples where the combination of self-aware capabilities are the same, but differ on how they interact with each others, e.g., the \emph{Basic Information Sharing Pattern} and \emph{Coordinated Decision-making Pattern}.

In the following, we elaborate on two patterns as examples, the more comprehensive specification can be found in our handbook~\cite{2014epicshandbook}.

\begin{figure}[!t]
  \centering 
   \includegraphics[width=\columnwidth]{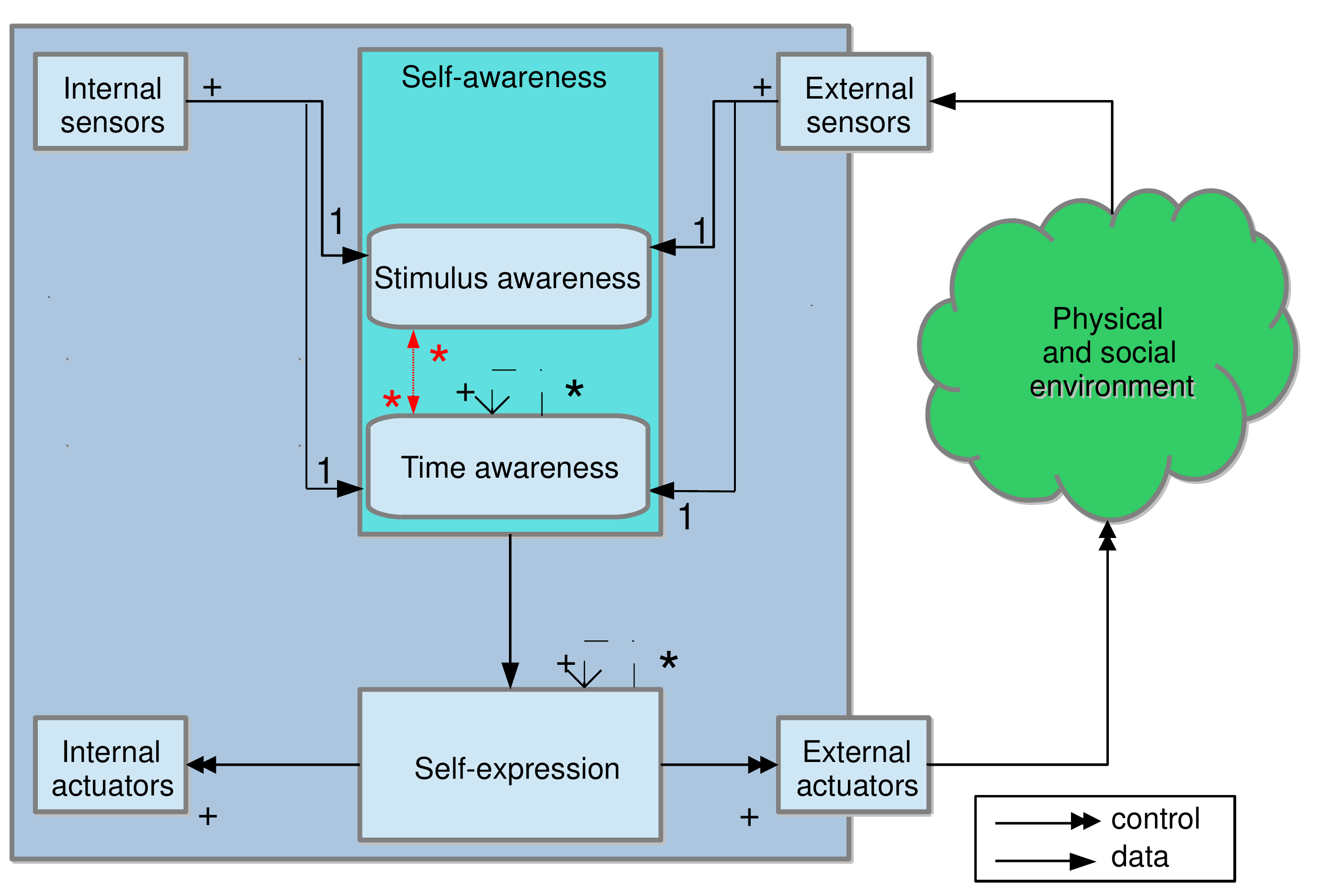}
    \caption{The temporal knowledge aware pattern.}
 \label{fig:p7}
  \end{figure}

\subsubsection{\underline{Temporal Knowledge Aware Pattern}}

\textbf{Problem/Motivation:} The knowledge of timing enables the capability of proactive adaptation and potentially, better adaptation quality. However, the other capability of awareness, e.g., interaction, might not be a necessity, therefore it could affect the self-aware system as it is suffering unnecessary overhead.

\textbf{Solution:} As shown in Figure~\ref{fig:p7}, in this pattern, the knowledge of timing enables the capability of proactive adaptation and potentially, better adaptation quality, which is specifically supported via time-awareness. The \emph{Temporal Knowledge Aware} pattern incorporates only time-awareness working in conjunction with stimulus awareness, which eliminate the unnecessary overhead introduced by the other capabilities of self-awareness, i.e., the goal, interaction and meta-self awareness may not be needed.

\textbf{Consequences:} When using this pattern, the key benefit is that the software system can be equipped with knowledge about historical data. The categories of data are vast, ranging from the internal states or the environment. However, this should not include data about the other nodes, as interactions has been omitted.  It should be noted that this pattern does not cater for changing goals and their related reasoning. That is, it assumes that the goal of the software system is known at design-time and statically encoded in the system, without the opportunity to modify and reason about at run-time.


\textbf{Example:} A concrete example of where this pattern is applied could be for the cloud environment where resource is sharing via Virtual Machine (VM) on each node of software system. In this context, by leveraging the historical usage of resources, time-series prediction would be able to predict the demand of VMs on a node of software system for the near future, which assists proactive provisioning of resource and potentially, prevents requirements violation and/or resource exhaustion.

\begin{figure}[!t]
  \centering 
   \includegraphics[width=\columnwidth]{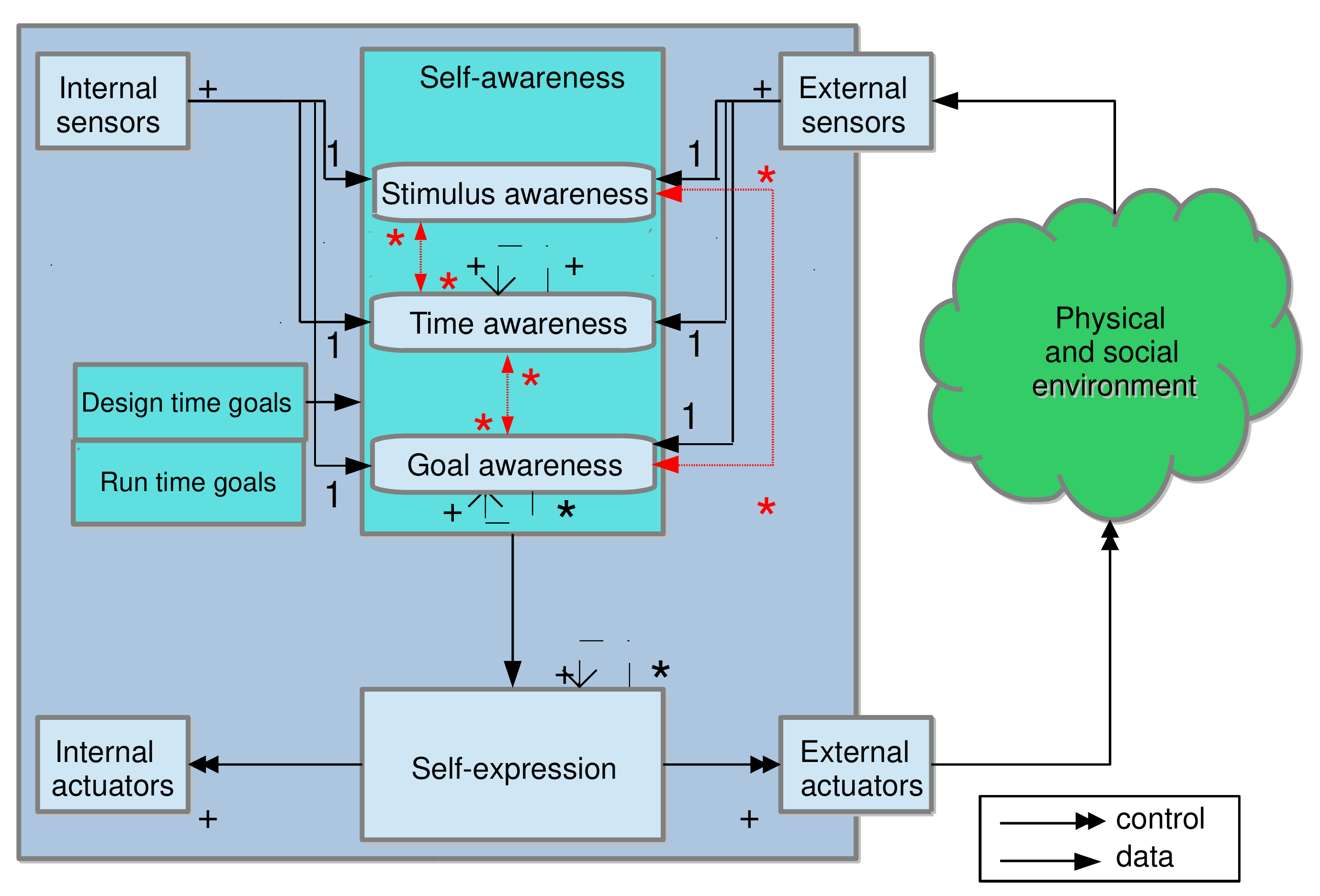}
    \caption{The temporal goal aware pattern.}
 \label{fig:p9}
  \end{figure}

\subsubsection{\underline{Temporal Goal Aware Pattern}}

\textbf{Problem/Motivation:} The knowledge of goals and time together might not necessarily to be shared amongst nodes, especially in cases where the optimization of local goals could lead to acceptable global optimum. 


\textbf{Solution:} As shown in Figure~\ref{fig:p9}, in the temporal goal aware pattern, the goal-awareness provides explicit capability to reason about and even modify the goal at runtime, which offers further guarantee on the optimality of certain goals. However, the knowledge of goals and time might not necessarily to be shared amongst different software systems, especially in cases where the optimization of local goals could lead to acceptable global optimum. Specifically, in this pattern there is no notion of `sharing' information as the software system is not aware of any interactions and, therefore, it does not aware of the presence of the other nodes. It is worth noting that the absence of interaction awareness does not mean there is no interaction\textemdash the software system and the environment could still interact with each other, but it merely does not aware of the details involved in the process.

\textbf{Consequences:} A key benefit of this pattern is that the knowledge of historical events can be used in conjunction with the ability to reason about goals. This often provides emergent adaptation behaviors~\cite{femosaa,seeding}. However, a major limitation is the removal of interaction awareness, especially when the goal-awareness is present, implies that different nodes of software systems could be in inconsistent state. The engineers should carefully verify that such situation would not result in violations of system requirements. In addition, the self-expression and self-adaptation on a software system could not use any information from others when making decisions.


\textbf{Example:} Example application domain of the pattern could be: for adaptive web application in a centralized mode, there is only a single software system exist, and thus no interaction is needed. Another more complex example is when orchestrating fully decentralized harmonic synchronization amongst different mobile devices, which requires each node of software system to aware of stimulus, time and goal but not necessarily interaction. In such case, each software system receives phase and frequency updates from the others or the environment, and reacts upon based on its own time and goal information. This is a typical example where there are occurrences of interaction, but no occurrences of interaction awareness; because a single software system only aware of the incoming phase and frequency updates but it has no knowledge of where they come from.

\begin{figure}[!t]
  \centering 
   \includegraphics[width=\columnwidth]{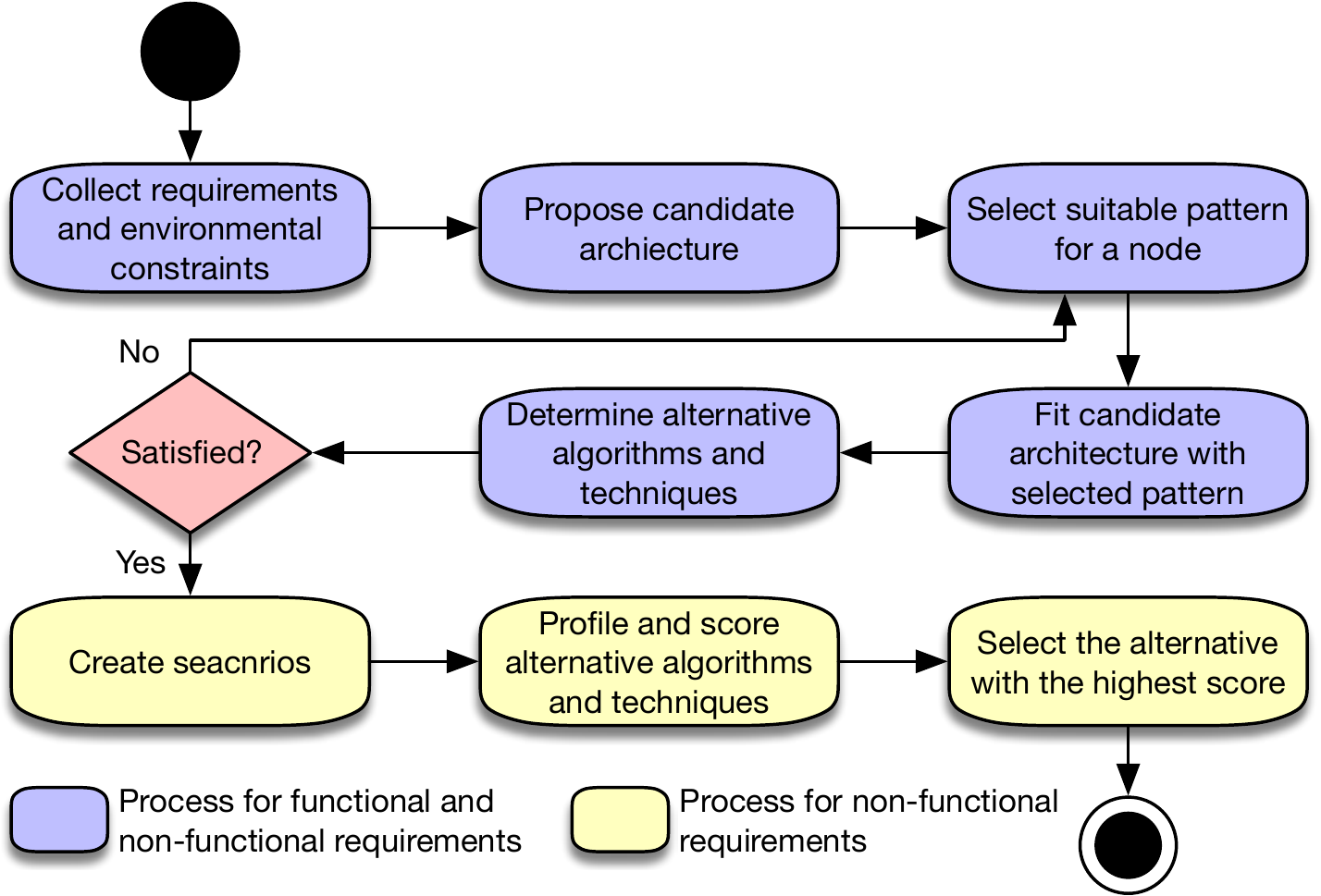}
    \caption{The guideline of selecting self-awareness architectural pattern and the underlying algorithms/techniques.}
 \label{fig:guid}
  \end{figure}

\subsection{Guideline on Selecting Patterns and Underlying Algorithms/Techniques}

In our handbook~\cite{2014epicshandbook}, we have codified a comprehensive guideline that assists the software engineers to select the self-awareness architectural patterns for a node, and the underlying algorithms/techniques\footnote{This may be a type of algorithms/techniques instead of a specific one.} that realize each capability. In a nutshell, the selection of patterns and algorithm/techniques follows the general processes of ATAM~\cite{paul2002evaluating}, which is a well-know methodology on design selection, such that the choice is made based on qualitative assessment and quantitative evaluation, supported by simulation and profiling.

Noteworthily, albeit that the patterns are alternative for a single node, different nodes can be based upon distinct patterns, or different instantiations of an identical pattern, under systems-of-system or distributed environment. Therefore, the patterns can be used in a composite manner. 

\begin{table*}[t!]
\centering
   \caption{Classification of the Representations on Domain Expertise with Possible Examples.}
\label{table:class}

\begin{tabularx}{\textwidth}{P{2.5cm}X}
\toprule 
\textbf{Category}&\textbf{Example Expertise Representations}\\ \midrule
Methodology& RUP~\cite{kruchten2004rational}, agile~\cite{martin2002agile}, SSADM~\cite{ashworth1988structured}, SCURM~\cite{schwaber1997scrum}, ...\\ 
Concept& technical debt~\cite{Cunningham:1992}, code smell~\cite{7194592}, software entropy~\cite{Jacobson:2004}, feature creep~\cite{1452719}, ...\\ 
Model&feature model~\cite{fm1990}, goal model~\cite{liu2004designing}, UML~\cite{booch2005unified}, Markov model~\cite{fine1998hierarchical}, Petri net~\cite{reisig1991petri}, queuing model~\cite{kendall1953stochastic}, I*~\cite{yu2009social}, viewpoints model~\cite{1702398}, design patterns~\cite{beck1987using}, ... \\ 
Documentation&SLA~\cite{keller2003wsla}, requirement documents~\cite{wiegers2013software}, user manual, configuration files, API documents, software and system specifications, ... \\ 
Program&source code of one (or more) programming language, library invocation and dependency, ... \\ 
Assumption&past problem instances and experiences, insights from peer and users discussions, ... \\

\bottomrule

\end{tabularx}
\end{table*}

As shown in Figure~\ref{fig:guid}, the overall guideline is an iterative process, in which the selection of pattern and the underlying algorithms/techniques can be continuously refined based on the profiling results. The final outcome for each node, after a satisfied number of iterations, would be the instantiation of a selected self-awareness architectural pattern with chosen underlying algorithms/techniques for self-awareness. Due to limited space, we advice interested readers to our handbook~\cite{2014epicshandbook} for detailed information.

\section{DBASES Foundations}
\label{sec:expertise}


\subsection{Representations of Engineers' Domain Expertise}
\label{sec:rep}
As mentioned in Section~\ref{sec:domain-expertise}, for domain information, it is important to distinguish between problem nature and domain expertise; the former is not necessarily equivalent to the latter. Domain expertise, particularly that from the software and system engineers, can be represented in various forms. For the simplicity of exposition, we use the following terminology to explain this concept in DBASES:


\begin{itemize}

\item \textbf{Expertise Representation:} Expertise representation is generally abstract, which can be further refined and customized for expressing the domain expertise that captures domain knowledge for a specific case. These are often the general skills and tools that are familiar to a software and system engineer. For example, feature model is a representation of the expertise, which is commonly used by software and system engineers. It can be applied to a wide range of application domains within each of which the representation would be specialized into a particular design instance.


\item \textbf{Category of Expertise Representation:} This refers to a group of expertise representations that share similar nature, e.g., feature model, UML models and goal model are all design models.

\end{itemize}


Clearly, an expertise representation can be specialized into different instances that share the same structure, rules and semantics, but each can capture/be tailored to handle different knowledge about the domain. Drawing on the recent survey about what expertise knowledge has been considered in practically engineering self-awareness~\cite{Chen:2018:STS,elhabbash2019self}, DBASES is underpinned by a classification, as shown in Table~\ref{table:class}, to categorize the most commonly used expertise representations when engineering software systems\footnote{The examples here do not intend to be exhaustive, but they serve as intuitive illustrations of the concepts.}. Each of the categories are explained as follows.



\textbf{Methodology:} This refers to the systematic specification and analysis methods that are applied to abstract the expertise and represent it to aid the development of software systems. An expertise representation can be considered in this category if all of the following criteria are met:

\begin{itemize}

\item[---] It covers all or nearly all the phases in engineering a software system.
\item[---] It contains specific methods, rules, postulates, procedures, or processes to manage a software or system project.
\item[---] It involves description about the roles of different stakeholders in the engineering process, e.g., analysts, designers and testers.

\end{itemize}

\textbf{Concept:} This includes the intents, drivers/forces and motivations that derive the knowledge/expertise capture and representation. An expertise representation stands as a \texttt{Concept} if all of the following criteria are met:

\begin{itemize}

\item[---] It represents an abstract idea or generic notion in mind that captures some common and justifiable phenomena of different instances in software and system engineering.
\item[---] It aims to describe an idea or notion in a ``plain" way that is intuitive and close to the general understanding of human.
\item[---] It is a widely recognized practice and truth in the engineering process.

\end{itemize}

\textbf{Model:} This involves the standard for abstracting the expertise; it can systematically capture at least certain aspects of a software system, which are mainly utilized during the analysis and design phases. An expertise representation belongs to \texttt{Model} if all of the following criteria are met:

\begin{itemize}

\item[---] It contains a formal notation or language to describe how knowledge about the software system can be captured.
\item[---] It can represent certain aspects of the software system and the relationships between them.
\item[---] It is a more formal way of representing concept(s).
\item[---] It is often illustrated in a graphical manner.

\end{itemize}

\textbf{Documentation:} This refers to artifacts that document and express the metadata for the representations of expertise, specifying scope, constrains, uses, anti-uses, etc., with an aim to be understandable for different stakeholders (e.g., end users, managers). An expertise representation belongs to \texttt{Documentation} if all of the following criteria are met:
\begin{itemize}

\item[---] It contains metadata provided on digital or analog media.
\item[---] It aims to illustrate data or represent agreement between parties for the software system.
\item[---] It is entirely (or mostly) based on ``plain" textual language of human.

\end{itemize}

\textbf{Program:} This involves the expertise representations that actually enable the software system to run. An expertise representation is related to \texttt{Program} if the following criterion is met:
\begin{itemize}

\item[---] It is related to the source code that enables the execution of the software system.

\end{itemize}

\textbf{Assumption:} This refers to the expertise representations that are directly derived from the subjective beliefs and experience of the software and system engineers, which may not be well-justifed. An expertise representation can be considered in this category if all of the following criteria are met:
\begin{itemize}

\item[---] It is a general belief about the software system derived from specific instances.
\item[---] It represents the sense of expectation on certain aspects of the software system, which is not guaranteed to be true.

\end{itemize}

The above classification in DBASES does not aim to be exhaustive, but they serve as a general guideline that covers majority of the cases, and thereby it can be flexibly extended. It may be possible that a given representations of expertise can fit more than one categories, in which case it is the engineer's decision on which one is more suitable. Similarly, it is also possible that a representation cannot be fitted into any category above. In such case, the representation can form an additional category (e.g., \texttt{Other} category), which can then be considered under the criteria of structurability and tangibility that we will elaborate below.

\subsection{Structurability and Tangibility}
\label{sec:sandt}

The expertise representation expresses knowledge can be a result of one's experience, which would be in various forms. Therefore, it is also important to understand whether these representations is structural and tangible, as studied in human cognition research~\cite{bradley2006analyzing}. In software and system engineering domain, it is not uncommon to see that more structural representations can be more beneficial~\cite{robillard1999role} and more tangible ones are better tools to express knowledge~\cite{andersen2002tangible}. Therefore, given the variety of different expertise representations, explicitly recognizing their category in terms of structurability and tangibility is important. A structural representation means that the organization of its information follows specific rules, or semantics; otherwise, it is said to be non-structural. Specifically, a given representations of expertise is structural if all of the following criteria can be satisfied:

\begin{itemize}

\item[---] Its organization and arrangement of the internal elements (and their relations) form some repeatable patterns.
\item[---] It can be specialized into as case dependent variants, which, although different, can still be derived from the same core.
\item[---] It contains explicit, step-by-step information about how itself can be `assembled'.

\end{itemize}

 
A tangible representation refers to the expertise representation that is perceptible by directly interacting and/or observing; or otherwise it is non-tangible. Again, a representation of expertise is tangible if all of the following criteria can be satisfied:

\begin{itemize}

\item[---] It can be directly seen or touched to understand the information it holds.
\item[---] It comes with a digital or analog media.

\end{itemize}


In Figure~\ref{fig:taxonomy}, we further taxonomize the aforementioned six categories of expertise representations depending on their nature with respect to the above criteria of structurability and tangibility, as part of DBASES. The taxonomy provides a more intuitive way for the engineers to understand how a category can be linked to these two properties. However, it is worth noting that any given expertise representation can be assigned using the above criteria.

 \begin{figure}[!t]
    \centering
   \begin{tikzpicture}[
box/.style={draw,rectangle,minimum size=3cm,text width=2.5cm,align=center}]
\matrix (conmat) [row sep=.1cm,column sep=.1cm] {
\node (tpos) [box,
    label=left:\( \textbf{structural} \),
    label=above:\( \textbf{tangible} \),
    ] {Model, \\ Program, \\ Documentation};
&
\node (fneg) [box,
    label=above:\textbf{non-tangible}] {Methodology};
\\
\node (fpos) [box,
    label=left:\( \textbf{non-structural} \)] {Documentation};
&
\node (tneg) [box] {Assumption, \\ Methodology,\\ Concept};
\\
};
\end{tikzpicture}
    \caption{Confusion matrix on the taxonomy of the expertise representations with respect to structurability and tangibility.}
        \label{fig:taxonomy}
  \end{figure}

\begin{table*}[t!]
\centering
   \caption{Possible Synergies between Expertise Representation and Self-Aware Capabilities.}
\label{table:transp}

\begin{tabularx}{\textwidth}{P{2.5cm}P{6.5cm}Y}
\toprule 
\textbf{Category}&\textbf{Example Expertise Representations}&\textbf{Self-Aware Capabilities}\\ \midrule
\multirow{3}{*}{Methodology}&SSADM~\cite{ashworth1988structured}&stimulus-, time-, interaction-, goal- and meta-self-awareness\\ 
&SCURM~\cite{schwaber1997scrum}&stimulus-, time-, interaction-, goal- and meta-self-awareness\\ 
&...&...\\
\midrule

\multirow{5}{*}{Concept}&technical debt~\cite{Cunningham:1992}&time- and goal-awareness\\ &code smell~\cite{7194592}&stimulus-, time- and goal-awareness\\ 
&software entropy~\cite{Jacobson:2004}&time- and goal-awareness\\ 
&feature creep~\cite{1452719}&stimulus-, time- and goal-awareness\\ 
&...&...\\\midrule

\multirow{8}{*}{Model}&feature model~\cite{fm1990}&stimulus-, time- and goal-awareness\\ 
&goal model~\cite{liu2004designing}&stimulus-, time- and goal-awareness\\
&UML~\cite{booch2005unified}&stimulus-, time-, interaction-, goal- and meta-self-awareness\\ 
&Petri net~\cite{reisig1991petri}&stimulus-, time-, interaction- and goal-awarness\\ 
&Markov model~\cite{fine1998hierarchical}&stimulus-, time-, interaction- and goal-awarness\\ 
&queuing model~\cite{kendall1953stochastic}&stimulus-, time- and goal-awareness\\ 
&design pattern~\cite{beck1987using}&stimulus- and goal-awareness\\ 
&...&...\\\midrule

\multirow{4}{*}{Documentation}&SLA~\cite{keller2003wsla}&stimulus-, time- and goal-awareness\\ 
&requirement documents~\cite{wiegers2013software}&stimulus-, time-, interaction-, goal- and meta-self-awareness\\ 
&API&stimulus- and goal-awareness\\ 
&...&...\\\midrule

\multirow{3}{*}{Program}&source code&stimulus-, time-, interaction- and goal-awareness\\ 
&library invocation and dependency&stimulus-, time- and goal-awareness\\ 
&...&...\\\midrule

\multirow{3}{*}{Assumption}&past problem instances and experiences&stimulus-, time-, interaction-, goal- and meta-self-awareness\\ 
&insights from peer and users discussions&stimulus-, time-, interaction-, goal- and meta-self-awareness\\
&...&...\\ 


\bottomrule

\end{tabularx}
\end{table*}
  
It is clear that expertise representations in the category of \texttt{Model} and \texttt{Program} are both structural and tangible, as they can easily meet all the criteria mentioned above. On the other extreme, representations in the category of \texttt{Assumption} and \texttt{Concept}, as the name suggests, are both non-structural and non-tangible. Because they cannot be derived from the same pattern, and are difficult to be seen or interacted with directly, which have failed to meet the criteria for being structural and tangible.

\texttt{Documentation} contains expertise representations that can be directly seen and comes with a media, thereby they are tangible, but could be structural or non-structural. For example, Service Level Agreement (SLA) and API documents also satisfy the three criteria of being structural. In contrast, requirement documents and user manuals are non-structural, whose content is documented by natural language without specific rules. Thereby they fail to meet the criterion that there are variants which can be derived from the same common ground.

The category of \texttt{Methodology} would have expertise representations that are non-tangible as they cannot be directly observed. Yet again, they could be structural or non-structural. For instance, SSADM is a rather structural methodology and it satisfies all three criteria. In contrast, SCRUM , which is a form of Agile methodology, does not contain explicit, step-by-step information about its internal structure due to the need of being flexible. Therefore, SCRUM is said to be non-structural.


%
%
%
%
%
%

\subsection{Relation Between Expertise Representations and Capabilities of Self-Awareness}

Expertise representations can be possibly synergized to inform, enrich and/or refine the capabilities of self-awareness depending on the domains, and guided by the specific design of expertise representations. Drawing on the work reviewed by recent survey on engineering self-awareness~\cite{Chen:2018:STS,elhabbash2019self} and our understandings form the EPiCS project\footnote{\url{http://epics.uni-paderborn.de/}}~\cite{epics}, in Table~\ref{table:transp}, we illustrate some examples of the possible synergies with respect to the categories presented previously.

Given the openness of certain categories of expertise representations (e.g., \texttt{Methodology} and \texttt{Assumption}), the domain expertise can be potentially synergized to benefit all the possible capabilities of self-awareness. For example, the SCURM methodology can help to better understand the engineering process of the algorithms that realize certain self-awareness. In addition, the methodology also covers the management between incremental development and operation phase. This, for instance, can assist the meta-self-awareness to collect suitable data about the applicability of other self-awareness as the software system runs, and thereby providing readily available information to be discussed again in the next phase of incremental development.

For other cases, on the other hand, domain expertise knowledge can be only useful to certain capabilities. For example, domain knowledge expressed using feature models would be useful for stimulus-, time- and goal- awareness, but can be of limited help for interaction and meta-self-awareness. This is because it neither expresses information on the interactions between nodes of software systems, nor provides foundations to reason about the needs of different self-awareness capabilities.



Another example is related to the goal modeling. In particular, goal modeling and its various refinements can be synergized with the benefit of goal-awareness. The aim, for example, is to dynamically analyze the satisfaction of that goal, areas and traces within the model that requires refinements and further elaboration to meet the goal. This can be supported by synergizing the goal model with the stimulus- and time-awareness which would enable better goal reasoning. However, the goal model itself does not often express information on interaction.

As mentioned, capturing and modeling the knowledge, expressed via domain expertise can take forms of structured or unstructured and tangible or non-tangible, which is heavily influenced by the available representations of domain expertise for the engineering of the self-aware and self-adaptive software system. Arguably, the structured and tangible expertise representations are often more systematic means and disciplined approaches, while unstructured and non-tangible ones can be naturally flexible for probing, learning and cross-fertilisation of expertise. In this regard, the structurability and tangibility can largely affect the design and maintenance difficulty of synergy, as we will discuss in Section~\ref{sec:diffculty-level}.



It is worth noting that the examples here are merely for guideline on the possible synergies in DBASES, they do not mean to restrict one to follow a specific synergy if both the expertise representation and the related capability of self-awareness are available. Whether a synergy is needed, as well as the level and form of such synergy (as we discuss in the following) are highly domain dependent.

\subsection{Levels of Domain Expertise Synergy}
\label{sec:leveloft}

Generally, the information possessed by an expertise representation can be synergized with a capability of self-awareness at different extents. However, given the complexity of expertise representation, as well as the underlying algorithms/techniques for self-awareness, the synergy of expertise with self-aware software system may required to be automatic depending on the level. 



In DBASES, we propose and distinguish four hierarchical levels of expertise synergy with a self-aware capability, which can be flexibly selected and reasoned about given the requirements. The hierarchical levels are derived from our experience on working with industry practitioners form the EPiCS project~\cite{epics}, together with the work in recent survey on engineering self-awareness~\cite{Chen:2018:STS,elhabbash2019self}. It is known that hierarchical analysis is highly beneficial for classifying concepts in engineering software systems, especially when dealing with requirements of the engineering problems~\cite{cheng2009goal}. Specifically, inspired by the work from Berry et al.~\cite{berry2005four}, we describe each level according to the aspects listed as below:

\begin{itemize}
\item \textbf{Motivation:} A scenario where the level is required
\item \textbf{Criteria:} A set of criteria classifies the synergy to a particular level.
\item \textbf{Description:} A general elaboration of the characteristics of the level
\item \textbf{Example:} An instance where the level has been used
\end{itemize}

The levels are structured in an incremental way, i.e., \emph{level 2} would retain all the properties of \emph{level 1} and \emph{level 0}.

\subsubsection{Level 0 of Synergy} 

\textbf{Motivation:} This is the level such that there is no actual domain expertise synergy, but could merely utilize the necessary information about the problem nature to achieve the most basic specialization of the AI algorithms. This is often the case when standard AI algorithms are directly applied. 

\textbf{Criteria:} Since this is the most basic level of synergy (i.e., no synergy at all), and thus there are no criteria for this level, as in essence, any realization of the self-awareness is at least \emph{level 0}.

\textbf{Description:} Here, the engineers may not (or only trivially) reason about the problem and thus there may be no expertise representations. The underlying algorithm and technique that realize a capability of self-awareness does not use any information derived from the domain expertise. At this level, the synergy is a manual process.


\textbf{Example:} Considering a distributed system, where there is a machine learning algorithm that learns what are the important nodes to be tuned, but if the nodes are simply taken from whatever nodes that are currently running, then here, information of the problem nature (the available nodes) is used in stimulus-awareness. However, there is a lack of human reasoning involved (thus no domain expertise). Therefore, in such case, we still have \emph{level 0} of domain expertise synergy.

\subsubsection{Level 1 of Synergy}  

\textbf{Motivation:} Apart from the problem nature, which is often naturally intuitive with the problems, software and system engineering involves many cases where the detailed information is not obvious, which can only be made available through the expertise of engineers together with various tools and methods.

\textbf{Criteria:} Specifically, the synergy is at \emph{level 1} if the following criterion is met:

\begin{itemize}

\item[---] The expertise representation is specialized through in-depth reasoning according to the software system to be built.

\end{itemize}

\textbf{Description:} This is the most common level where there is a limited synergy between domain expertise and self-awareness. Here, the engineers do reason about the problem and there are certain expertise representations. However, there is no, or only trivial, machine reasoning on the reasoned expertise representation that aims to extract more meaningful information for a capability of self-awareness (and the underlying algorithm/technique), which is the key step to sufficiently synergize the expertise.  At this level, the synergy can be either a manual or automatic process.


\textbf{Example:} For example, the produced feature model design is a representation of expertise after careful human reasoning, but if the goal-awareness simply embed all the features form the model to optimize, then it is clearly a \emph{level 1} of domain expertise synergy, as some information about the human reasoning is used (the features) while there is no further, non-trivial reasoning about the feature model itself.

\subsubsection{Level 2 of Synergy} 

\textbf{Motivation:} The expertise representation produced by extensive human reasoning is likely to be complicated and large, which may be an inevitable result for the software system that is built and evolved over years. In such case, the useful information contained in the expertise representation is blur and difficult to be used directly. 


\textbf{Criteria:} The synergy is at \emph{level 2} if all of the following criteria are met:

\begin{itemize}

\item[---] The expertise representation is specialized through in-depth human reasoning according to the software system to be built.

\item[---] There is a non-trivial automatic process that extracts information from the expertise representation for the software system.

\end{itemize}

\textbf{Description:} In this level, the engineers are required to reason about the problem and produce certain representations of their expertise. There is also a need of further automatic machine reasoning, which extracts and synergizes the useful information of the reasoned expertise representation with the underlying algorithm and techniques for realizing self-awareness. However, the underlying algorithms and techniques do not need to be aware of the information about the expertise; they may operate as if there is no such information.

\textbf{Example:} For example, an engineer may reason about and produce a feature model, then, the model would be further reasoned and extracted, such that the irrelevant features for optimizing the software system are ruled out in the capability of goal-awareness. However, from the perspective of the search algorithm, it does not aware that the given features to tune have been tailored by the experts' specialized knowledge; it would merely operate as if those features were selected arbitrarily.

\subsubsection{Level 3 of Synergy} 

\textbf{Motivation:} While most algorithm/techniques would work without changing their internal structure, it is often the case that when their internal components are tailored specifically with the extracted domain expertise, the expected results can be largely improved. Such a process is not essential, but desirable.

\textbf{Criteria:} In particular, the synergy is at \emph{level 3} if all of the following criteria are met:

\begin{itemize}

\item[---] The expertise representation is specialized through in-depth human reasoning according to the software system to be built.

\item[---] There is a non-trivial automatic process that extracts information from the expertise representation for the software system.

\item[---] The internal components of the algorithm are tailored, such that it can actively and directly exploit the information extracted from the expertise representation.

\end{itemize} 

\textbf{Description:} This is the highest level of domain expertise synergy. Here, both human reasoning and automatic machine reasoning on the representation of expertise are needed. In addition, the underlying algorithm and technique for realizing self-awareness need to be tailored in a way that they can be aware of the experts' specialized knowledge, and thus promote more explicit reactions and exploitation of the expertise. This often implies a non-trivial consolidation to the internal components of the algorithm and techniques, which would make them less general but being more specific to the given problem. 

\textbf{Example:} Considering a queuing model, which is analyzed and designed by the engineers, used to synergize with a tailored machine learning algorithm to offer better awareness of goal. In this case, the queen model has some parameters that can be tuned automatically. More importantly, the machine learning algorithm is aware of the expertise expressed in the model, such that the training and updating mechanism can be tailored by the queuing model, which will clearly influence the accuracy of learning.

It is worth noting that for all levels, self-awareness and self-adaptation are still achieved through the underlying algorithms and techniques, but their behaviors are guided by varying the amount of information about the engineers' domain expertise, as constrained by the corresponding level of domain expertise synergy. 

\subsection{Benefit Score on Synergy}

Generally, it is expected that a higher level of synergy would lead to better quality of self-awareness, and eventually better results of self-adaptation. This is because the underlying algorithm and technique can be guided, or even consolidated, with the information of domain expertise to fit with the domain problem better. To support quantitative reasoning on the potential benefit for different levels of synergy in DBASES, each level can be assigned a numeric score as below:

\begin{itemize}

\item \textbf{Level 0:} benefit score=1.25
\item \textbf{Level 1:} benefit score=1.5
\item \textbf{Level 2:} benefit score=1.75
\item \textbf{Level 3:} benefit score=2

\end{itemize} 

\noindent where the value are normalized into the range between 1 and 2 to assure numeric stability. Noteworthily, the scores are fairly flexible as they serve as indication on the relative rank between different levels. Hence, the above scores are default settings in DBASES where the margin between different levels are equivalent. It is however perfectly acceptable to ask the stakeholders and engineers to assign the relative benefits score depending on the needs, similar to what have been done in CBAM~\cite{DBLP:conf/icse/KazmanAK01}, as long as the ranking remains unchanged. For example, if one consider that \emph{level 3} is likely to obtain much higher benefits than the others, then one may assign the benefit scores from \emph{level 0} to \emph{level 3} as: 1.1, 1.2, 1.3 and 2, respectively.

Indeed, the proficiency would have a definite impact on the likely benefit, as immature expertise, e.g., that from a naive or inexperienced engineer, would likely to mislead the algorithms and techniques for self-awareness and self-adaptation. To reflect on this, within the methodology we introduce in Section~\ref{sec:method}, the engineers are asked to weight the proficiency on the expertise representation and the underlying algorithms for self-awareness, based on which a more informed-decision of the synergy can be made. 


\subsection{Difficulty Score on Designing Synergy}
\label{sec:diffculty-level}

In DBASES, the design of the synergizing domain expertise with a self-aware capability can be of either \emph{specific} or \emph{general} forms. In the \emph{specific} case, one needs to analyze and reason about a particular instance of expertise representation (e.g., a design of feature model), and synergize it with a specific algorithm/technique (or any algorithms/techniques of the same type) that realizes self-awareness and self-adaptation. In the \emph{general} case, the  synergy needs to operate on different instances of expertise representation, e.g., it works on any design instance of the feature model, and any algorithms/techniques of the same type. Undoubtedly, these forms do not applied on the \emph{level 0} of synergy.

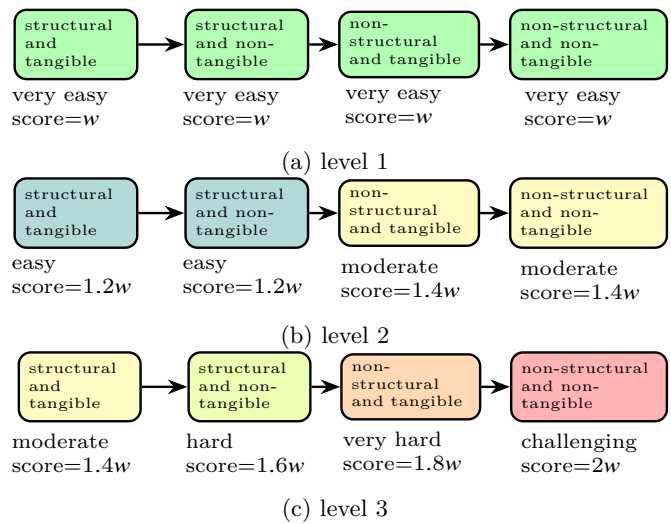
\begin{figure}
  \centering
  \begin{subfigure}[t]{\columnwidth}
 \begin{tikzpicture}
  [every label/.append style={font=\scriptsize}]
\begin{scope}[every node/.style={rectangle,rounded corners,thick,draw}]
    \node (A) at (0,0) [fill=green!30!white,font=\tiny,text width=1.2cm,label=below:{\parbox[c]{1.5cm}{very easy \\ score=$w$}}] {structural and tangible};
    \node (B) at (2,0) [fill=green!30!white,font=\tiny,text width=1.2cm,label=below:{\parbox[c]{1.5cm}{very easy \\ score=$w$}}] {structural and non-tangible};
    \node (C) at (3.9,0) [fill=green!30!white,font=\tiny,text width=1.4cm,label=below:{\parbox[c]{1.5cm}{very easy \\ score=$w$}}] {non-structural and tangible};
    \node (D) at (6,0) [fill=green!30!white,font=\tiny,text width=1.6cm,label=below:{\parbox[c]{1.5cm}{very easy \\ score=$w$}}] {non-structural and non-tangible};
\end{scope}

\begin{scope}[>={Stealth[black]},
              every node/.style={fill=white,circle},
              every edge/.style={draw=black, thick}]
              \path [->] (A) edge (B);
               \path [->] (B) edge (C);
                \path [->] (C) edge (D);
\end{scope}
\end{tikzpicture}
\subcaption{level 1}
    \end{subfigure}
  
    \begin{subfigure}[t]{\columnwidth}
 \begin{tikzpicture}
 [every label/.append style={font=\scriptsize}]
\begin{scope}[every node/.style={rectangle,rounded corners,thick,draw}]
    \node (A) at (0,0) [fill=teal!30!white,font=\tiny,text width=1.2cm,label=below:{\parbox[c]{1.5cm}{easy \\ score=$1.2w$}}] {structural and tangible};
    \node (B) at (2,0) [fill=teal!30!white,font=\tiny,text width=1.2cm,label=below:{\parbox[c]{1.5cm}{easy \\ score=$1.2w$}}] {structural and non-tangible};
    \node (C) at (3.9,0) [fill=yellow!30!white,font=\tiny,text width=1.4cm,label=below:{\parbox[c]{1.6cm}{moderate \\ score=$1.4w$}}] {non-structural and tangible};
    \node (D) at (6,0) [fill=yellow!30!white,font=\tiny,text width=1.6cm,label=below:{\parbox[c]{1.6cm}{moderate \\ score=$1.4w$}}] {non-structural and non-tangible};
\end{scope}

\begin{scope}[>={Stealth[black]},
              every node/.style={fill=white,circle},
              every edge/.style={draw=black, thick}]
              \path [->] (A) edge (B);
               \path [->] (B) edge (C);
                \path [->] (C) edge (D);
\end{scope}
\end{tikzpicture}
\subcaption{level 2}
    \end{subfigure}
    
      \begin{subfigure}[t]{\columnwidth}
 \begin{tikzpicture}
  [every label/.append style={font=\scriptsize}]
\begin{scope}[every node/.style={rectangle,rounded corners,thick,draw}]
    \node (A) at (0,0) [fill=yellow!30!white,font=\tiny,text width=1.2cm,label=below:{\parbox[c]{1.6cm}{moderate \\ score=$1.4w$}}] {structural and tangible};
    \node (B) at (2,0) [fill=lime!30!white,font=\tiny,text width=1.2cm,label=below:{\parbox[c]{1.5cm}{hard \\ score=$1.6w$}}] {structural and non-tangible};
    \node (C) at (3.9,0) [fill=orange!30!white,font=\tiny,text width=1.4cm,label=below:{\parbox[c]{1.6cm}{very hard \\ score=$1.8w$}}] {non-structural and tangible};
    \node (D) at (6,0) [fill=red!30!white,font=\tiny,text width=1.6cm,label=below:{\parbox[c]{1.6cm}{challenging \\ score=$2w$}}] {non-structural and non-tangible};
\end{scope}

\begin{scope}[>={Stealth[black]},
              every node/.style={fill=white,circle},
              every edge/.style={draw=black, thick}]
              \path [->] (A) edge (B);
               \path [->] (B) edge (C);
                \path [->] (C) edge (D);
\end{scope}
\end{tikzpicture}
\subcaption{level 3}
    \end{subfigure}

    \caption{Design difficulty and the related score of expertise synergy with respect to the levels, structurability and tangibility. ($w$ denotes the weight ($w\in [1,2]$) that distinguishes the difficulty between \emph{general} and \emph{specific} form of synergy for all cases. \emph{level 0} always has a difficulty score of 1)}
  \label{fig:difficulty}
  \end{figure}

It is clear that designing the \emph{general} synergy would impose greater difficulty than the \emph{specific} one, as wider range of the possible instances under the expertise representation needs to be considered. Here, the difficulty also serves as a general indicator of the cost in terms of labour, time and resource for both implementation and maintenance, therefore it is a crucial factor to consider when synergizing domain expertise. Within each of the two forms of synergies, the relative degrees of design difficulty varies depending on the levels of expertise synergy, as well as the structurability and tangibility of the expertise representation involved. Depending on different situations, the relative level of difficulty and the associated numeric scores have been illustrated in Figure~\ref{fig:difficulty}\footnote{The illustration shows only relative degrees of design difficulty, i.e., a `very easy' does not means it is easy in an absolute sense, but it is relatively easier comparing with the others. Similarly, a `very easy' in the \emph{general} synergy form is not equivalent to the `very easy' in the \emph{specific} synergy form.}. Note that the design difficulty for \emph{level 0} of synergy is constantly set as 1, i.e., they are at most as hard as \emph{level 1} synergy even considering different forms, since there is no actual synergy at all.


More specifically, in Figure~\ref{fig:difficulty}, the difficulty is ranked in a way that it is consistent with the general understanding. At \emph{level 1}, the synergy shares similar design difficulty regardless to the structurability and tangibility, because at this level the main difficulty is related to the human reasoning of the domain knowledge, which is part of the tasks that the engineers have to do regardless whether there is a synergy. In particular, the algorithms and techniques for realizing self-awareness and self-adaptation are directly exploited to the domain, rendering the actual synergy relatively straightforward. At \emph{level 2}, the synergy becomes more difficult in general. Particularly, the design difficulty becomes higher as the related expertise representation turns into a non-structural form, but remain unchanged with respect to the tangibility. This is because here, the underlying algorithms and techniques do not required to be aware of the domain expertise, thus the tangibility is less important. However, machine reasoning on the given expertise representation is necessary, therefore the domain expertise needs to be made structural for the automatic reasoning and synergy to take place. Such an extra processing of structuring could impose additional design difficulty. Finally, at \emph{level 3}, the expertise representation needs to be both structural and tangible, and thereby for expertise representation that belongs to the category of \texttt{Assumption} or \texttt{Concept}, additional efforts need to be conducted on both structuralization and tangibilization, rendering it as the most difficult case of synergy. Relatively, structuralization is more complex and difficult than tangibilization, as the former often requires in-depth and high proficiency on the expertise representation, while the latter, can be as simply as translating and documenting the concepts. 

Based on the ranking, each case is assigned a numeric score to add quantitative values in the design process. The scores have been normalized into the range between 1 and 2, which can be used directly in the methodology discussed in Section~\ref{sec:method}. $w$ is the normalized weight (between 1 and 2) that distinguishes the difficulty between \emph{general} and \emph{specific} form of synergy (e.g., 2 for \emph{general} and 1.5 for \emph{specific}) as provided by the engineers. Such a weight is applied to all possible synergy under consideration when engineering a self-aware and self-adaptive software system. 

Noteworthily, similar to the benefit scores, the default margin between the difficulty scores of different cases are almost identical. However, it is perfectly possible that if one consider a case to be more difficult than the others and hence amend the margin, as long as the ranking is preserved.

Indeed, the actual synergy approach is highly domain dependent, relying on the selected underlying algorithms/techniques for self-awareness, the expertise representation that is available and the other constraints as well as requirements. Nevertheless, given the information about the expertise representation and the expected level of domain expertise synergy, the degree of design difficulty offers the engineers with intuitive guidelines and information on the likely barriers, in addition to the likely benefits. This arises the opportunity for them to rethink and even refine the level of expertise synergy at the design stage, considering the trade-off between efforts and the expected quality. To demonstrate such in details, in Section~\ref{sec:case-study}, we elaborate examples of the synergy approaches within the contexts of three diverse case studies.

\section{Enriched Self-Awareness Patterns in DBASES}
\label{sec:transposition}

We now illustrate how the notions of domain expertise representations and their synergies in DBASES's foundation can be embedded with the capabilities of self-awareness, which are collectively expressed using the self-awareness patterns.

\begin{figure}[!t]
  \centering 
   \includegraphics[width=\columnwidth]{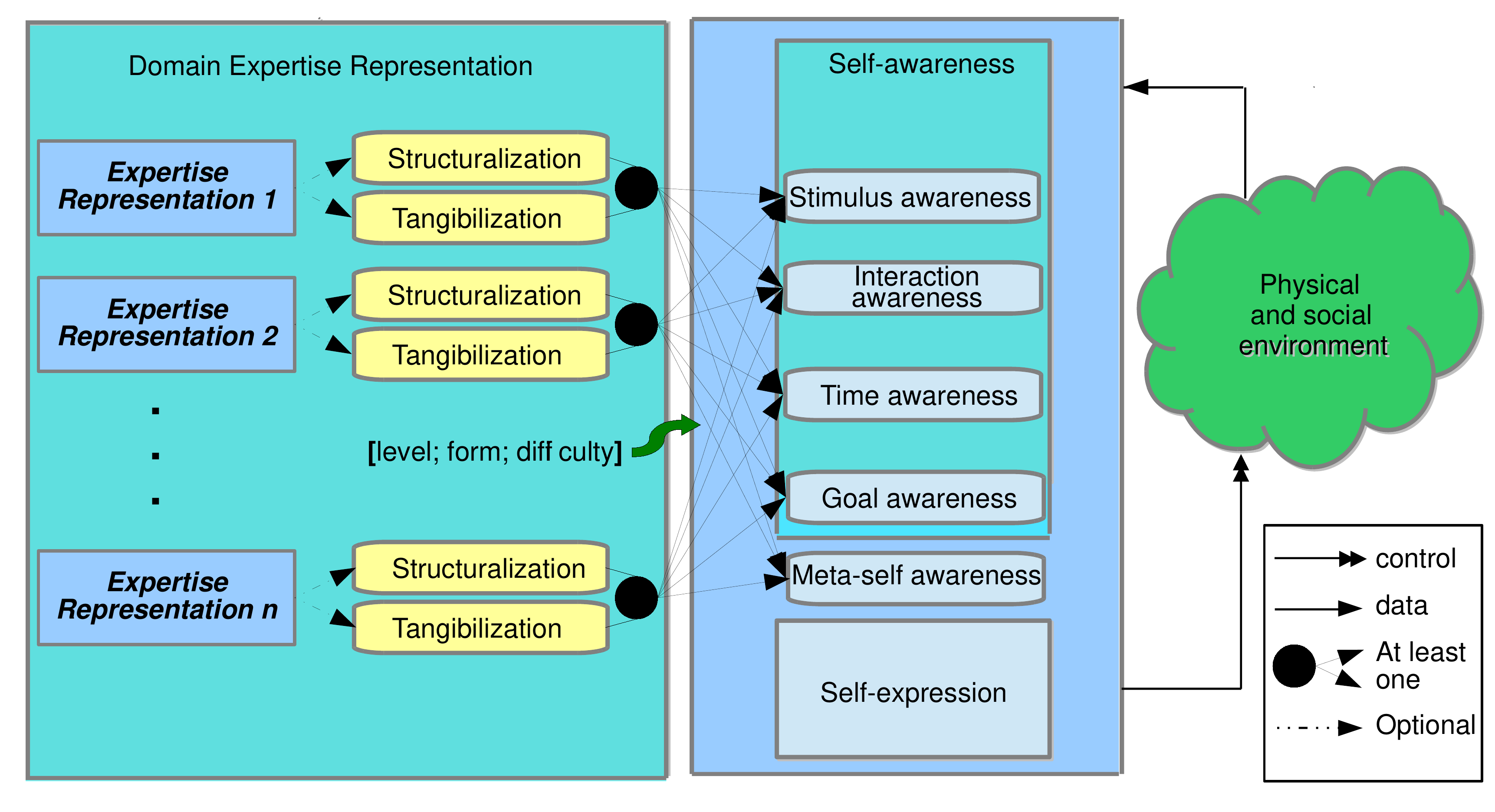}
    \caption{The possible capabilities of self-awareness in the self-awareness architectural patterns with explicit synergy between domain expertise and self-awareness.}
 \label{fig:gen-transp}
  \end{figure}

\subsection{Capabilities of Self-Awareness in the Patterns with Explicit Domain Expertise}
\label{sec:new-pa-exp}

The proposed self-awareness architectural patterns, as discussed in Section~\ref{sec:patterns}, can be enriched based on the proposed synergy framework in Section~\ref{sec:expertise}. Figure~\ref{fig:gen-transp} shows the general capabilities of self-awareness, which underpins the self-awareness architectural patterns, with explicit links to different expertise representations. Such a general enrichment can be instantiated into diverse instances, depending on the available expertise representation, the selected pattern and the required synergy. Clearly, for a particular domain, there can be more than one expertise representation (from the same or different categories), but only one specialized instance of an expertise representation exists at a time. Those expertise representations, depending on their categories, may or may not undergo structuralization and tangibilization. Importantly, an expertise representation needs to be synergized with at least one capability of self-awareness (e.g, time, goal) and its underlying algorithm/technique. On the other hand, there is no cap on the maximum number of self-awareness capabilities that it can synergize with; it is possible that an expertise representation may be synergized with all the capabilities of self-awareness. According to Figure~\ref{fig:difficulty}, each synergy expresses the expected level involved, as well as the form and the design difficulty, which are separated by semicolon. 

Noteworthily, it is important to distinguish \emph{level 0} of synergy and no domain information is required. The former has no synergy but the information of the problem nature may still be used. The latter refers to no information is used in a self-aware capability at all. With the enriched pattern, \emph{level 0} is still expressed, but without showing the selection of form and the difficulty level.  This becomes much more intuitive when instantiating the enriched self-awareness pattern with explicit domain expertise, which we elaborate in the following section.


\begin{figure}[!t]
  \centering 
   \includegraphics[width=\columnwidth]{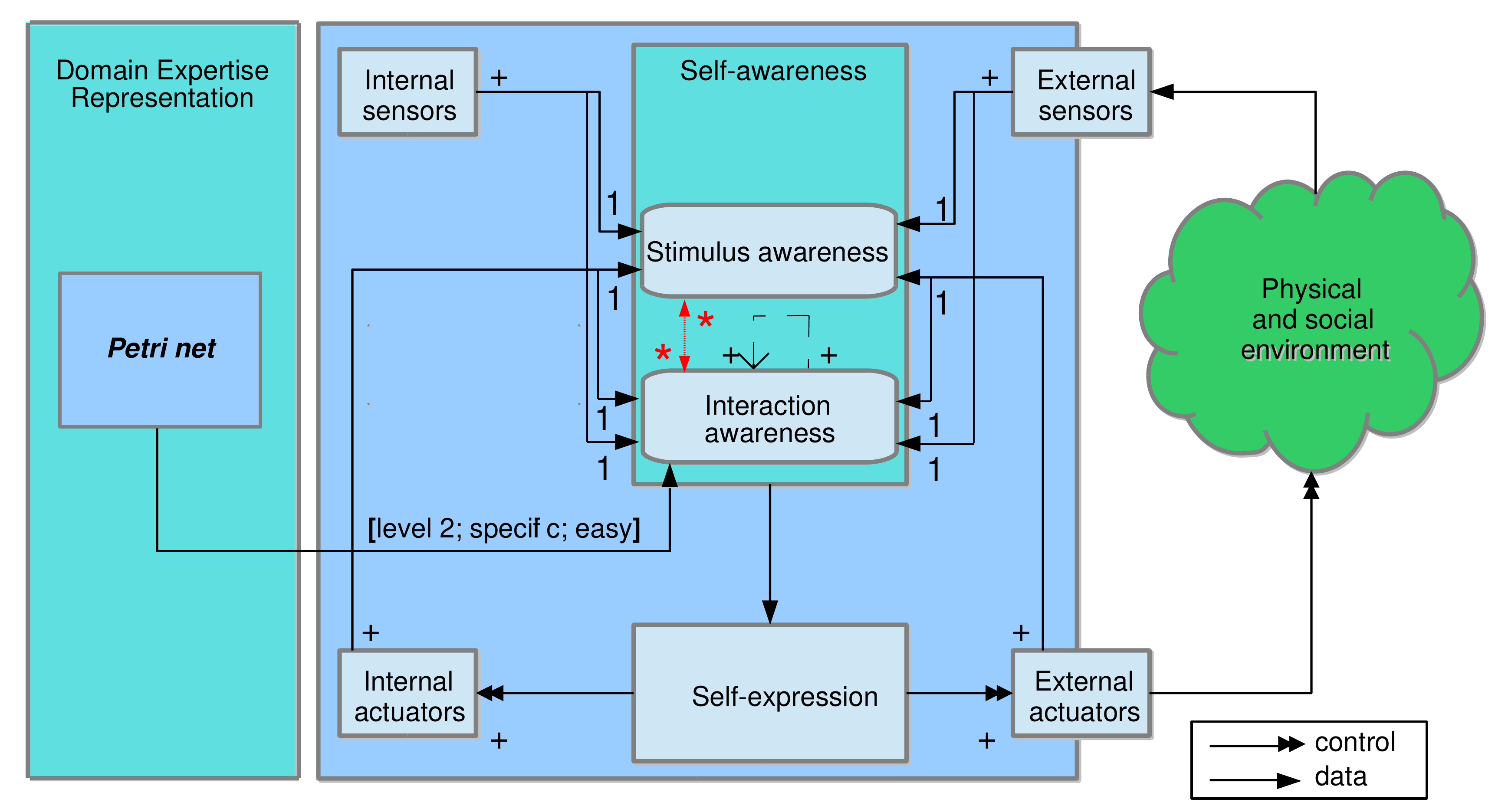}
    \caption{Instantiating \emph{Information Sharing Pattern} with synergy between Petri net and self-awareness.}
 \label{fig:inf-s}
  \end{figure}

\subsection{Examples of Instantiating the Patterns with Explicit Domain Expertise}

In Figure~\ref{fig:inf-s}, we illustrate an example where the \emph{Information Sharing Pattern} and the related algorithms and techniques have been chosen. Then, following the general pattern from Figure~\ref{fig:gen-transp}, the \emph{Information Sharing Pattern} can be instantiated with explicit domain expertise and the related synergies in different ways, among which Figure~\ref{fig:inf-s} is one candidate. In this example, the expertise representation is a design of the Petri net that contains rich domain expertise about the concurrency and transitions between conditions etc. This is particularly useful for the interaction-awareness and the underlying algorithm/technique, which enables a \emph{level 2} synergy between domain expertise and self-awareness. Specifically, the actual synergy can vary, for example, suppose a machine learning algorithm underpins the interaction-awareness to learn the likely under-utilized node for assigning more workloads. Here, the designed Petri net provides strong domain expertise about the features (conditions), which can be further parsed automatically to form a more relevant set of features. Finally, the resulted feature set is learned by the machine learning algorithm. This is clearly a \emph{level 2} of synergy, as there are both human and machine reasoning on the expertise representation, yet the machine learning algorithm itself does not know the fact that the given feature set was derived from domain expertise. There is no link between a design of the Petri net and stimulus awareness, which means no information has ever been used for stimulus.

  \begin{figure}[!t]
  \centering 
   \includegraphics[width=\columnwidth]{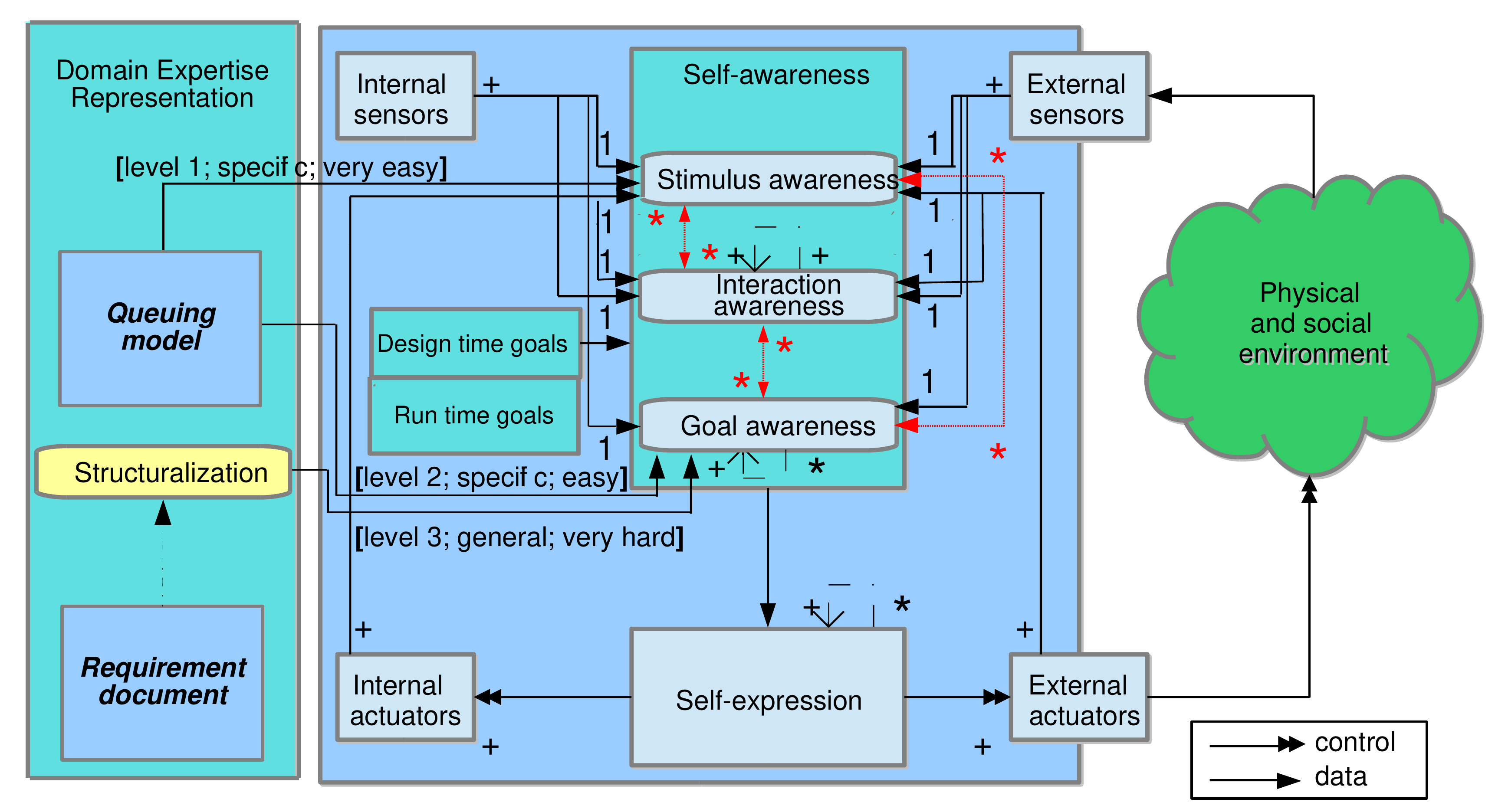}
    \caption{Instantiating \emph{Goal Sharing Pattern} with synergies of queuing model and requirement document.}
 \label{fig:g-s}
  \end{figure}

As expressed in the figure, the form of synergy is specific, which means only a design of Petri net needs to be synergized with the capability of self-awareness. Given that the Petri net belongs to the category of \texttt{Model} which is both structural and tangible (as seen from Figure~\ref{fig:taxonomy}), there is no additional structuralization and tangibilization. The design difficulty of synergy is therefore `easy' according to Figure~\ref{fig:difficulty}. In contrast, if the expected level of domain expertise synergy was \emph{level 1} or \emph{level 3}, then the design difficulty would becomes `very easy' or `moderate', respectively.

Figure~\ref{fig:g-s} shows a slightly more complicated example, in which the \emph{Goal Sharing Pattern} and the related algorithms and techniques have been selected. In this example, the \emph{Goal Sharing Pattern} can be instantiated with two aspects of domain expertise that are of different expertise representations and from distinct categories. Again, there could be different ways of synergies depending on the form and level, within which Figure~\ref{fig:g-s} illustrates only one candidate. Specifically, the design of queuing model is clearly a type of model while the requirement document belongs to the \texttt{Documentation} category. There are three synergies of domain expertise, each of which belongs to a different level. At the simplest form, the queuing model can create a \emph{level 1} of synergy with stimulus-awareness. This can be, for example, the feature components of the model serves directly as the detection points of any stimulus from the software systems, and therefore no extra reasoning and analysis conducted on the produced queuing model. Another synergy is between the queuing model and the goal-awareness, which can be of \emph{level 2}. Here, certain parameters in the designed queuing model may be changed dynamically, either by a deterministic or machine learning algorithm. The tailored model, in turn, acts as the function to evaluate an adaptation solution within a search algorithm that optimizes toward the optimality of a goal. Such extra reasoning conducted on the queuing model has promoted the synergy to \emph{level 2}.

In this example, the requirements document requires a relatively more complex, \emph{level 3} synergy with the goal-awareness. For example, the negotiated requirement document may be further analyzed using techniques for natural language processing, then, the results are synergized with the internal structure of a search algorithm, e.g., to form tailored operators. In this way, the synergized expertise is fully awared by the underlying algorithm that realizes goal-awareness, which can explicitly react to the knowledge of expertise. This is aligned with the criteria of \emph{level 3}. Again, those missing links between an expertise representation and a capability of self-awareness implies that there is no information to be used at all.

In this case, the queuing model can be linked with \emph{specific} form of synergies while the requirements document requires the \emph{general} form, in which case any given formats and designs of the requirements document needs to be synergized with the self-awareness capability, and thus it is relatively harder. The relative design difficulty for all three synergies can be distinguishable using Figure~\ref{fig:taxonomy} and~\ref{fig:difficulty}. A queuing model is both structural and tangible, and thus no extra processes are needed, therefore the \emph{level 1} synergy has a design difficulty of `very easy' while the \emph{level 2} one is classified as `easy'. The synergy related to the requirements document is more complex, as it belongs to the \texttt{Documentation} category and it is tangible but non-structural. As a result, given the required synergy of \emph{level 3}, the relative design difficulty is `very hard'. Note that since the requirements document requires \emph{general} forms for its two synergies, they are likely to be more difficult than the \emph{specific} one for the queuing model.

\section{Methodological Analysis in DBASES}
\label{sec:method}

Drawing on the aforementioned notions and enriched patterns, in this section, we codify a detailed methodology, as part of DBASES, that can assist the quantitative design on the synergy when engineering self-aware and self-adaptive software systems. In a nutshell, the methodology contains the steps below, whose details will be explained in the following sections:

\begin{enumerate}[start=1,label={Step~\arabic*:},leftmargin=5\parindent]

\item \textbf{\underline{Patterns and Algorithms.}} Selecting patterns and algorithms.
\item \textbf{\underline{Representations of Expertise.}} Determining the available representations of expertise.
\item \textbf{\underline{Candidates Creation.}} Creating design candidates by instantiating the selected enriched pattern with synergy between domain expertise and self-awareness.
\item \textbf{\underline{Difficulty and Benefit Scores.}} Calculating the overall difficulty and benefit scores for all the candidate synergies of expertise under the chosen pattern.
\item \textbf{\underline{Further Investigation.}} Selecting the suitable candidate(s) for further investigation.

\end{enumerate}

\subsection{Patterns and Algorithms}

The first step is to determine which is the suitable architectural pattern for self-awareness and the underlying algorithms/techniques\footnote{One may only need to decide the type of algorithms, rather than a specific one.} that realize the self-aware capabilities. As mentioned in Section~\ref{sec:patterns}, we have proposed a handbook, together with a comprehensive guideline to guide the engineer to make such selections.  A more thorough explanation and case studies can be found in the handbook~\cite{2014epicshandbook}.

\subsection{Representations of Expertise}

The actual representation of domain expertise is highly depending on the case, and thus their diversity can vary. However, arguably any given software and system engineering would require at least one formal representation of expertise. In this step, we ask the engineers to create a list of all available representation of the expertise based on their existing knowledge, some of which could be taken from the examples in Table~\ref{table:class}.

\subsection{Candidates Creation}

According to the available representations of expertise identified in step 2, this step aims to answer the following questions for each of these representations:

\begin{enumerate}

\item Which category does the expertise representation belong to? (using the criteria in Section~\ref{sec:rep})
\item If such a representation structural? is it tangible? (using the criteria or classification in Section~\ref{sec:sandt})
\item The expertise representation can be synergized with which algorithm/technique that realizes the self-aware capability? What are the possible levels of synergy? (using the criteria in Section~\ref{sec:leveloft})
\item What is the possible form for each synergy?
\item What is the difficulty level for each synergy? (using the Figure~\ref{fig:difficulty})

\end{enumerate}

Noteworthily, the different synergies of expertise representations and their combinations form the possible alternative instantiations of the enrichment for the selected pattern, as shown in Section~\ref{sec:transposition}. In this way, step 3 aims to create a candidate set of instantiations for the enriched patterns with information about all possible ways of synergies. For example, suppose that there are two expertise representations and the chosen pattern is \emph{Information Sharing Pattern}, which has two self-aware capabilities. If both representations need to be synergized with all self-aware capabilities while the synergy can be at all levels and under both forms, then considering all possible combinations, the outcomes of step 3 would be $2\times 4^{4}=512$ candidates. The final selection would be made based on the quantitative scores on both the difficulty and benefits for all the alternative candidates.

\subsection{Difficulty and Benefit Scores}

In this step, we aim to visualize the difficulty and benefits score for all the candidates identified from step 4 using the synergy framework. In particular, the overall difficulty of a candidate $C_n$ that has a total of $n$ synergies is calculated as:
\begin{equation}
\varmathbb{D}_{C_n}= \sum_{i=1}^{n}{{d_i} \over p_i}
\end{equation}

\noindent whereby $d_i$ is the original difficulty score for synergizing the corresponding expertise representation in the $i$th synergy. As mentioned in Section~\ref{sec:diffculty-level}, the original difficulty score has been pre-defined according to the structurability and tangibility of the representation. The $w$ is a normalized weight given by the engineers and it is applicable to all other synergies. $p_i$ is the proficiency on the $i$th synergy (normalized between 1 and 2), which covers both the expertise representation and the underlying algorithms/techniques that realize the corresponding self-aware capability. The higher proficiency, the less difficulty for achieving the synergy.

The overall expected benefit of a candidate $C_n$ can be computed as:
\begin{gather}
\varmathbb{B}_{C_n}=
\begin{cases}
\sum_{i=1}^{n}p_i \times b_i & \text{\emph{at level 0}} \\ 
\sum_{i=1}^{n}w \times p_i \times b_i & \text{\emph{otherwise}}
\end{cases}
\end{gather}

\noindent where $b_i$ is the original expected benefit score for the $i$th synergy, as discussed in Section~\ref{sec:leveloft}. Again, $w$ and $p_i$ is the actual form (i.e., \emph{general} or \emph{specific}) of the synergy and the proficiency, respectively. The higher proficiency, the larger the expected benefit.

As mentioned, the value of  $w$ and $p_i$ are entirely depends on the domain, and therefore it is difficult to draw any general guidelines. However, their relative settings can be discussed case by case. For example, the relative $w$ between \emph{general} or \emph{specific} form of synergy may be small for structural and tangible expertise representation, as it is more straightforward to generalize it from specific cases; in some situations, the $w$ may be identical as the two forms may not differ too much, such as queuing model. In contrast, for non-structural and/or non-tangible expertise representations, their margin of $w$ can be amplified. As regards to $p$, the category of domain expertise representation and the selected algorithms, together with the engineer's own experience, can provide indication about how its value for different synergies can be relatively set. For example, an engineer who works on software variability management and machine learning algorithms for years would likely to rank a high proficiency for the synergy between feature model and learning algorithm, but for other synergies, the proficiency can be given a relatively low value.

 \begin{figure}[!t]
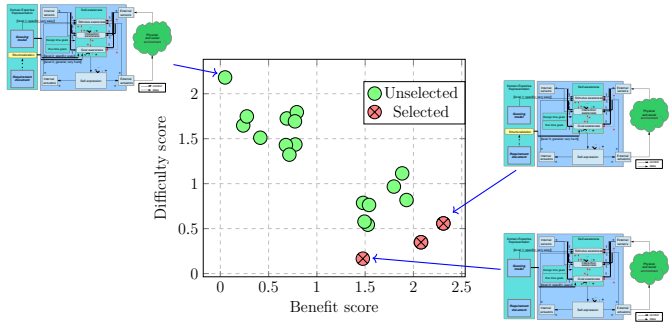

    \centering
   \includestandalone[width=\columnwidth]{tikz/exp-decision}
   \caption{Example of visualizing the difficulty and benefit scores for all candidates.}
        \label{fig:exp-dec}
  \end{figure}

\subsection{Further Investigation}

As we can see from the example of benefit/difficult plot in Figure~\ref{fig:exp-dec}, each candidate is an instantiation of the selected and enriched patterns with a particular way of synergizing domain expertise. While some of the candidates are clearly dominated by the others, there can be a trade-off between the difficulty and the expected benefit.

Indeed, to physically validate whether the achieved benefits and incurred difficulty (in terms of both implementation and maintenance) by the candidates are truly acceptable, it is an ideal case that if all the candidates can be subject to further investigation and profiling, i.e., the actual implementation, profiling and evaluation. However, given the time/resource constraint in real-world software and system development scenario, it is often the case that only a handful of them can be prototyped~\cite{DBLP:conf/icse/KazmanAK01}. This is in fact what we seek to provide with the engineers: an intuitive and principled guideline to extract the candidates for further investigation. In DBASES, the intuitive visualization of benefit/difficult plot provides the necessary foundation for the engineers to select only the most desirable ones, ruling out those that are clearly unneeded and thus saving the valuable human efforts in investigating them. As an example, Figure~\ref{fig:exp-dec} shows three selected candidates for further investigation.

Noteworthily, despite that there may be more than one way to implement the prototype of a particular candidate, it is generally possible to use a representative in the comparison process during further investigation, as what has been done in the domain of architecture profiling~\cite{DBLP:conf/icse/KazmanAK01}.

After further investigation, the final selection for production would inevitably involve not only the engineers, but also other stakeholders of the software systems. However, the methodology in DBASES, supported by the framework about synergy between domain expertise and self-awareness, their levels of difficulty and the enriched patterns, have enabled a more intuitive and quantitative visualization of all the possible alternatives in the trade-off. This, in turn, provides better informed decision making when synergizing domain expertise with the self-awareness in software systems. 



\section{Case Studies: Practical Applications of DBASES Framework}
\label{sec:case-study}

In this section, we illustrate the practical applications of DBASES framework on three tutorial case studies, each of which is recent research effort that seeks to engineer self-awareness into different types of software systems. Those studies are the collaborations in a team of researchers and engineers from China and UK, under the funding grants from their research councils.

As part of the methodology in DBASES, for all case studies, a set of desirable and representative candidates was selected for further investigation. This includes the actual prototype implementation of these candidates, deploying the resulted self-aware and self-adaptive system(s), running them and measuring their behaviors according to various quality indicators with real-world benchmarks. Eventually, the most promising one with the verified results would be chosen for production.

Each case study covers a type of software systems that may be applied to different scenarios, e.g., a highly constrained software system may run as a web-based systems or service-based systems. Therefore, for the candidates that are subject to further investigations, their prototypes were run on one or more scenarios. All the quantitative experiments are done in real environments, using the actual software system that fit under a given scenario. The data and source code used for all three case studies are publicly available on GitHub\footnote{\url{https://github.com/taochen/ssase/tree/master/experiments-data}}. 


\subsection{Self-Awareness and Self-Adaptation for Highly Constrained Software Systems}

\textbf{Context:} Self-adaptive software systems often have several non-functional quality attributes ( e.g, latency and throughput), which are difficult to manage due to the changing environment, such as workload. Those software systems are centralized, but structurally complex, i.e., there is a large number of features and complex dependency constraints. A typical example could be the multi-layered web applications, in which the actual software is often rely on a stack of third party libraries and frameworks, each of which own different adaptable features that can interplay together to influence the behaviors of the entire software system.

\textbf{Problem:} The aim of the first case study is, at runtime, to achieve more effective multi-objective optimization on the non-functional qualities of software systems. Clearly, in such context, self-awareness offers stronger capability for a software system to conduct more informed optimization and reasoning.

\textbf{Challenges:} The challenges here are two-folds: (i) it is difficult to effectively and systematically convert the design of self-adaptive systems, expressed as a feature model, to the context of a search algorithm while considering the right encoding of features in the solution representation. This is even more complex in the presence of feature dependency constraints, e.g., the cache size can only be adapted when the cache feature has been `turned on', or, the size of a thread pool needs to be equal or greater than the number of spare threads in the pool. (ii) Optimizing multiple conflicting objectives and managing their trade-offs are complex and challenging in self-adaptive systems, especially at run time. This is attributed to the huge number of alternative adaptation solutions that can vary with their quality for the said requirements. Moreover, the dynamic and uncertain nature of self-adaptive systems further complicates the conflicting relations between objectives, rendering the trade-off surface difficult to explore.


\subsubsection{Patterns and Algorithms}

After analyzing the requirements and following the handbook~\cite{2014epicshandbook}, it has been identified that there is no need to have knowledge about the interactions. This is because the target software system was not aimed for distributed environment, and that it is considered as satisfactory to optimize the local goal for a single self-adaptive system. Further, the environment is not expected to actively react on the adaptation of the software system, and thus no interaction between it and the environment. There is also no need for a meta-self-awareness, because the extra overhead on reasoning about the different capabilities of self-awareness is unnecessary, as the requirements on the capabilities are clear. In contrast, goal-awareness is the essential part as it permits capability to reason about goal and search toward an optimal (or near-optimal) solution. Time-awareness is also important in the modeling of goal, which consolidates the capability to thoroughly evaluate, and even predict, the effectiveness of a solution during the optimization process. As a result, these have led to the conclusion that the \emph{Temporal Goal Aware Pattern} is the most appropriate pattern for the design. The pattern has been illustrated in Figure~\ref{fig:p9}.

The primary goal is to optimize non-functional quality, and thus a vast of search algorithms are available. However, there may be an explosion of the search space for the self-adaptive system, which renders the problem as intractable. Further, it is difficult, if not possible, to obtain a precise understanding on the nature of the optimization problem beforehand and there are often multiple conflicting quality to be optimized. Drawing on these and as guided by the handbook~\cite{2014epicshandbook}, it has been concluded that the metaheuristic algorithms, particularly the evolutionary algorithms, are promising to realize the capability of goal-awareness in the software systems. However, given the wide range of possible domains, it is expected that the solution does not tie to a specific evolutionary algorithm, rather, it should support a diverse set of evolutionary algorithms. In addition, machine learning algorithms and other modeling techniques can be used to support the knowledge of time, which form the objective model that is essential in the reasoning of goal. Finally, stimulus-awareness, which is the simplest capability of self-awareness, can be realized by periodic detection. A complete list of the algorithms and techniques that realize the capabilities of self-awareness involved are show in Table~\ref{table:1st-alg}.

\begin{table}[t!]

   \caption{The algorithms and techniques that realize the capabilities of self-awareness for the first case study.}
\label{table:1st-alg}
\centering
\begin{tabularx}{0.9\columnwidth}{p{2.5cm}X}
\toprule 
\textbf{Self-Awareness}&\textbf{Algorithms and Techniques}\\ \midrule
stimulus-awareness&periodic detection \\
time-awareness&machine learning/analytical model \\
goal-awareness&evolutionary algorithm \\
\bottomrule

\end{tabularx}
\end{table}

\subsubsection{Representations of Expertise}

\begin{figure}[!t]
  \centering 
   \includegraphics[width=0.6\columnwidth]{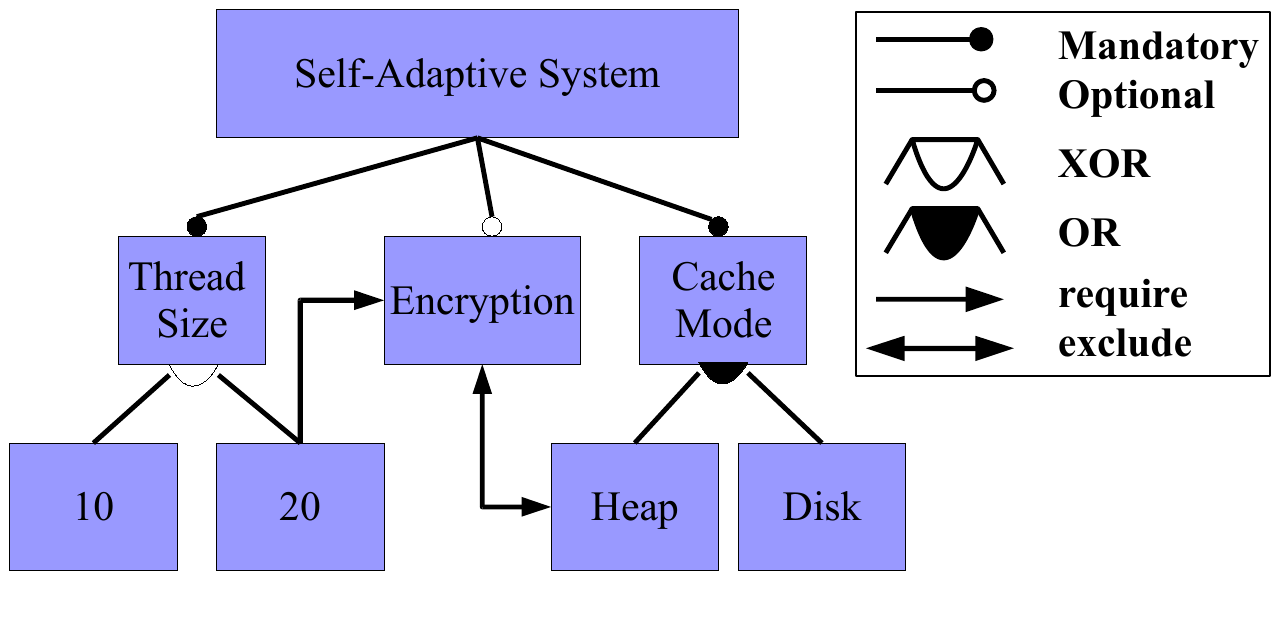}
    \caption{A feature model.}
 \label{fig:fm}
  \end{figure}

In this case, the only available representation of expertise is the feature model~\cite{czarnecki2004staged}, which is expressed as the tree structure. Such a model is widely used for software and system engineers to represent the functional variability of a software. In the context of self-adaptive software systems, the inherited concept of a feature model allows it to define the extent to which the software system is able to adapt at runtime (i.e., a range of variations that the software system can achieve). In particular, there is no definite constraint about the level that the feature model can cover, i.e., the features define the prominent or distinctive aspects between different variations of a software system~\cite{fm1990}, which range from high-level architectural elements (an entire component) to low-level configurations (a specific parameter). Figure~\ref{fig:fm} shows an example of the feature model, where there are four in-branch dependencies and two cross-branch dependencies:

\begin{itemize}
\item \textsc{Optional} refers to the feature that might be `turned off'.
\item \textsc{Mandatory} denotes core features that cannot be `turned off'.
\item \textsc{XOR} represents the feature in a group such that exactly one group member can be `turned on'.
\item \textsc{OR} means a group in which at least one group member needs to be `turned on'.
\item $F_i$ \textsc{require} $F_j$  means the former can only be `turned on' if the latter is `turned on'.
\item $F_i$ \textsc{exclude} $F_j$ denotes two features that are symmetrically mutually exclusive.
\end{itemize}

  \subsubsection{Candidates Creation}

At this step, all the possible ways of synergy can be created by instantiating the enriched self-awareness architectural pattern. In particular, answers to the questions presented in Section~\ref{sec:method} are shown as follows:

\begin{enumerate}

\item Which category does the expertise representation belong to? 
\begin{itemize}
\item[---] \textbf{\emph{Answer:}} Feature model belongs to the \texttt{Model} category.
\end{itemize}
\item If such a representation structural? is it tangible?
\begin{itemize}
\item[---] \textbf{\emph{Answer:}} It is both structural and tangible.
\end{itemize}
\item The expertise representation can be synergized with which algorithm/technique that realizes the self-aware capability? What are the possible levels of synergy?
\begin{itemize}
\item[---] \textbf{\emph{Answer:}} It needs to be synergized with all three self-aware capabilities in the enriched \emph{Temporal Goal Aware Pattern}. However, the synergy can only be at \emph{level 1} to the stimulus awareness but \emph{level 1} and \emph{level 2} are allowed for time awareness. For goal awareness, all levels except \emph{level 0} are possible, but \emph{level 2} and \emph{level 3} would required the synergy with time awareness to be at  \emph{level 2} .
\end{itemize}
\item What is the possible form for each synergy?
\begin{itemize}
\item[---] \textbf{\emph{Answer:}} Only the synergy with goal awareness can be of both \emph{specific} or \emph{general} form. The others are to be realized in a \emph{general} form.
\end{itemize}
\item What is the difficulty level for each synergy?
\begin{itemize}
\item[---] \textbf{\emph{Answer:}} According to Figure~\ref{fig:difficulty}, the difficulty level ranges between \emph{very easy} to \emph{moderate}.
\end{itemize}

\end{enumerate}

The above answers have led to six different candidates of synergizing domain expertise represented as the enriched \emph{Temporal Goal Aware Pattern}. 


 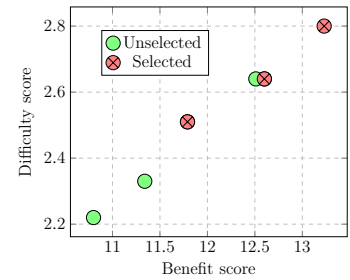
\begin{figure}[!t]
    \centering
   \begin{tikzpicture}
    \begin{axis}[
        grid = major,
        grid style={dashed},
       ylabel=Difficulty score,
        xlabel=Benefit score,
           label style={font=\fontsize{12}{12}\selectfont},
                    tick label style={font=\fontsize{12}{12}\selectfont},
      legend style={font=\fontsize{12}{12}\selectfont,at={(0.5,0.8)}, anchor=east,legend columns=1}]

 \addplot[only marks,mark=*,mark options={scale=2.5,fill=green!50}] plot coordinates  {
(10.8,2.22)
(11.34,2.33)
(11.79,2.51)
(12.6,2.64)
(12.51,2.64)
 };
 
  \addplot[only marks,mark=otimes*,mark options={scale=2.5,fill=red!50}] plot coordinates  {
(13.23,2.8)
(12.6,2.64)
(11.79,2.51)
  };

\addlegendentry{Unselected}
\addlegendentry{Selected}
    \end{axis}

    \end{tikzpicture}
    \caption{Difficulty and benefit scores for all candidates in the first case study.}
        \label{fig:exp-d1}
  \end{figure}

\subsubsection{Difficulty and Benefit Scores}

For all the six candidates, their overall scores with respect to both the difficulty and benefit are illustrated in Figure~\ref{fig:exp-d1}. In particular, the $w$ between \emph{specific} and \emph{general} form of synergy is set as 1.2 and 1.4, respectively, the proficiency is set as 1.8 for all synergies in a candidate.

\begin{figure}[!t]
  \centering 
    \begin{subfigure}[t]{\columnwidth}
   \includegraphics[width=\columnwidth]{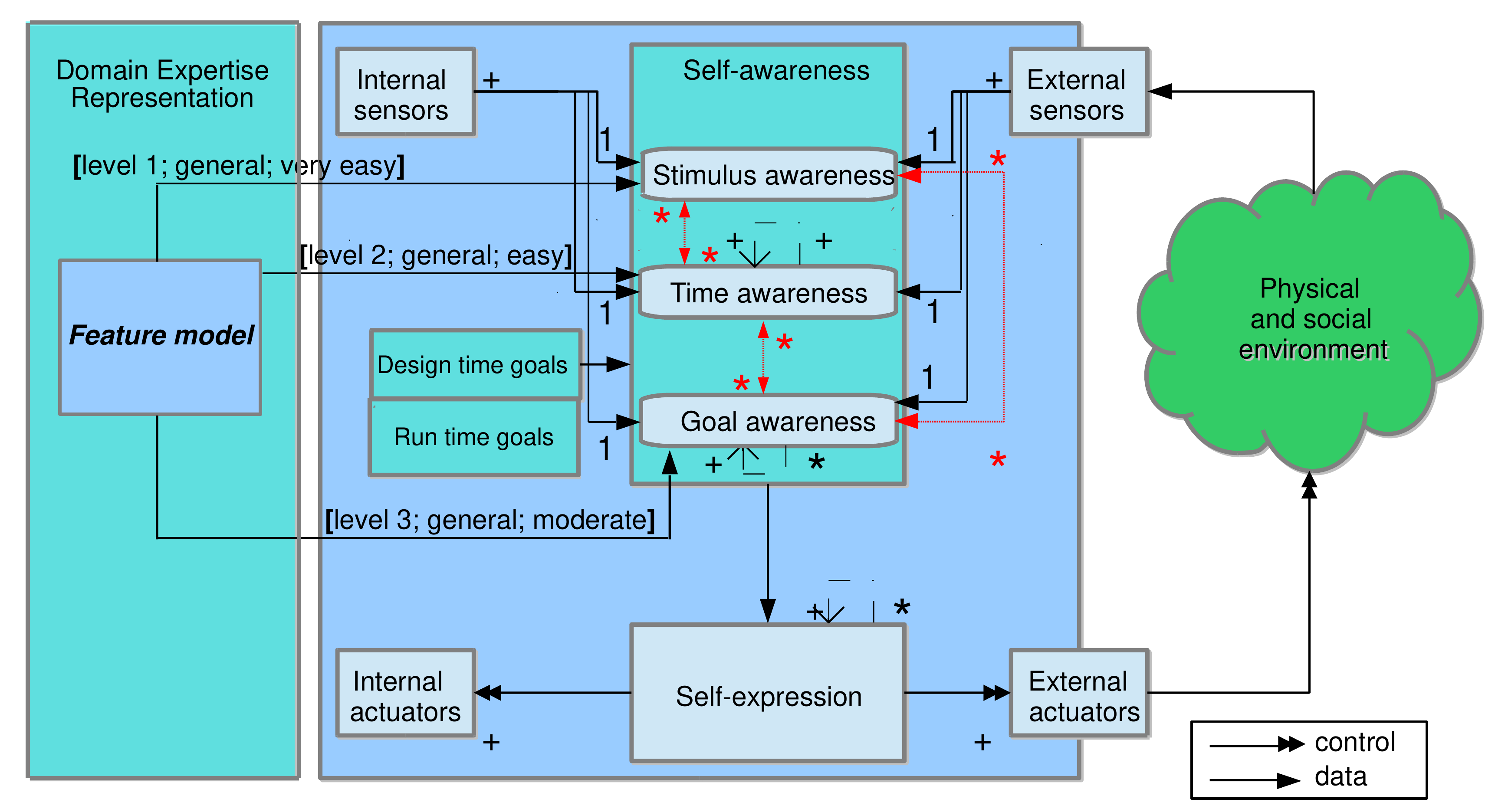}
\subcaption{candidate $C_1$}
 \label{fig:1st-arch}
    \end{subfigure}
    
        \begin{subfigure}[t]{\columnwidth}
   \includegraphics[width=\columnwidth]{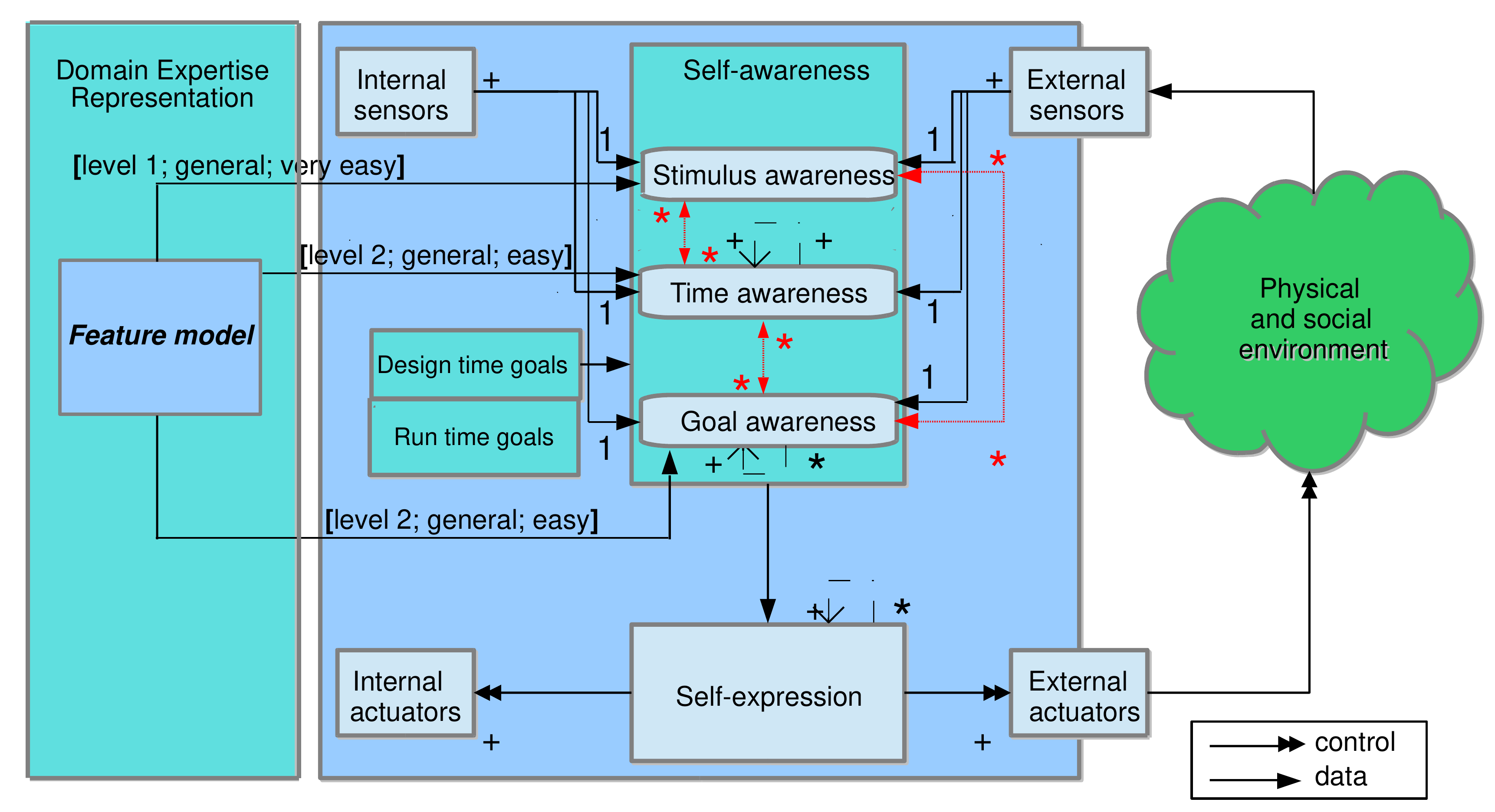}
\subcaption{candidate $C_2$}
 \label{fig:1st-arch-2}
    \end{subfigure}

        \begin{subfigure}[t]{\columnwidth}
   \includegraphics[width=\columnwidth]{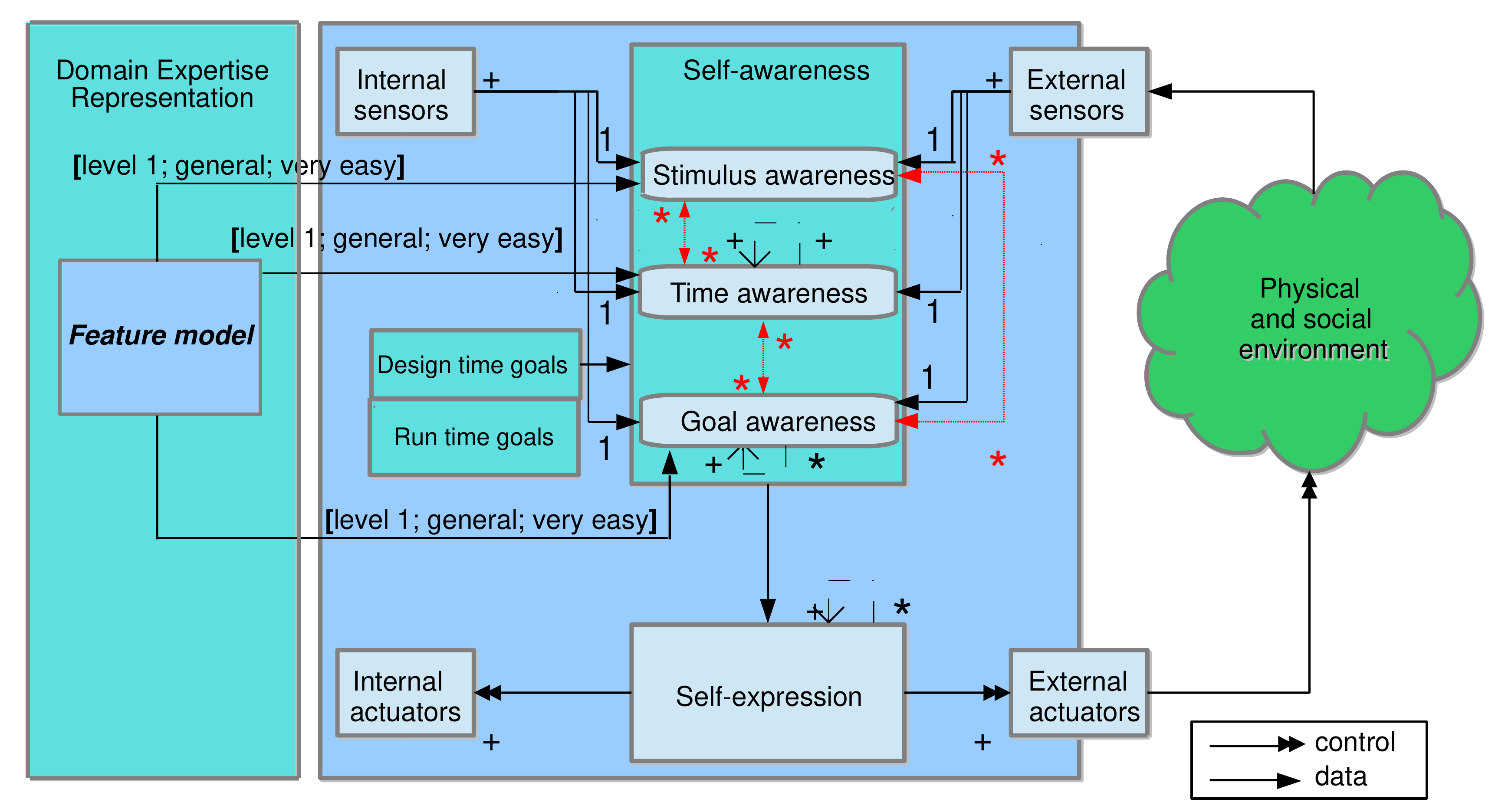}
\subcaption{candidate $C_3$}
 \label{fig:1st-arch-3}
    \end{subfigure}
    
    \caption{Possible candidates selected for further investigation in the first case study.}

  \end{figure}

\subsubsection{Further Investigation}

As shown in Figure~\ref{fig:exp-d1}, after discussions, three candidates have been selected for further investigation, as they are either desirable or serve as representatives for the others. Briefly, each of the candidates is specified as below:

\begin{itemize}
\item $C_1$ ($\varmathbb{B}_{C_1}=13.23$, $\varmathbb{D}_{C_1}=2.8$): As shown in Figure~\ref{fig:1st-arch}, the candidate automatically extracts only important features to create synergies in time- and goal-awareness, at \emph{level 2} and \emph{level 3}, respectively. In particular, dependency constraints are also injected and synergized with the evolutionary algorithms that underpins the goal-awareness. No machine reasoning is required for stimulus-awareness, which senses directly on the features at \emph{level 1} of synergy.
 
\item $C_2$ ($\varmathbb{B}_{C_2}=12.6$, $\varmathbb{D}_{C_2}=2.64$): The candidate, illustrated in Figure~\ref{fig:1st-arch-2}, achieves the synergy of the feature model with goal-awareness at \emph{level 2} (\emph{general} form), such that the evolutionary algorithm does not aware of the dependency constraints; all the other synergies remain the same as that of $C_1$.
 
\item $C_3$ ($\varmathbb{B}_{C_3}=11.79$, $\varmathbb{D}_{C_3}=2.51$): Figure~\ref{fig:1st-arch-3} illustrates the candidate in which the feature model is synergized with time- and goal-awareness at \emph{level 1} (\emph{general} form), i.e., all possible features are selected to be tuned without further parsing of the feature model, and no dependency constraint is captured by the evolutionary algorithm. All the other synergies remain the same as that of $C_1$.
\end{itemize}

More technical details on the actual synergy approaches can be found in~\cite{femosaa}.

\begin{table}[t!]

   \caption{The real subject software systems for the experiments of the first case study.}
\label{table:1st-exp}
\begin{threeparttable}
 
\begin{tabularx}{\columnwidth}{P{0.8cm}P{1.5cm}YYYY}
\toprule 
&\textbf{Objective}&\textbf{\#Feat.}&\textbf{\#Dep.}&\textbf{Env.}&\textbf{Space}\\ \midrule
\textsc{RUBiS}&latency; power&1,151&89,736&workload&1.3$\times 10^{16}$\\ \midrule
\textsc{SOA}&throughput; cost&221&255&services&5.6$\times 10^{18}$\\
\bottomrule

\end{tabularx}
\begin{tablenotes}
        \item[*] Feat. denotes features; Dep. denotes dependencies; Env. denotes environment; Space denotes search space.
    \end{tablenotes}
\end{threeparttable}
\end{table}

\subsubsection{Further Investigation Setup}

Since all the synergies are in \emph{general} form, the candidates are evaluated on two different real subject software systems, namely \textsc{RUBiS}~\cite{rubis} (a web system) and \textsc{SOA}~\cite{wada2012e3} (a service system), under two distinct categories of evolutionary algorithms, i.e., Non-dominated Sorted Genetic Algorithm II~\cite{nsgaii} (NSGA-II) and Indicator-Based Evolutionary Algorithm~\cite{ZitzlerK04} (IBEA), for realizing goal-awareness that optimizes different conflicting quality objectives. The details of the two subject software systems can be found in Table~\ref{table:1st-exp}. Given that the optimization occur at runtime, the setup of both algorithms have been carefully tuned, such that the mutation rate is 0.1 and crossover rate to be 0.9, with 100 population size for 10 generation. Each experiment is repeated 100 tuns to cater for the stochastic nature of the optimization. For the time-awareness, machine learning model~\cite{Chen:2015:tse,Chen:2013,Chen:2014:ucc} is used for \textsc{RUBiS} and the analytical model~\cite{wada2012e3,seeding} is adopted for \textsc{SOA}. The results are statistically significant as confirmed by the Wilcoxon Signed Rank test ($p<$0.05) with non-trivial effect sizes, following the guideline provided by Kampenes et al.~\cite{kampenes2007systematic}.

The quality indicators for benefit and difficulty are shown in Table~\ref{table:1st-qi}. As can be seen, the benefits are assessed by various performance attributes, which are scenario dependent, as well as the percentage of valid solutions found for adaptation. The difficulty is evaluated by using Lines-Of-Code (LOC) of the implemented prototypes for the candidates, as it is a common metric to measure the complexity in software engineering. A higher LOC implies higher complexity in implementation and maintenance, hence higher difficulty\footnote{Note that we do not include LOC for any third parity libraries}.

\begin{table}[t!]

   \caption{The quality indicators for benefit and difficulty for the first case study}
\label{table:1st-qi}
\centering
\begin{tabularx}{0.7\columnwidth}{p{1.5cm}X}
\toprule 
\textbf{Attribute}&\textbf{Quality Indicators}\\ \midrule
Benefit&latency, power, throughput, cost and \% of valid solutions\\ 
Difficulty&Lines-Of-Code (LOC)\\
\bottomrule

\end{tabularx}
\end{table}

\subsubsection{Results}

\begin{figure}
  
  \begin{subfigure}[t]{0.3\columnwidth}
\begin{tikzpicture}
    \begin{axis}[
      width=5cm,
    height=5cm,
    ymax=100,
    xmax=20,
        xlabel=Power (watt),
        ylabel=Latency (ms),
       legend style={at={(0.5,0.95),font=\fontsize{7}{8}\selectfont},
      anchor=north west,legend columns=1}]
           \addplot[only marks,mark=o,blue] plot coordinates {
(3.5713661434677504,58.23333333333334)
(2.2612590242935755,22.97402597402597)
(2.766469305777989,66.94142259414225)
(3.28417370226597,17.211604095563114)
(2.1597181094963687,7.011764705882337)
(0.9301991103361478,14.279835390946491)
(2.242167865870191,17.716981132075457)
(2.5419879772208014,7.938118811881152)
(5.394175468670917,183.39137931034577)
(10.24245659083614,258.7177497575167)
(3.447365715872904,13.61882129277559)
(3.469398493067439,28.384288747346112)
(1.2669007290094043,25.082332761578026)
(3.8949532005463876,17.610373944511412)
(15.719743032884827,42.4170708050437)
(16.628072528847028,18.15099457504525)
(11.020345961260599,15.718614718614685)
(6.692941252334177,44.85106382978702)
(3.8468969997049043,36.23714585519414)
(5.345881643514195,14.564315352697108)
(3.8474589609672347,12.778523489932816)
(1.938978602837379,6.103044496487106)
(3.961485252754642,8.324009324009277)
(3.036996887762802,19.501154734411056)
(2.0332053274138215,11.432584269662879)
(1.3350845526615116,5.9082352941176275)
(2.066219491630903,8.072681704260608)
(2.297254813378331,15.73165137614679)
(1.9957502535291787,6.577464788732397)
(2.8551838861677963,114.50670241286926)
(4.6069611731823965,11.18965517241377)
(2.9643283980459443,259.28870292887046)
(8.524818170103128,38.82458233890227)
(5.250533402701891,13.79760319573897)
(3.1491373021822104,12.086956521739086)
(1.8051037516560244,10.358792184724667)
(1.9572375290257586,12.808510638297866)
(0.3604280119231487,5.703999999999979)
(2.525705227273467,9.706586826347262)
(3.349261152729564,8.64542936288087)
(2.2739490165053304,17.634042553191485)
(1.704099470728879,13.19090909090908)
(3.8700405578955968,10.503355704697984)
(3.315029119578536,10.011799410029495)
(2.851009183909968,7.378870673952598)
(0.858549332691293,6.183574879227047)
(1.6701552865037947,10.33527696792999)
(1.7300444345148862,8.796610169491478)
(8.03041122326991,304.59136212624725)
(5.91976284651673,91.68264840182641)
(2.2702307899550194,84.70226537216826)
(1.9386047995163744,35.046610169491544)
(3.643429236795671,27.4483870967742)
(4.217756613963548,21.927864214992915)
(2.4224197906226608,8.19152542372879)
(2.2403367233454383,32.6566037735849)
(6.927315580084202,66.42594296228148)
(6.098666604345755,28.92330677290844)
(1.6295625598451835,8.95959595959595)
(6.911628549267834,29.14117647058826)
(9.520724226566347,40.5608108108108)
(11.020992131771653,51.926931106471905)
(1.0853607610930658,17.75172413793099)
(2.487589470798081,6.759740259740255)
(2.9475775718048594,16.725581395348833)
(2.1588382480294177,16.267489711934196)
(3.0116433023050213,10.756457564575609)
(1.6153113069819915,7.2919708029197)
(5.795878723658887,14.670360110803303)
(2.993260906540964,10.499999999999961)
(2.5291962241348442,13.326848249027202)
(4.892395042018238,212.02893518518582)
(8.055344117151318,30.20585967617582)
(6.82219752261981,40.39362490733867)
(6.7357719789977475,24.492753623188477)
(20.402288732394364,316.04586593034594)
(0.8174722490116202,4.530026109660545)
(6.456776482423126,22.353191489361677)
(5.321091475175104,25.598615916954994)
(5.962293922607371,32.87845303867407)
(2.1604368745646902,10.285714285714288)

          };

%
          
         \addlegendentry{NSGA-II}
       \end{axis}

    \end{tikzpicture}
\subcaption{$C_1$}
    \end{subfigure}
    ~
      \begin{subfigure}[t]{0.3\columnwidth}
\begin{tikzpicture}
    \begin{axis}[
      width=5cm,
    height=5cm,
    ymax=100,
        xmax=20,
          xlabel=Power (watt),
        ylabel=Latency (ms),
       legend style={at={(0.5,0.95),font=\fontsize{7}{8}\selectfont},
      anchor=north west,legend columns=1}]
           \addplot[only marks,mark=o,blue] plot coordinates {
(3.8840604311075793,253.63235294117698)
(3.003389832060982,69.12969283276466)
(2.01309537285287,33.16896551724132)
(0.9038915336104572,10.818565400843868)
(3.5219591794796252,9.733532934131697)
(4.3496148210246055,16.315668202764957)
(2.4373957347137134,21.43307086614173)
(0.8836194991862769,297.1078998073221)
(8.264337954837591,23.332782824112453)
(13.304042699577833,22.498729889923908)
(18.242442106629234,94.99777530589564)
(6.78966738255271,128.8297872340437)
(3.534045939509615,29.70808080808105)
(3.735159428078876,19.875796178343936)
(2.1694977273911347,19.8867313915858)
(2.3693584561563865,8.659909909909867)
(7.377160517769478,13.615555555555565)
(11.310026646345193,10.786743515850118)
(5.508627662012135,2.924242424242426)
(2.15484830420483,9.02457002456998)
(1.7450895807372042,9.209090909090868)
(1.9450920676184076,12.44207317073167)
(15.551511721584273,149.71641791044888)
(9.631456005248177,11.607171314740986)
(12.120931583718999,22.175965665236117)
(1.040411264060409,41.46693386773557)
(3.0039080572291046,7.148459383753466)
(10.603146858602425,18.606699751861)
(18.375,161.97289377289377)
(3.818025159414032,12.085201793721941)
(5.869421676312577,29.359177215189913)
(6.9286004424341,19.936350777934884)
(8.30446649778002,83.92256341789036)
(6.881150021008801,24.732905982905915)
(1.0714053011029887,10.32692307692306)
(3.0624453758091237,11.844961240310043)
(2.208867787307626,15.10970464135017)
(3.471445693597084,14.672222222222208)
(1.856534269740629,8.074324324324289)
(4.737460026527152,20.977443609022554)
(1.7933247147520603,8.078224101479849)
(4.179611199458176,15.073446327683584)
(3.4989114778403985,10.392535392535418)
(3.2387097430134277,12.968164794007528)
(1.1033534904734958,5.235483870967709)
(6.2723204347663195,53.88984881209463)
(6.597990617494557,180.9650084317025)
(8.805695055997008,79.23316582914639)
(7.299759994139855,27.35221421215246)
(1.4423406713665958,13.251968503936995)
(5.664966668553352,317.2057761732807)
(7.8933275320235285,27.587047939445032)
(7.326444904637591,43.67431192660528)
(2.2475292001095095,10.41379310344824)
(2.9275046257031256,103.1746817538902)
(9.133972701752043,36.22306108442)
(10.014687936204888,31.675580577058362)
(3.47701080207761,22.63404255319144)
(1.6409641657223857,41.558988764044884)
(4.364723494669335,9.02892561983467)
(1.2866730760130973,6.720279720279707)
(8.583204802639662,29.541554959785508)
(19.6,25.430913348946085)
(20.045454545454547,17.20600858369096)
(2.1488326650467693,225.54861944777934)
(8.029060244317614,46.276703111858545)
(5.860049795839027,58.79182156133837)
(2.8403542369337464,16.98732840549099)
(1.0332953652831045,109.2154377880189)
(11.536259238141708,287.8835136855474)
(1.3125895570404966,34.852150537634365)
(3.349434021753137,11.841758241758225)
(5.506019253502612,14.086522462562415)
(3.897681742184129,11.094474153297687)
(0.934960754571656,7.0652173913043494)

          };

 %
          
         \addlegendentry{NSGA-II}
       \end{axis}

    \end{tikzpicture}
\subcaption{$C_2$}
    \end{subfigure}
~
  \begin{subfigure}[t]{0.3\columnwidth}
\begin{tikzpicture}
    \begin{axis}[
      width=5cm,
    height=5cm,
    ymax=100,
        xmax=20,
         xlabel=Power (watt),
        ylabel=Latency (ms),
       legend style={at={(0.5,0.95),font=\fontsize{7}{8}\selectfont},
      anchor=north west,legend columns=1}]
           \addplot[only marks,mark=o,blue] plot coordinates {
(9.515088236113186,81.64102564102559)
(0.8756904282461151,25.065476190476183)
(3.3857623986812566,44.11538461538456)
(8.939448447978968,201.50882352941193)
(1.5855869113581784,9.089743589743561)
(0.5654454420649586,4.100000000000001)
(4.606187928182192,34.149758454106234)
(2.575419077970241,14.711956521739111)
(0.9012436138847069,104.66606170598892)
(5.465222912579841,55.496018202502746)
(3.741543474124577,20.085603112840555)
(4.952610728798932,12.585365853658484)
(3.827876391414666,14.852976913730206)
(4.079626857225554,12.74600188146753)
(4.321037028217834,23.644531250000057)
(1.1878021204398674,157.4300847457638)
(3.3913483809020133,12.18297872340425)
(9.947220869299839,322.3658088235305)
(7.981752842010636,94.52771084337334)
(1.293417903552469,18.64374999999995)
(1.9210371478064339,11.743589743589737)
(12.077136205051351,286.6105331599393)
(7.56844371636166,428.86239999999907)
(6.122834004484807,344.91334488734833)
(3.6050004311862813,43.413436692506515)
(14.616710446170247,414.88209285187475)
(16.756625923297886,87.30000000000113)
(1.064613472252959,442.73972602739843)
(12.885867623861108,297.4744027303756)
(3.1982187222409304,35.347826086956466)
(2.22632201398219,355.4663461538464)
(6.165857375905784,56.8850931677019)
(4.773227728262989,19.517711171662068)
(13.874776386404296,48.17977528089886)
(9.64934908498456,62.404761904761905)
(7.753369483322625,150.50750220653168)
(2.530967955296948,4.893280632411066)
(2.7415808719586394,24.893150684931477)
(3.706956512725528,13.81104033970277)
(5.07493695502854,408.4920273348528)
(9.67144690236194,63.443691786621805)
(10.876719990036241,96.29111969112004)
(10.985799380853727,68.7051926298163)
(1.438563147068367,9.299435028248585)
(15.165857375905784,2.0897435897435597)
(17.773227728262988,3.100000000000001)
(16.793436047567507,3.2335115864610002)
(17.250328404076825,3.0677618069836003)

          };

%
          
          \addlegendentry{NSGA-II}
       \end{axis}

    \end{tikzpicture}
\subcaption{$C_3$}
    \end{subfigure}

      \begin{subfigure}[t]{0.3\columnwidth}
\begin{tikzpicture}
    \begin{axis}[
      width=5cm,
    height=5cm,
    ymax=100,
    xmax=20,
        xlabel=Power (watt),
        ylabel=Latency (ms),
       legend style={at={(0.5,0.95),font=\fontsize{7}{8}\selectfont},
      anchor=north west,legend columns=1}]
           \addplot[only marks,mark=o,blue] plot coordinates {
(11.997293252156267,128.05673758865237)
(0.791885230759674,5.124293785310737)
(2.3915107358412753,59.97549019607838)
(1.9168616282888757,35.671052631578874)
(2.1628718150842685,18.27559055118111)
(0.8429841891884917,3.507518796992483)
(2.9364765815581926,13.808955223880583)
(5.72969612123412,7.986449864498605)
(6.102366227155386,8.918749999999974)
(1.1648811526682652,21.845890410958845)
(4.499393353955428,8.828106852497097)
(1.897484865220256,254.60967741935514)
(4.060150149986262,12.293440736478674)
(5.346665977724874,18.940327237728685)
(5.5064652962480425,14.602898550724579)
(1.229759928880421,4.582733812949643)
(3.178792053904275,12.766636280765708)
(2.0851897877897025,198.5000000000005)
(6.210870813599988,163.50674373795786)
(2.4674702949973604,31.891615541922263)
(2.66820925329778,28.725888324873107)
(1.2803498482193376,8.197222222222205)
(3.8242740750212483,16.382198952879563)
(1.8139102296499474,28.580090497737725)
(5.686818732364569,15.886130136986296)
(4.3845556661788505,16.76206604572395)
(1.7603249166465476,7.299003322259117)
(4.58414697522872,9.434782608695665)
(3.759536026105414,9.85656836461126)
(4.322540818594508,13.858627858627859)
(0.9644443169749547,6.006109979633382)
(4.386965159148854,11.74468085106384)
(4.912030893494157,11.059840425531899)
(3.052367059986856,8.483455882352931)
(1.0228242072209703,96.09821428571405)
(2.7673861610563297,14.964653902798231)
(3.2241674607128137,14.238888888888875)
(2.733787098782113,12.690476190476183)
(0.9747052831826362,8.570765661252869)
(2.095821047468088,8.631672597864748)
(2.3742291036487106,6.808988764044929)
(2.4139846886299146,9.587628865979353)
(1.00516371831701,3.507109004739337)
(1.588955876549979,7.570397111913338)
(1.5830896676420112,7.969788519637419)
(1.9843925627347894,13.41887905604718)
(0.830629955715392,15.405405405405402)
(2.260656252336572,8.535519125683013)
(0.8128337928320626,73.74742268041241)
(1.488234132155977,28.2528363047002)
(2.5213372213757888,18.865410497981117)
(2.110701171108296,8.977900552486204)
(9.811251438205204,62.82893081760999)
(0.7546208240742819,67.8130217028377)
(11.05306134141381,55.60667340748218)
(10.532409754226485,70.33417085427136)
(8.69120168199787,31.43586956521744)
(2.4222439627029932,217.7113997114001)
(8.984318923353635,68.50738150738195)
(10.136718499955553,31.825788751714526)
(7.7439799863093555,36.86053199137307)
(1.7113968469967742,14.020242914979727)
(3.3589875816602235,8.621564482029576)
(5.160813579712126,13.858468677494212)
(5.66882457887794,9.766439909297024)
(1.5732223990056364,8.070000000000002)
(2.164955192736492,17.81733021077285)
(4.843983433315685,11.997481108312318)
(3.9120445490608646,21.146814404432142)
(1.3055327348507888,8.145833333333334)
(4.369001946620976,31.786221590908998)
(5.066737536390209,22.31365576102435)
(2.2630164392541046,74.81626506024202)
(7.664941048731779,26.75546975546968)
(1.6712605276321655,10.715469613259643)
(13.231840847614356,28.72897196261679)
(13.700026087371352,34.55737704918042)
(4.049869244424138,16.415274463007147)

          };

%
          
          \addlegendentry{IBEA}
       \end{axis}

    \end{tikzpicture}
\subcaption{$C_1$}
    \end{subfigure}
    ~
      \begin{subfigure}[t]{0.3\columnwidth}
\begin{tikzpicture}
    \begin{axis}[
      width=5cm,
    height=5cm,
    ymax=100,
    xmax=20,
     xlabel=Power (watt),
        ylabel=Latency (ms),
       legend style={at={(0.5,0.95),font=\fontsize{7}{8}\selectfont},
      anchor=north west,legend columns=1}]
           \addplot[only marks,mark=o,blue] plot coordinates {
(9.484525864895827,165.26785714285728)
(12.008048456115707,486.491103202847)
(5.164053382336497,127.85472972972967)
(3.4003016759606366,15.422480620154984)
(1.8136201180963387,488.0312500000006)
(6.617205258902133,258.69882352941295)
(6.302228271964279,20.185915492957744)
(1.2310409664102324,29.532289628179953)
(6.139346861216671,13.232214765100661)
(10.843683463295873,20.792156862745234)
(8.514887377296958,25.679961089494192)
(12.25419140443562,12.194805194805177)
(2.165586059902634,25.703958691910557)
(4.716043706809033,22.85291396854774)
(3.946384887101259,23.382242990654277)
(1.7435694081858841,35.089347079037765)
(1.6531998478735346,40.14399092970516)
(10.402121899535242,26.798345398138576)
(7.1806073875009035,48.24938875305621)
(3.2626848986328767,12.937269372693683)
(3.0159865961473646,13.872972972973004)
(4.607929599474889,14.520383693045583)
(3.184403651521207,16.153846153846164)
(2.826968024881029,15.04782608695649)
(7.096420626610721,241.39086294416293)
(1.0906804409501554,216.3217550274229)
(4.639268432146526,19.95558375634512)
(5.133517568186627,81.07248157248137)
(2.737092244656743,17.389830508474553)
(5.9301195671982825,32.86748633879801)
(5.884357718014131,15.642076502732225)
(5.693160560797219,10.721189591078083)
(0.7759688215299371,65.76430205949667)
(3.5684883099887204,11.429983525535473)
(3.796190369498471,26.129427792915575)
(5.991375730722305,15.12709030100336)
(2.139637231120285,87.45502645502638)
(3.385596641040846,14.629032258064536)
(3.229482145889349,18.522342064714895)
(2.8908725683801473,22.159055118110217)
(2.7167237899251413,11.725099601593614)
(3.025227991416275,7.085820895522364)
(2.454015387492294,8.243107769423514)
(3.6208162017568144,9.379061371841116)
(1.3941028928359005,2.3497757847533642)
(1.8362507500908518,9.969111969111932)
(2.589002166280141,7.185303514376976)
(7.222519759282426,489.7846607669621)
(2.463253194100788,167.802784222738)
(5.883978741970197,42.464342313787796)
(4.131479489617087,12.698324022346394)
(1.1421959874794163,38.214559386973264)
(4.127136873307813,25.992164544564314)
(1.6617826119816026,8.100628930817596)
(2.378050707280716,65.75903614457853)
(9.578180213470713,124.04588744588865)
(1.5261425521788219,21.37755102040814)
(3.9959501557750072,145.1879914224456)
(15.095878787878789,47.25296167247404)
(17.831334211283608,45.2082595870209)
(4.677694092843033,21.380180180180172)
(1.6220909847142395,30.03353658536579)
(3.351056251278375,17.77293577981653)
(2.5411056775539254,10.783908045976986)
(2.227942638431475,21.481651376146758)
(4.109185617297619,16.88851351351349)
(11.240070709224634,18.83068783068784)
(3.9690746783927993,16.40000000000001)
(6.344176946739703,29.74999999999999)
(2.278578794968079,290.8599640933534)
(6.404592822868301,12.778726708074528)
(5.664323263777224,156.07072135785054)
(6.06744352841,265.12942008486624)
(8.425192027222204,69.23593380614834)
(7.200646952071126,27.697604790419152)
(2.9700034441355325,23.15384615384617)

          };

%
          
          \addlegendentry{IBEA}
       \end{axis}

    \end{tikzpicture}
\subcaption{$C_2$}
    \end{subfigure}
~
  \begin{subfigure}[t]{0.3\columnwidth}
\begin{tikzpicture}
    \begin{axis}[
      width=5cm,
    height=5cm,
    ymax=100,
    xmax=20,
      xlabel=Power (watt),
        ylabel=Latency (ms),
       legend style={at={(0.5,0.95),font=\fontsize{7}{8}\selectfont},
      anchor=north west,legend columns=1}]
           \addplot[only marks,mark=o,blue] plot coordinates {
(1.366493658337927,14.499999999999945)
(3.0719635347629826,363.05439330544)
(3.407462927319833,66.25806451612898)
(1.1857394649048412,9.067510548523188)
(0.4664697685213931,17.188524590163954)
(3.326725530408635,54.30000000000015)
(2.352037560237435,12.12857142857138)
(0.8040237566895972,39.3554757630162)
(3.319671470104313,12.714285714285696)
(3.806782126996827,136.06798029556685)
(4.767230241521281,17.713333333333306)
(1.4955698508936957,4.25958702064894)
(5.624304089379926,76.12513144058937)
(1.5942567757183914,10.639423076923057)
(2.647027446519551,55.863498483316256)
(4.632737711450905,21.6030680728668)
(4.930445578967744,21.995073891625697)
(1.411704552568231,8.954983922829546)
(1.321329852590702,124.05855855855856)
(2.625433281051507,23.376086956521718)
(2.5991354545567846,15.087356321839094)
(4.252697760383626,131.13229571984428)
(2.948034405654596,196.87966804979314)
(7.281553455411698,59.18535825545185)
(8.32634584361899,106.69009314140575)
(5.290741733515987,121.72608978529705)
(3.160901490294135,17.16219512195117)
(0.960341432342903,77.48842592592578)
(3.606708004330614,20.001426533523492)
(3.385162908802985,16.517663043478226)
(3.908080048928738,11.983471074380176)
(1.152772270094814,8.205250596658695)
(3.614671097422753,15.415094339622645)
(3.118737377606707,10.51685393258427)
(1.1352830053775327,94.68894601542405)
(2.730808725423421,12.21442125237192)
(4.159877271678249,10.875177304964545)
(3.117928355742518,20.461904761904744)
(0.9397804194439564,5.231707317073172)
(1.9241181716178346,11.226463104325678)
(2.3948247555882887,11.213836477987376)
(2.3833446660615127,12.982558139534868)
(4.99760268629834,5.7685589519650575)
(9.042068611594324,201.17102137767247)
(1.9091698337632736,180.39968404423422)
(10.523714064762316,75.66159695817515)
(7.518522474507724,147.99746192893392)
(7.681617890592843,55.75172413793102)
(2.8013002382072782,38.427872860635716)
(10.933895462409172,72.36607142857147)
(10.539847709837016,22.373913043478275)
(10.565502971436018,18.04615384615385)
(11.064134838226462,28.789473684210492)
(11.156979161250149,29.5164319248826)
(10.739058665711243,11.630136986301375)
(5.778479070493592,329.98503740648226)
(11.647175035570832,162.23608670181466)
(7.912897879293132,377.2144288577152)

          };

%
          
          \addlegendentry{IBEA}
       \end{axis}

    \end{tikzpicture}
\subcaption{$C_3$}
    \end{subfigure}

    \caption{Benefits on \textsc{RUBiS} under further investigated candidates over all runs.}
  \label{fig:rubis-switch}
  \end{figure}
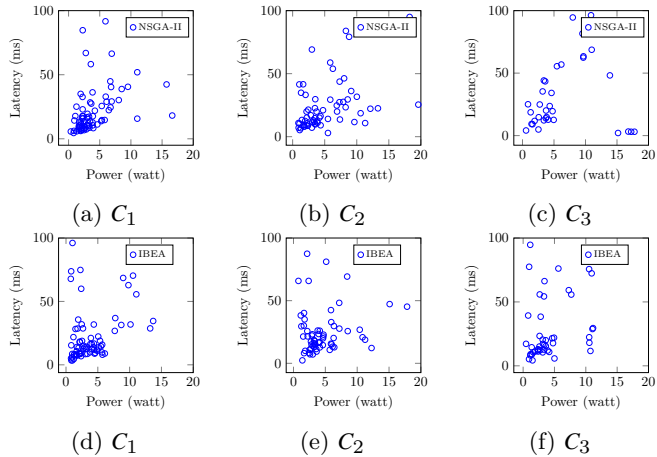

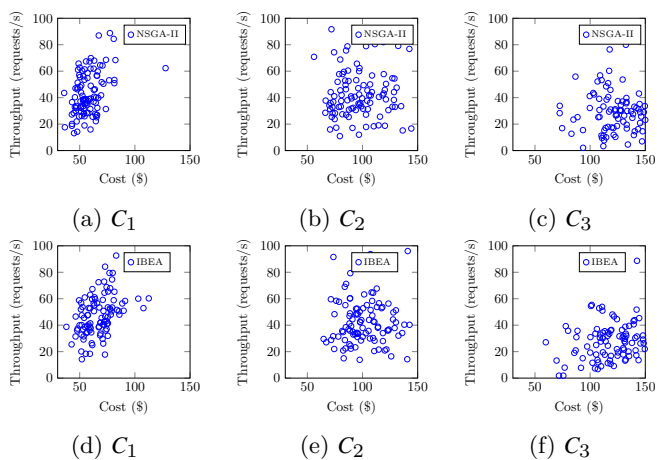
\begin{figure}
  
  \begin{subfigure}[t]{0.3\columnwidth}
\begin{tikzpicture}
    \begin{axis}[
      width=5cm,
    height=5cm,
    ymin=0,
    ymax=100,
    xmin=30,
    xmax=150,
        xlabel=Cost (\$),
        ylabel=Throughput (requests/s),
       legend style={at={(0.5,0.95),font=\fontsize{7}{8}\selectfont},
      anchor=north west,legend columns=1}]
           \addplot[only marks,mark=o,blue] plot coordinates {
(61.67050690086489,25.55554237261638)
(54.40249899501286,25.918066482830422)
(49.41716542909026,25.757398993576697)
(65.76984036529151,26.9158480100817)
(42.914224212486225,18.71321273631719)
(65.76989084248967,56.511437189138036)
(57.116943073166766,58.60322326534449)
(56.35863991776982,28.471615416804692)
(64.77043485326571,40.18115903816779)
(53.27532927264305,56.137038193091506)
(82.35860443337222,68.3822166703732)
(51.808018624876944,26.12924535887842)
(47.92599403874603,46.30638934476806)
(42.984424756382936,20.122349353324836)
(66.3047176617148,59.81922678272473)
(58.91937974908118,67.30457749058667)
(44.58349442503953,28.414164534712285)
(68.54410072610064,64.59036106735866)
(63.70108493931967,46.27769202055562)
(70.66475925215114,64.25048519745866)
(51.301376973308194,33.33409910539166)
(127.81404661353918,62.302384319675575)
(43.06731444992616,36.58642975963664)
(52.46801190205825,41.43882258842589)
(53.21844971860348,38.08719529622917)
(53.167182880555245,37.370583398877955)
(69.37518974331257,41.781251089070686)
(53.55509336231129,31.45531631290926)
(57.60920564360729,37.462212318462846)
(36.50183970225961,17.59449640538134)
(63.05727057292709,40.71837683189125)
(50.73853850232873,37.21141273740295)
(53.81393937528105,38.24636595770417)
(56.903461832401696,34.739401863215875)
(54.970683013854014,38.96297785505538)
(52.994920282553494,33.48943590768324)
(46.8754735860852,33.676527606036444)
(60.05344820473367,39.35660023717831)
(53.504261442625086,38.83801110160398)
(46.13977858829658,47.05518733686109)
(48.434189705460966,61.08105994146566)
(54.4165011458832,61.375106170005004)
(48.949429137061514,65.74869810330641)
(51.205186656925655,43.903696555881645)
(48.157511994038714,51.319729178442884)
(35.90535307534438,43.54033224204164)
(59.752759008788146,61.91683866086532)
(54.30485211025844,53.49769382409064)
(50.19811331962203,55.595305702231194)
(50.01782389950351,62.73495473825902)
(49.61119020301691,57.15118508951144)
(46.363761889507984,30.02749480408494)
(62.66119051204072,61.91683866086532)
(51.44853699782547,49.291435385603016)
(55.69596968775945,67.40259290009978)
(59.98632800289417,48.99738915706367)
(50.479093200624426,47.913924175343055)
(46.81697952566006,32.303474859245796)
(51.33383994508391,29.387746903711523)
(76.68601756894148,53.12329482804412)
(42.858523036751585,27.272584364122284)
(62.77613312168676,45.9143277067156)
(81.40496725643594,53.39967841457752)
(81.37121113843826,50.927667540390445)
(62.530282425811535,36.87309761157344)
(65.17934030556407,26.817720620011332)
(55.302736142294606,35.4152336338063)
(71.46979387312753,50.48395045904325)
(54.00392901075148,27.901185601731946)
(64.04190997360827,27.73385210691816)
(54.6460497643962,35.8589507151535)
(57.726767223216314,28.37360000729158)
(64.51837573600545,36.05498153417972)
(65.18955204224542,25.158000021283883)
(66.793293591098,39.807517320104424)
(73.79988264691822,68.55887589244011)
(47.743504213527785,14.244429857255087)
(80.81168360963215,84.41524873066494)
(48.00262984085825,47.65031136505061)
(52.122819862924295,60.44262501098971)
(63.140634909833445,67.96813084940754)
(67.25214236112697,86.98658798426258)
(57.01262578916655,54.85754239234466)
(77.37354195742142,88.76939833043949)
(59.62046006956702,31.451107488713184)
(58.23330567835039,38.94201216550156)
(60.09141266991194,69.89593841840609)
(45.642417228921765,40.306924676314814)
(65.99092335654021,32.274287508666134)
(55.68240889306972,38.397808860313)
(64.6072587397662,22.210975940823314)
(45.60328955483319,32.76478132695252)
(51.10361115309975,32.3257830995105)
(54.31603610668817,18.957135801081634)
(47.23114057237042,22.023481705002293)
(53.65372965334785,27.99836589622379)
(65.40110401728867,56.74982595437814)
(73.70005563694721,31.397398001811016)
(69.82537378518576,51.69633099014736)
(59.76786697038503,15.943392436034248)
(56.698308935170836,36.11010394563756)
(44.84904717883451,13.148583278356194)

          };

%
          
         \addlegendentry{NSGA-II}
       \end{axis}

    \end{tikzpicture}
\subcaption{$C_1$}
    \end{subfigure}
    ~
      \begin{subfigure}[t]{0.3\columnwidth}
\begin{tikzpicture}
    \begin{axis}[
      width=5cm,
    height=5cm,
  ymin=0,
    ymax=100,
    xmin=30,
    xmax=150,
        xlabel=Cost (\$),
        ylabel=Throughput (requests/s),
       legend style={at={(0.5,0.95),font=\fontsize{7}{8}\selectfont},
      anchor=north west,legend columns=1}]
           \addplot[only marks,mark=o,blue] plot coordinates {
(136.43727597211966,15.195833928970341)
(87.48741768327422,25.380878565465565)
(115.8344281847515,25.61609889082351)
(107.45044840852077,41.592576452912596)
(96.79132806332602,11.98970483341239)
(96.94389639851012,42.37678632582146)
(128.46596178982642,38.30820204761948)
(110.18061131314353,18.36450972630143)
(79.48508736668813,10.92683057623296)
(144.22144424284664,16.670611090081508)
(84.27040067005632,52.910403524877296)
(120.3206628119141,18.762052077633935)
(103.1616887062095,27.160364719996206)
(87.73674073266032,30.932591608785263)
(81.73010491566353,27.442919197117746)
(127.04538631540758,54.684495076154036)
(104.17870204758827,37.70914816398071)
(108.4444064845026,44.2321926549211)
(119.22853195564846,57.78510718795077)
(77.42780384308747,51.92495395266979)
(104.81342893375657,31.306917349185373)
(86.24162359782204,38.002504215211964)
(91.56836441068761,32.31523844367123)
(130.94274763862654,47.55101693071433)
(72.12489804704728,25.47250748505046)
(80.74926284327067,37.462212318462846)
(127.11583729630043,33.121702917196764)
(74.77213932776012,23.769454128817422)
(121.31585525332491,50.39187283202147)
(66.29756607964839,28.54541415930908)
(128.53692303628654,44.17973505376102)
(99.6275763610949,39.496331913236816)
(80.26655236733922,33.30007109421706)
(76.21599026172264,17.28687054651491)
(114.55043146539597,19.078784592247953)
(104.44090093185702,45.48237420979383)
(128.94785370905507,54.42593398407693)
(73.51619549642348,36.79267158841984)
(106.92814497274156,36.11257475996581)
(94.33848767501918,53.19172415867547)
(97.28443613016064,62.30254417212598)
(96.95558870258566,59.98656027753854)
(79.90850478741251,45.52889402846259)
(84.96383149590139,46.354467025705645)
(83.11448464587556,59.525180554185425)
(106.07588458578394,41.87307688642671)
(123.47850000492502,30.563940577634714)
(118.20692183551223,41.763187098294104)
(96.37549207138649,35.51290395466537)
(97.17462001080713,34.69369584397993)
(69.95592461571303,34.09291809065059)
(120.20156558467347,45.86780788804686)
(119.18411737334755,43.16311734301282)
(72.77131278821972,32.66683917308581)
(87.3687243327267,52.581562337183804)
(134.9847615671006,33.528916084436425)
(93.89490443668133,30.59290973485478)
(97.92893701935157,35.94968828604958)
(71.00718614691027,35.95696612466661)
(74.42528035531906,41.07392439503733)
(75.07635437520287,28.264549915571962)
(121.20780076566041,38.697948142402346)
(103.23419529396631,17.84086698151336)
(106.85160771697494,50.01153605348361)
(98.87110070549087,33.45404403020929)
(129.4150422162682,33.01615275079619)
(92.91641807619023,40.637396981923565)
(90.11795636005267,39.41442657106119)
(70.64222894129881,31.286857148497603)
(89.5855281264127,40.99371656524334)
(78.55002103218675,32.93347410667394)
(72.86839701495657,28.27558459777847)
(66.60195265053143,37.7698199629733)
(81.04941251040408,41.97581114984501)
(90.10259711836741,31.756188399459024)
(71.14214683571716,44.165010628640104)
(100.52078499927917,31.544671295937032)
(96.41658414358264,91.26010095762324)
(85.10781433633693,75.6268911543057)
(96.10694449748104,47.16773626280856)
(142.47568288524195,76.83084708276186)
(84.61840230481843,49.591607479256325)
(114.86002355121967,40.5292148167892)
(71.85422813824181,91.68134250174802)
(56.06014957050888,70.83440617730504)
(90.88608815143535,17.099258657653756)
(81.89912691531514,57.83082012177051)
(85.17845794019843,27.069980798524302)
(100.9871133020074,49.44501233653932)
(86.31408421776345,78.76853790907337)
(109.15787291896113,109.88429363876791)
(74.23722919991654,74.26692027214897)
(70.61450004029882,15.943392436034248)
(108.65017733582751,56.74735514004989)
(118.8218654680836,81.94844106426739)
(86.8778962620538,60.46378757831467)
(132.34750271050126,79.10420119812777)
(94.92546760148029,65.66562347412825)
(79.84016396742874,68.81581714544663)
(78.81690915442218,57.775006441913405)
(103.20371087451134,78.91600594622328)
(111.23414433626162,80.43853133063249)

          };

%
          
         \addlegendentry{NSGA-II}
       \end{axis}

    \end{tikzpicture}
\subcaption{$C_2$}
    \end{subfigure}
~
  \begin{subfigure}[t]{0.3\columnwidth}
\begin{tikzpicture}
    \begin{axis}[
      width=5cm,
    height=5cm,
    ymin=0,
    ymax=100,
    xmin=30,
    xmax=150,
        xlabel=Cost (\$),
        ylabel=Throughput (requests/s),
       legend style={at={(0.5,0.95),font=\fontsize{7}{8}\selectfont},
      anchor=north west,legend columns=1}]
           \addplot[only marks,mark=o,blue] plot coordinates {
(139.6360562653287,11.692703693665475)
(151.29959890296692,19.005970053793686)
(121.33282786632944,11.909006977251167)
(114.52807801128554,10.015459262161672)
(112.67671467512032,7.762294982389449)
(133.6397772937927,29.986305672378368)
(122.43364213803014,30.26119243113304)
(155.17042232671196,18.151937641257547)
(122.91752950052467,10.301847598466638)
(130.08915594740188,21.856493298842544)
(133.0130482282041,36.68189253840496)
(170.2237369216899,23.068038352753486)
(172.9389579222997,34.39313950338766)
(74.61806356395913,16.933726581786544)
(157.26521157002054,20.998131912151052)
(116.07351252563676,17.87705088250287)
(146.3284419483341,25.72714092529378)
(133.13665015331006,19.533153221386986)
(118.52946001755768,31.049179929557795)
(140.07028828527882,33.143173547855014)
(143.80857395338631,31.204516731849367)
(113.95698152810361,17.08906338407812)
(121.15098662594244,17.08906338407812)
(112.02685083653982,33.33026524620823)
(138.873575452463,27.506627079824426)
(99.68557477591168,32.175762069443905)
(72.39720916816674,28.29461457824918)
(112.12285539275828,11.265425217694315)
(110.26216199890034,11.360887996462637)
(138.27390756762827,24.692187704992563)
(129.3376521730276,37.11978381781806)
(125.39751605728866,24.095125764104452)
(108.51737948500295,25.47250748505046)
(158.95691416751313,49.786993146448616)
(117.12397607671294,29.421196718135295)
(118.39396679975098,45.879804580569086)
(108.43774966503105,30.674996532851367)
(144.81754725830416,35.19651652870539)
(83.0851958168883,12.717247794187273)
(93.50617549302227,15.471813145495885)
(137.11938699618372,29.653095808038426)
(110.4145275575947,53.22131023755722)
(109.00604400932747,56.80548341767735)
(102.8306447213999,43.54033224204164)
(116.3916923972992,40.49789155278181)
(117.25031232611516,60.26294386407194)
(106.73153137855084,52.23586066534972)
(86.68305769182811,55.88935193077053)
(152.1598809317136,59.72121137321163)
(114.9093550024233,38.498295084154364)
(84.84866757924067,23.41854357671092)
(103.59004177016836,42.42816993610858)
(132.88251331806634,39.7777908849012)
(108.21511407300713,39.10934566031535)
(72.48286248059561,33.8196222401071)
(151.78729324853197,39.84710897020188)
(124.3596521893814,18.57253723784988)
(88.56911450690052,25.250806550524576)
(145.37601160271916,28.834979730644708)
(155.88539802616464,41.431685681694574)
(119.56651721090971,17.47140961412333)
(140.4241237468662,14.597171678217713)
(107.44306056074966,25.222109226312142)
(159.98002184531242,52.5705276549773)
(125.14988670142021,37.40379542022725)
(99.63043690667155,41.16885206856052)
(166.9931875059159,46.554075607089025)
(162.1232533890543,55.88935193077053)
(152.9346992337145,24.611058650151154)
(148.66097063312083,28.984650583452567)
(156.3559512715218,43.51163491782919)
(134.21445467036037,4.569622752327636)
(129.43785819155386,53.69372464311685)
(144.24657400068034,43.090716103113934)
(143.4248023526289,16.174708240581854)
(127.81196348360363,30.626820520731343)
(149.61194807396566,33.65542095814785)
(166.86948285058236,14.084408344577472)
(132.7597493870838,35.724815205097734)
(93.6037043638783,2.0842513193189967)
(117.69194840609029,76.41789087605476)
(149.98105542840577,23.125638439719623)
(168.81740707191386,17.536042989037963)
(122.73568719401038,31.238483437162287)
(147.5246491428358,14.576009110892752)
(145.24572067391622,20.829870016878633)
(147.056070123307,6.954764879765362)
(133.1983921620714,16.66026043021175)
(130.86466116099734,26.419465467560297)
(126.95421701421384,30.52050849495589)
(116.3108425641244,50.9430314448624)
(114.51095026094389,48.50567168320747)
(123.2697539858016,26.52577749333574)
(111.95999225248555,3.345322618685049)
(141.97453628424256,32.27318056063723)
(112.88024833225496,15.24138151422588)
(124.12622960839558,26.85735674697341)
(132.3725294460997,79.96481981541177)
(127.97070977750928,30.13632670244496)
(166.55526105690612,29.379499345705508)
(127.80279472599969,34.463743905731675)
(110.73207740498711,9.545474037768766)

          };

%
          
         \addlegendentry{NSGA-II}
       \end{axis}

    \end{tikzpicture}
\subcaption{$C_3$}
    \end{subfigure}

      \begin{subfigure}[t]{0.3\columnwidth}
\begin{tikzpicture}
    \begin{axis}[
      width=5cm,
    height=5cm,
    ymin=0,
    ymax=100,
    xmin=30,
    xmax=150,
        xlabel=Cost (\$),
        ylabel=Throughput (requests/s),
       legend style={at={(0.5,0.95),font=\fontsize{7}{8}\selectfont},
      anchor=north west,legend columns=1}]
           \addplot[only marks,mark=o,blue] plot coordinates {
(74.72563811263966,40.8246845640571)
(72.8748399045323,17.77775424455112)
(57.98463195383679,40.47458834504545)
(75.46765922982351,31.94056761461708)
(54.56166791329311,60.97112109507672)
(81.41380197238306,65.2646320938786)
(80.79637754088726,58.868572169671395)
(53.18271163196355,36.83336560515449)
(71.49846330430059,50.78347740086263)
(78.80778839629049,50.47291577683674)
(63.279599097439274,55.05357321137088)
(74.35572499623122,68.8201079497863)
(51.47208681249322,20.122349353324836)
(75.75204279514986,65.47231451677301)
(51.911857247832856,47.47020709399587)
(75.48139382308436,46.16802500717433)
(63.47568170508066,61.08105994146566)
(57.39784436389468,47.73555599832275)
(58.13641685975328,52.98493049045039)
(48.148103252458284,37.4023382949396)
(107.75901690494206,52.8752634770691)
(70.93888416372718,59.127750183410164)
(91.11570885721035,58.314842986288866)
(67.42851305654577,38.96297785505538)
(51.78136072930573,40.12514875018657)
(49.86992624945421,39.934223192649924)
(74.46807714994371,43.190387706078326)
(56.876606348240976,39.24936619136035)
(53.624135886696756,31.550779091677583)
(53.441971498411746,29.170397137075398)
(103.06320213341596,60.038055868382905)
(68.77888263223626,41.781251089070686)
(59.89102000673996,37.80847467829106)
(112.57424564308057,60.25677297348382)
(51.625031675408785,32.7969111880268)
(60.454594408304644,34.55231016486267)
(58.931376774603585,38.8713489354705)
(56.85045806699943,34.769771912056044)
(72.21768270536546,42.66404474211098)
(37.929748669007076,38.723023227393014)
(72.11553301897817,72.94000258262899)
(62.65009385725645,56.511437189138036)
(49.394838036190116,58.80507988630481)
(77.57798214188041,67.94432539096007)
(51.302751462428226,56.80548341767735)
(47.61958022167468,36.05498153417972)
(64.02277041613061,59.623195963698514)
(53.93534885440063,45.53992871066909)
(51.07202760662947,51.94181443681039)
(46.32183365187019,46.525378282876595)
(54.63104093414588,38.723023227393014)
(59.53045099453863,63.00030364258593)
(52.49370873295286,53.76304272841754)
(28.08287102498799,24.345709745824255)
(61.48272252643888,67.30457749058667)
(65.74312158301103,56.60945259865113)
(57.09587372906137,47.47020709399587)
(67.76462643480865,53.39967841457752)
(75.57792691101314,41.44272036390108)
(78.14031197221058,47.913924175343055)
(89.81540538258129,50.927667540390445)
(73.49925793540888,35.94968828604958)
(73.1095282270517,46.554075607089025)
(88.5703844539918,56.41342177962491)
(83.76732149917957,51.561244550175736)
(85.46663105152746,50.55326854434392)
(79.48698965230196,53.39967841457752)
(85.9435603306942,47.95316615539477)
(67.26561510736684,38.42897699885369)
(71.12764752526444,40.42857346748114)
(69.8284823056472,45.53992871066909)
(77.64442335349436,51.94181443681039)
(72.0248116193533,40.42857346748114)
(55.460104383442,18.38754110103016)
(77.34930194814669,52.690612428903414)
(75.47466784253685,53.490159067203166)
(70.71497626219067,29.659188861588433)
(59.468585130485934,52.74180115057468)
(51.91038428886727,14.244429857255087)
(72.40946857715637,64.52634481255053)
(61.214278465415475,59.739935937772316)
(62.94411290702925,50.59937746321956)
(79.32855487623064,74.54836780124158)
(68.88289453019598,41.96590036699497)
(79.90455745240688,79.50093646616644)
(76.96323211157352,79.47122698229079)
(83.01318946892793,92.62748428276475)
(72.99880583648311,84.23065876665464)
(70.65936565811523,102.90211821110572)
(67.06635912523937,72.48410992597206)
(71.32257018318948,26.087886213922634)
(46.20724281149106,31.451107488713184)
(74.36784065420886,37.00926296784654)
(42.611689724541314,25.475116349462787)
(60.878632016282516,26.305713473263076)
(68.47752602307165,32.32689004753941)
(59.56042597348001,41.414023039688644)
(55.9376988428825,34.02585262631857)
(59.22374462993881,18.360329957019808)
(59.04051369665047,36.162706484510835)
(68.83599839450898,33.64056388577873)
(61.41067482656637,24.932019992303132)
          };

%
          
          \addlegendentry{IBEA}
       \end{axis}

    \end{tikzpicture}
\subcaption{$C_1$}
    \end{subfigure}
    ~
      \begin{subfigure}[t]{0.3\columnwidth}
\begin{tikzpicture}
    \begin{axis}[
      width=5cm,
    height=5cm,
    ymin=0,
    ymax=100,
    xmin=30,
    xmax=150,
        xlabel=Cost (\$),
        ylabel=Throughput (requests/s),
       legend style={at={(0.5,0.95),font=\fontsize{7}{8}\selectfont},
      anchor=north west,legend columns=1}]
           \addplot[only marks,mark=o,blue] plot coordinates {
(123.78757223088856,36.70133152382145)
(126.01036288808706,16.427565337414617)
(123.23007328819949,36.00880680416502)
(95.3359064299313,31.13942400192121)
(121.679486433789,21.682988913440624)
(118.45759567651997,34.02278996586465)
(105.45922809577078,35.70338080476514)
(98.5878231933608,28.414164534712285)
(106.74278775957198,29.94124297899726)
(118.48628389543256,27.993030120656933)
(95.24269924823726,52.50051939236854)
(135.9873336578023,34.07880025384588)
(97.45711337550169,13.832898870649712)
(84.0029230843862,28.7604268945405)
(100.58175539146615,52.8175969956366)
(107.34389342286563,51.51603183047675)
(129.10740460529738,37.82969000711813)
(78.04066563872841,31.39544228938601)
(142.61028683584695,40.29587507937111)
(97.45864925740398,50.16139214249514)
(89.7248938216523,39.24936619136035)
(71.55268938329682,28.852055814125393)
(89.21609224591901,24.50126214745592)
(109.91445350812906,42.82571228744108)
(81.59150657863516,35.634407808118496)
(95.43989138545834,34.07880025384588)
(100.13464130274112,31.237599263884697)
(110.76876172181976,42.46280504281236)
(93.5255994690913,23.255130051106704)
(130.26154927111975,42.14040657871463)
(105.95312416228745,43.97266955252829)
(93.33735061128834,33.15649787635092)
(127.65729694647585,44.34158783529476)
(117.95617960418429,53.7165229097488)
(74.35848960521918,22.005164095574063)
(113.89285327911884,38.037027341358495)
(94.4743002681781,34.20516789891738)
(73.38493032733746,31.658013137495544)
(109.60813425031593,30.31529688638974)
(137.91903933404365,40.38205364881239)
(74.58786918391753,48.31729232860963)
(102.983410440059,33.91531177300516)
(88.67733815904481,48.92807107176299)
(78.74972354719876,57.69291758037176)
(76.1808231030703,58.43068089025829)
(95.98736868220692,33.451316297610596)
(88.16437745312606,25.564845476059038)
(114.69474941933602,21.55525740206977)
(132.59404633343104,32.08887460112559)
(127.95805114351965,49.68109789139399)
(88.15594249251355,41.30276938525669)
(101.48882769856996,63.276687229119325)
(84.48096081251344,55.151588620884)
(110.65534831923812,20.448159038695945)
(98.92052038855465,43.902254852544175)
(140.560073085304,14.262014662222905)
(64.76893973873335,29.583209859332392)
(91.06918321965131,35.10287488444342)
(103.91534491656758,55.18665083165407)
(95.74950507222685,48.26358284170746)
(110.89007960869594,28.24726650573921)
(97.83588560070503,42.62420075513481)
(108.74136308988508,58.36136280495761)
(114.08792076015888,38.80848899571575)
(94.41763741459621,48.15974163026026)
(123.49484795187792,56.333069012117726)
(88.57705655507738,35.87592317985135)
(100.07613465234628,37.95656259329405)
(120.22409485013874,20.903635123076867)
(94.51684548522026,48.824007085564475)
(97.85847801610113,42.12842021953788)
(82.48214915536929,20.855974212860083)
(108.88204289141508,47.35636031133293)
(84.82736802509379,38.47157396651124)
(90.87743445654237,44.44892897206099)
(88.2144314205803,46.915998217591564)
(91.71989230789967,32.9519089414002)
(114.89716624543391,56.56857969912898)
(83.89823023551554,69.01447390426779)
(107.1912445287056,93.72196699536431)
(112.88453989027306,67.68174169165843)
(122.65917477169792,36.05750140676429)
(71.31152491419977,59.612032548052014)
(99.30066334706373,83.72604820265472)
(73.62834985492171,91.53634527892636)
(88.61154972148101,60.13420039148272)
(82.35564675117102,34.78651932454821)
(88.76196136526997,60.40463018969642)
(113.06921308395465,53.72593775288472)
(85.12466257522948,71.23968836189438)
(82.94852412017116,14.962404799461488)
(67.12196343740192,27.0688738504954)
(89.44430063600153,22.250962828922308)
(94.16325627799739,52.37971699506817)
(109.64632784346196,64.42367272437879)
(117.27208968014271,52.85214726086421)
(108.21682171103434,64.80779234164318)
(84.53812206200922,22.408770445542125)
(140.98523025615825,96.04403901807322)
(87.357091065904,36.98588650446378)
(130.25995545608393,51.253926878697634)
(88.94209792985575,79.31027039993369)

          };

%
          
          \addlegendentry{IBEA}
       \end{axis}

    \end{tikzpicture}
\subcaption{$C_2$}
    \end{subfigure}
~
  \begin{subfigure}[t]{0.3\columnwidth}
\begin{tikzpicture}
    \begin{axis}[
      width=5cm,
    height=5cm,
    ymin=0,
    ymax=100,
    xmin=30,
    xmax=150,
        xlabel=Cost (\$),
        ylabel=Throughput (requests/s),
       legend style={at={(0.5,0.95),font=\fontsize{7}{8}\selectfont},
      anchor=north west,legend columns=1}]
           \addplot[only marks,mark=o,blue] plot coordinates {
(143.88973501184785,13.546812941807746)
(71.76112091016452,1.8470278964207554)
(102.10003327178363,11.57133864944192)
(124.22254045748353,21.891016424989065)
(148.55268492864695,26.35595776224353)
(130.27875649556452,31.83333356879912)
(170.51325663154586,17.964845942904333)
(76.8973332573427,7.921465643864451)
(114.05058075983945,27.765094379251178)
(142.40444764663448,31.483237349787473)
(139.51816520880828,40.618705228926856)
(151.90438438614507,15.676092907887039)
(134.79373841909336,20.098262175630055)
(141.87672517368875,17.24823404555312)
(75.70738298081288,1.8470278964207554)
(131.79359435153395,26.977106880826426)
(141.20755286102093,20.97404473445627)
(131.0633635029546,27.944518359237534)
(154.7089063651921,15.775389545838792)
(111.57943908567557,19.533153221386986)
(145.58310600891332,26.1929532415851)
(85.64468743614812,19.059475096145405)
(100.26180943980741,31.00094768377023)
(111.35367543911757,30.070266873596395)
(118.61982454287343,10.735905018696316)
(113.28633031384449,17.65033847913776)
(139.36155263451775,43.13723592178268)
(125.24416643862884,19.40573745597362)
(86.89285507383659,20.942289838394633)
(118.66848549554575,36.05690956063863)
(106.68826886282895,13.741269951064822)
(132.35094738600537,33.48943590768324)
(113.2631373303563,35.47255502658476)
(79.98907137927279,35.774355083517094)
(115.96472436474059,11.360887996462637)
(135.28481082428254,39.478030834985034)
(109.67455891638946,27.605923717776175)
(107.36651765388304,11.841465235361056)
(100.09375609135981,19.297846786002903)
(111.14886146846682,50.9960968710218)
(77.9203183134779,39.26475571825333)
(59.81265716581652,27.08306952433823)
(106.11928826391501,37.06912843059966)
(69.571419671169,13.265151175635708)
(100.73067302271494,54.50080603830407)
(108.84271074766858,30.54052997073282)
(142.0908719047668,51.76344625979009)
(101.46136583383108,55.347619439910225)
(129.91337471131624,22.680780266824385)
(128.89962295146944,36.389648523807296)
(93.26271229919712,14.999398739936254)
(109.18703248193106,32.86286999211204)
(124.31587234676172,44.55447913846157)
(88.55862606969876,35.749900623433874)
(126.54445302799,31.623994952453437)
(119.21436942596512,36.41834584801973)
(121.4628349656742,9.139245504655271)
(115.94485396507949,32.011531732570205)
(140.15632569894575,43.805681146368535)
(131.5930501988117,29.693716569126664)
(117.88324651916162,13.36316658514882)
(137.9140171560217,26.620705166858937)
(87.53643351541199,30.94362629099177)
(91.70150502526178,7.779396936401245)
(131.87740483455786,38.83296207378194)
(183.33253563953528,36.11326493727389)
(124.87889699741125,16.93923714497614)
(154.13719064355865,46.75010642611526)
(144.47632286811285,45.36156053364879)
(166.51052011431474,47.45917241178935)
(116.20918595483855,34.59711755641258)
(116.06305311348433,24.502008558431537)
(111.3961489269907,34.47040482268703)
(108.22291700254192,52.823422797570686)
(166.09900294841128,52.823422797570686)
(129.37943757016723,30.626820520731343)
(156.54976915591254,41.10178660422849)
(151.2170908577515,13.050818169131439)
(118.85235110151069,16.820281942889366)
(140.90424127271027,30.414196469180443)
(149.4870100560412,22.023481705002293)
(129.95070813189554,17.868729190704528)
(148.0729220743807,29.154232117843296)
(120.20999572047394,48.07181448830844)
(110.81211519768738,53.8810798121524)
(130.61949530698706,21.53298788671591)
(97.2993453595526,23.897322868828194)
(147.89367527078014,32.220578021763956)
(179.10179102384146,15.624456358707919)
(105.49417915059067,24.98572947920531)
(142.48196172902436,88.66512832237238)
(131.66844605873354,13.209732733780156)
(132.29148687901818,13.978096318802029)
(106.94244142030625,6.796957263145544)
(121.61129901807706,46.57539329988069)
(117.35564152054422,33.48275626915891)
(104.75852408214024,33.64056388577873)
(109.93890232677455,8.933811121337813)
(104.39550730049773,7.340053620305198)
(132.7759461783551,32.0517526742652)
(135.34054662612022,16.661367378240648)
(105.46068931244409,17.816126651831254)

          };

%
          
          \addlegendentry{IBEA}
       \end{axis}

    \end{tikzpicture}
\subcaption{$C_3$}
    \end{subfigure}

    \caption{Benefits on \textsc{SOA} under further investigated candidates over all runs.}
  \label{fig:soa-switch}
  \end{figure}

As shown in Figure~\ref{fig:rubis-switch} and~\ref{fig:soa-switch}, clearly, we see that for all cases, in contrast to $C_3$, $C_2$ finds more solutions that are condensed to the bottom-left (top-left for \textsc{SOA}) corner of the objective space. This means that more advanced synergy helps to enable more promising results in the optimization. When comparing $C_1$ and $C_2$, the solutions are even more condensed to the ideal corner under $C_1$, and is of particular significance in the case of \textsc{SOA} due to its stronger extents of conflicts. This proves that allowing the underlying algorithm for goal-awareness to be aware of the domain expertise, although impose higher design difficulty, can be very beneficial in terms of the results.

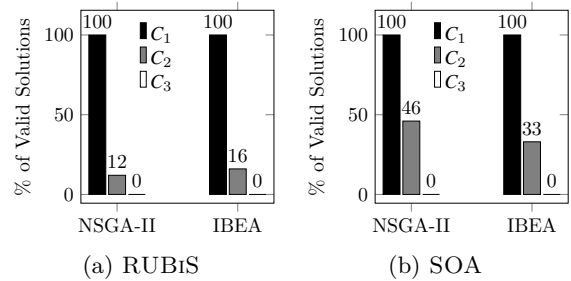
\begin{figure}
  
  \begin{subfigure}[t]{0.4\columnwidth}
\begin{tikzpicture}
\begin{axis}[
    ybar=0.05cm,
    ymax=110,
    width=5cm,
    height=5cm,
    bar width=0.3cm,
 enlarge x limits=0.3,
 enlarge y limits=0.05,
    legend style={draw=none,at={(0.4,0.96)},
      anchor=north,legend columns=1},
    ylabel={\% of Valid Solutions},
    nodes near coords,
    symbolic x coords={NSGA-II,IBEA},
    xtick=data
    ]

\addplot[fill=black] coordinates {(NSGA-II,100) (IBEA,100)};
\addplot[fill=gray] coordinates {(NSGA-II,12) (IBEA,16)};
\addplot[fill=white] coordinates {(NSGA-II,0) (IBEA,0)};



\legend{$C_1$,$C_2$,$C_3$}

\end{axis}
\end{tikzpicture}
\subcaption{\textsc{RUBiS}}
    \end{subfigure}
    ~
      \begin{subfigure}[t]{0.4\columnwidth}
\begin{tikzpicture}
\begin{axis}[
    ybar=0.05cm,
    ymax=110,
    width=5cm,
    height=5cm,
    bar width=0.3cm,
 enlarge x limits=0.3,
 enlarge y limits=0.05,
    legend style={draw=none,at={(0.4,0.96)},
      anchor=north,legend columns=1},
    ylabel={\% of Valid Solutions},
    nodes near coords,
    symbolic x coords={NSGA-II,IBEA},
    xtick=data
    ]

\addplot[fill=black] coordinates {(NSGA-II,100) (IBEA,100)};
\addplot[fill=gray] coordinates {(NSGA-II,46) (IBEA,33)};
\addplot[fill=white] coordinates {(NSGA-II,0) (IBEA,0)};



\legend{$C_1$,$C_2$,$C_3$}

\end{axis}
\end{tikzpicture}
\subcaption{\textsc{SOA}}

        \end{subfigure}

    \caption{Mean percentage of valid solutions found under further investigated candidates over all runs.}
  \label{fig:valid}
  \end{figure}

Figure~\ref{fig:valid} illustrates the mean percentage of valid solutions found, and we see that the $C_1$ achieves 100\% valid solution as the evolutionary algorithm is aware of the expertise about the dependency during the evolution, which promotes the ability to actively repair the solutions that violate dependency. $C_2$, on the other hand, do not have such benefits but it is more likely to result in valid solutions than that of $C_3$. This is because $C_2$ encodes a set of automatically extracted and more elitist features to be tuned.  $C_3$, in contrast, encodes all the features in the feature model, which can hardly find valid solutions given the high number of features in the subject software systems.

For the difficulty shown in Table~\ref{table:1st-bd}, as expected, $C_1$ has the highest LOC which implies higher difficulty in implementation and maintenance. $C_2$ is ranked the second but its differences to $C_3$ is small, which suggests that the difficulty related to automatically extracting the important features is considerably low.

\begin{bclogo}[couleur=gray!10,arrondi=0.1,epBord=1.5,couleurBord=black!70,logo=\bctrombone]{{\normalsize Final choice for the first case study:}}According to the verified results from the further investigation, the team has decided that $C_1$ is a more preferred and promising choice for production. 
\end{bclogo}

\begin{table}[t!]
\centering
   \caption{The LOC for the first case study}
\label{table:1st-bd}

\begin{tabularx}{0.4\columnwidth}{YY}
\toprule 
\textbf{Candidate}&\textbf{LOC}\\ \midrule
$C_1$&80,718\\ 
$C_2$&74,092\\ 
$C_3$&73,466\\
\bottomrule

\end{tabularx}
\end{table}

\subsection{Self-Awareness and Self-Adaptation for Change Expensive Software Systems}

\textbf{Context:} Self-adaptive software systems may subject to financial contracts with respect to its performance and resource consumption to perform adaptation. For example, a software system deployed on the Cloud Computing platform are charged on the amount of resources it consumes, and it may incur monetary penalty (or reward) for violating (or exceeding) some agreed threshold of performance. In particular, the adaptations in the target self-adaptive software systems are often expensive, or the reasoning process related to the adaptation is resource consuming, and therefore in certain cases, it could be more beneficial to not adapting.

\textbf{Problem:} In the second case study, the aim is to dynamically determine when and whether to adapt those critical software systems for which adaptations can impose expensive cost. This again exhibits a strong requirement of self-awareness. 

\textbf{Challenge:} The key challenge is how to model and reason about the dynamic and uncertain cost-benefit between adapting the software system and not adapting it, then deciding on when and whether to adapt. It is required to measure the software systems not only on the achieved quality of non-functional attributes, but also, in terms of the monetary values that it generates, or carry as debts.

\subsubsection{Patterns and Algorithms}

After analyzing the requirements and following the handbook~\cite{2014epicshandbook}, it has been concluded that there is no interaction awareness required, as the target software system was not aim for distributed environment, and that it is considered as satisfied to optimize the local goal for a single self-adaptive system. Further, the environment is not expected to actively react on the adaptation of the software system, and thus no interaction between it and the environment. There is also no need for a meta-self-awareness, because the requirements on the required capabilities is clear and that the problem itself aims to reduce the extra computations involved in the self-adaptation process. Therefore the overhead produced by meta-self-awareness, which could be potentially high, should be better avoid. Indeed, the self-adaptive software system itself is often goal-aware due to the need of explicitly reasoning on the goals and objectives. However, for the problem that should be dealt with (i.e., when and whether to adapt), extensive reasoning on the goals is not the key purpose; rather, it is more related to track and make a binary decision: to adapt or not to adapt, drawing on insights about their time-varying cost-benefits. As a result, these have led to the conclusion that the \emph{Temporal Knowledge Aware Pattern} as the appropriate pattern for the design. The pattern has been illustrated in Figure~\ref{fig:p7}.

In this case study, the primary goal is to model the time-varying cost-benefit on the decision of adapting and not adapting the software system. Therefore, by following the steps in the handbook~\cite{2014epicshandbook}, machine learning algorithm has been identified as the promising way to handle the problem. This is because they are often effective in producing fast prediction in acceptable time, given that sufficient amount of past samples. Since there are only two decisions to model, the problem can be rendered as a binary classification problem, where, given a set of features, (e.g., software system status, environment changes, etc) the model aims to predict whether it is better to adapt or not. Again, given the generality of the target software system, the solution should not be specific to a particular machine learning algorithm, and thus it should support a wide range of the types, allowing for better flexibility on customization. As for the stimulus-awareness, it can be easily realized by periodic detection. A complete list of the algorithms and techniques that realize the capabilities of self-awareness involved are show in Table~\ref{table:2nd-alg}.

\begin{table}[t!]

   \caption{The algorithms and techniques that realize the capabilities of self-awareness for the second case study.}
\label{table:2nd-alg}
\centering
\begin{tabularx}{0.9\columnwidth}{p{2.5cm}X}
\toprule 
\textbf{Self-Awareness}&\textbf{Algorithms and Techniques}\\ \midrule
stimulus awareness&periodic detection \\
time-awareness&machine learning \\
\bottomrule

\end{tabularx}
\end{table}

\subsubsection{Representations of Expertise}

Here, there are two representations of expertise, namely the Service Level Agreement (SLA) and the technical debt concept.

 \lstset{
language=XML,
  basicstyle=\ttfamily,
  columns=fullflexible,
  showstringspaces=false,
  commentstyle=\color{gray}\upshape
}

\definecolor{maroon}{rgb}{0.5,0,0}
\definecolor{darkgreen}{rgb}{0,0.5,0}
\lstdefinelanguage{XML}
{
  basicstyle=\ttfamily,
  morestring=[s]{"}{"},
  morecomment=[s]{?}{?},
  morecomment=[s]{!--}{--},
  commentstyle=\color{darkgreen},
  moredelim=[s][\color{black}]{>}{<},
  moredelim=[s][\color{red}]{\ }{=},
  stringstyle=\color{blue},
  identifierstyle=\color{maroon}
}

\begin{center}
\tiny\linespread{0.8}
\begin{minipage}{\columnwidth}
\begin{minipage}{.48\columnwidth}
\begin{lstlisting}[language=XML]
<wsag:GuaranteeTerm Name="Latency">
  <wsag:ServiceScope ServiceName="Adaptive System"/>
  <wsag:QualifyingCondition>
       {"function" : "AVG EVERY 100s"}
  </wsag:QualifyingCondition>
  <wsag:ServiceLevelObjective>
       <wsag:KPITarget>
          <wsag:KPIName>MeanTime</wsag:KPIName>
          <wsag:CustomServiceLevel>
               {"constraint" : "MeanTime LESS THAN 0.12s"}
           </wsag:CustomServiceLevel>
       </wsag:KPITarget>
  </wsag:ServiceLevelObjective>
  <wsag:BusinessValueList>
       <wsag:Penalty>
          <wsag:AssessmentInterval>
              <wsag:TimeInterval>100s</wsag:TimeInterval>
          </wsag:AssessmentInterval>
          <wsag:ValueUnit>USD_PER_SECOND</wsag:ValueUnit>
          <wsag:ValueExpression>2.76</wsag:ValueExpression>
        </wsag:Penalty>
        <wsag:Reward>
          <wsag:AssessmentInterval>
              <wsag:TimeInterval>100s</wsag:TimeInterval>
          </wsag:AssessmentInterval>
          <wsag:ValueUnit>USD_PER_SECOND</wsag:ValueUnit>
          <wsag:ValueExpression>1.13</wsag:ValueExpression>
       </wsag:Reward>
  </wsag:BusinessValueList>
</wsag:GuaranteeTerm>
\end{lstlisting}
\end{minipage}
~\hspace{0.7cm}
\begin{minipage}{.48\columnwidth}
\begin{lstlisting}[language=XML]
<wsag:GuaranteeTerm 
        Name="CPUTime">
    <wsag:ServiceScope 
        ServiceName="Engine"/>
    <wsag:ServiceLevelObjective>
        <wsag:KPITarget>
            <wsag:KPIName>
                CPUTime
            </wsag:KPIName>
            <wsag:CustomServiceLevel>
                {"constraint" : 
                      "CPUTime LESS THAN 0s"}
            </wsag:CustomServiceLevel>
        </wsag:KPITarget>
    </wsag:ServiceLevelObjective>
    <wsag:BusinessValueList>
        <wsag:Penalty>
            <wsag:AssessmentInterval>
                <wsag:Count>1</wsag:Count>
            </wsag:AssessmentInterval>
            <wsag:ValueUnit>
                  USD_PER_SECOND
            </wsag:ValueUnit>
            <wsag:ValueExpression>
                  0.0788
            </wsag:ValueExpression>
        </wsag:Penalty>
    </wsag:BusinessValueList>
</wsag:GuaranteeTerm>
\end{lstlisting}
\end{minipage}
\captionof{figure}{The fragment of a SLA.}
\label{fig:sla}
\end{minipage}
\end{center}

In general, SLA is a formal legal binding negotiated between the software company and the end users before the software system is built~\cite{skene2009service}. An example fragment of the typical SLA, derived from the well-known WS-Agreement~\cite{andrieux2007web}, is shown in Figure~\ref{fig:sla}, which states the rate of reward and penalty on the mean latency (\$/s) and rate of CPU time of planning (\$/s) for a software system. Specifically, the SLA states that the rate for the cost of adaptation is \$0.345 per CPU second; and an adaptation that utilizes 2s would lead to a total cost of \$0.69. Similarly, the SLA may contain a penalty rate of mean latency violation as \$0.043/s for a requirement of 2s. Therefore, if there is a mean latency of 2.5s for a period, then the penalty for it would be (2.5 $-$ 2) $\times$ 0.043 = \$0.0215.  

Technical debt for software engineering was coined by Cunningham~\cite{Cunningham:1992}, to help deciding whether to improve the software, considering the costs and benefits of improvement versus that of not improving it. In general, when software faces bugs or requires improvement, the engineers have two options: (i) improve the software, in which case the quality of the software may be improved, but extra rework cost would needs to be paid for the human and resources spent, or (ii) leave it as it is, and thereby the software remain as flawed, which could accumulate the interests incurred by the bugs. The benefit of technical debt concept is that it offers an intuitive way for software and system engineers to make decision about whether to improve or not, and to track the debt over time.

  \subsubsection{Candidates Creation}

At this step, the team creates all the possible ways of synergy by instantiating the enriched self-awareness architectural pattern. In particular, they answer the questions presented in Section~\ref{sec:method} as follows:

\begin{enumerate}

\item Which category does the expertise representation belong to? 
\begin{itemize}
\item[---] \textbf{\emph{Answer:}} SLA belongs to the \texttt{Documentation} category but technical debt belongs to the \texttt{Concept}.
\end{itemize}
\item If such a representation structural? is it tangible?
\begin{itemize}
\item[---] \textbf{\emph{Answer:}} SLA is both structural and tangible while technical debt is neither structural nor tangible.
\end{itemize}
\item The expertise representation can be synergized with which algorithm/technique that realizes the self-aware capability? What are the possible levels of synergy?
\begin{itemize}
\item[---] \textbf{\emph{Answer:}} SLA needs to be synergized with both self-aware capabilities in the \emph{Temporal Knowledge Aware Pattern}, but information and expertise related to technical debt needs to be used with the time awareness only. The synergy between SLA and stimulus awareness needs to be at \emph{level 1}, while for time awareness, it can be of any level (including \emph{level 0}). Similarly, the technical debt can be synergized with time awareness at any level (including \emph{level 0}). 
\end{itemize}
\item What is the possible form for each synergy?
\begin{itemize}
\item[---] \textbf{\emph{Answer:}} All the synergies need to be realized in a \emph{general} form.
\end{itemize}
\item What is the difficulty level for each synergy?
\begin{itemize}
\item[---] \textbf{\emph{Answer:}} According to Figure~\ref{fig:difficulty}, the difficulty level ranges between \emph{very easy} to \emph{challenging}.
\end{itemize}

\end{enumerate}

The above answers produce 16 different candidates of synergizing domain expertise represented as the enriched \emph{Temporal Knowledge Aware Pattern}.


   \begin{figure}[!t]
    \centering
   \begin{tikzpicture}
    \begin{axis}[
        grid = major,
        grid style={dashed},
       ylabel=Difficulty Score,
        xlabel=Benefit Score,
           label style={font=\fontsize{12}{12}\selectfont},
                    tick label style={font=\fontsize{12}{12}\selectfont},
      legend style={font=\fontsize{12}{12}\selectfont,at={(0.5,0.8)}, anchor=east,legend columns=1}]

 \addplot[only marks,mark=*,mark options={scale=2.5,fill=green!50}] plot coordinates  {
(7.905,2)
(9.18,2.26)
(9.705,2.64)
(10.23,3.2)
(9.435,2.22)
(10.71,2.48)
(11.235,2.86)
(11.76,3.42)
(10.065,2.37)
(11.34,2.63)
(12.39,3.57)
(10.695,2.53)
(11.57,2.79)
(12.445,3.17)
(13.02,3.73)
 };
 
  \addplot[only marks,mark=otimes*,mark options={scale=2.5,fill=red!50}] plot coordinates  {
(11.865,3.01)
(11.235,2.86)
  };

\addlegendentry{Unselected}
\addlegendentry{Selected}
    \end{axis}

    \end{tikzpicture}
    \caption{Difficulty and benefit scores for all candidates in the second case study.}
        \label{fig:exp-d2}
  \end{figure}
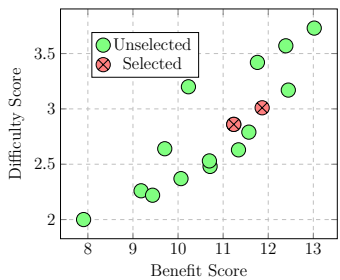

\subsubsection{Difficulty and Benefit Scores}

For all the 16 candidates, their overall scores with respect to both the difficulty and benefit are shown in Figure~\ref{fig:exp-d2}. In this context, the $w$ between \emph{specific} and \emph{general} form of synergy is set as 1.2 and 1.4, respectively. For each candidate, the proficiency is set as 1.8 for all synergies related to SLA and 1.5 for those related to technical debt.

\begin{figure}[!t]
  \centering 
    \begin{subfigure}[t]{\columnwidth}
   \includegraphics[width=\columnwidth]{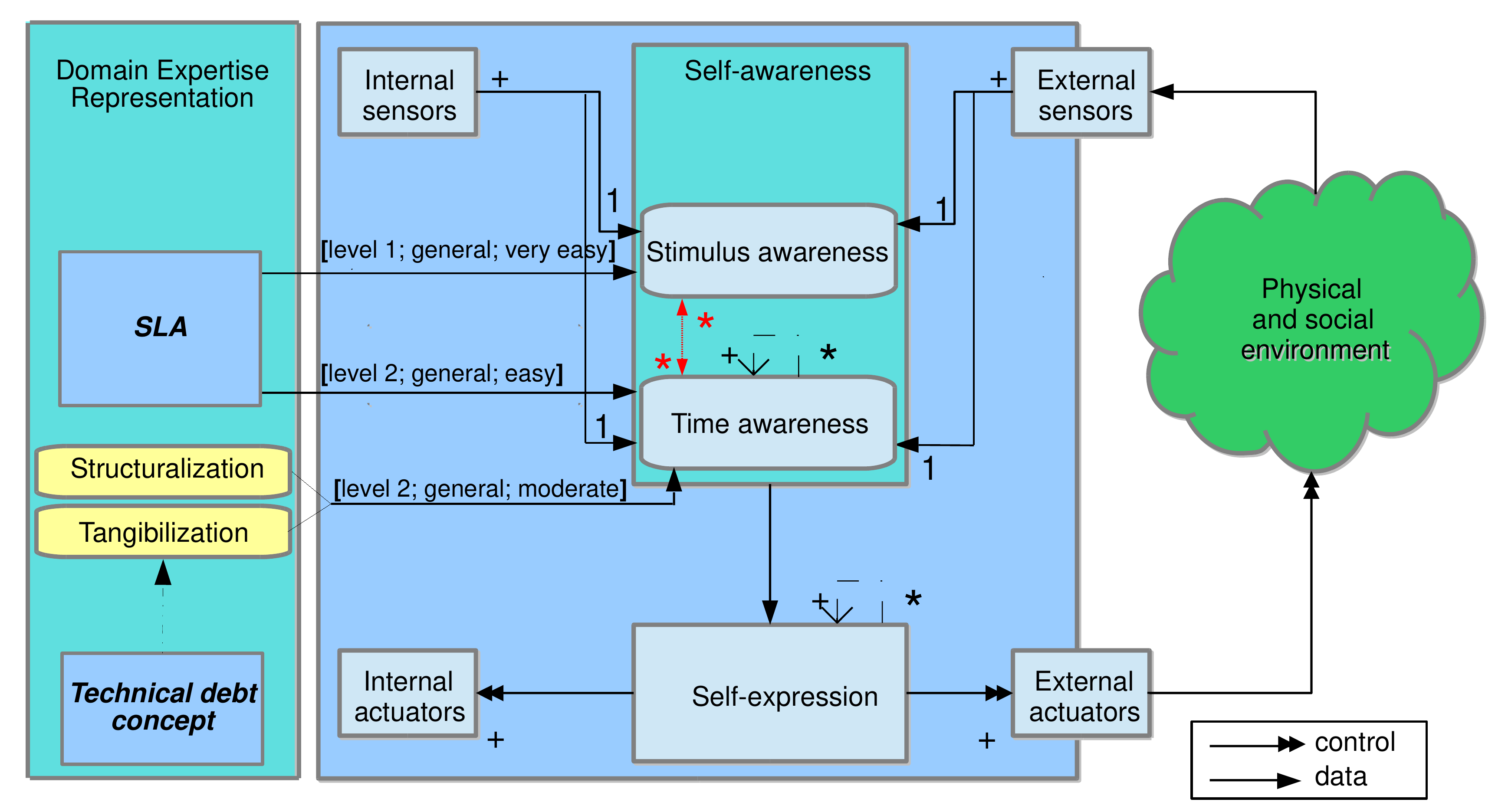}
\subcaption{candidate $C_1$}
 \label{fig:2nd-arch}
    \end{subfigure}
    
        \begin{subfigure}[t]{\columnwidth}
   \includegraphics[width=\columnwidth]{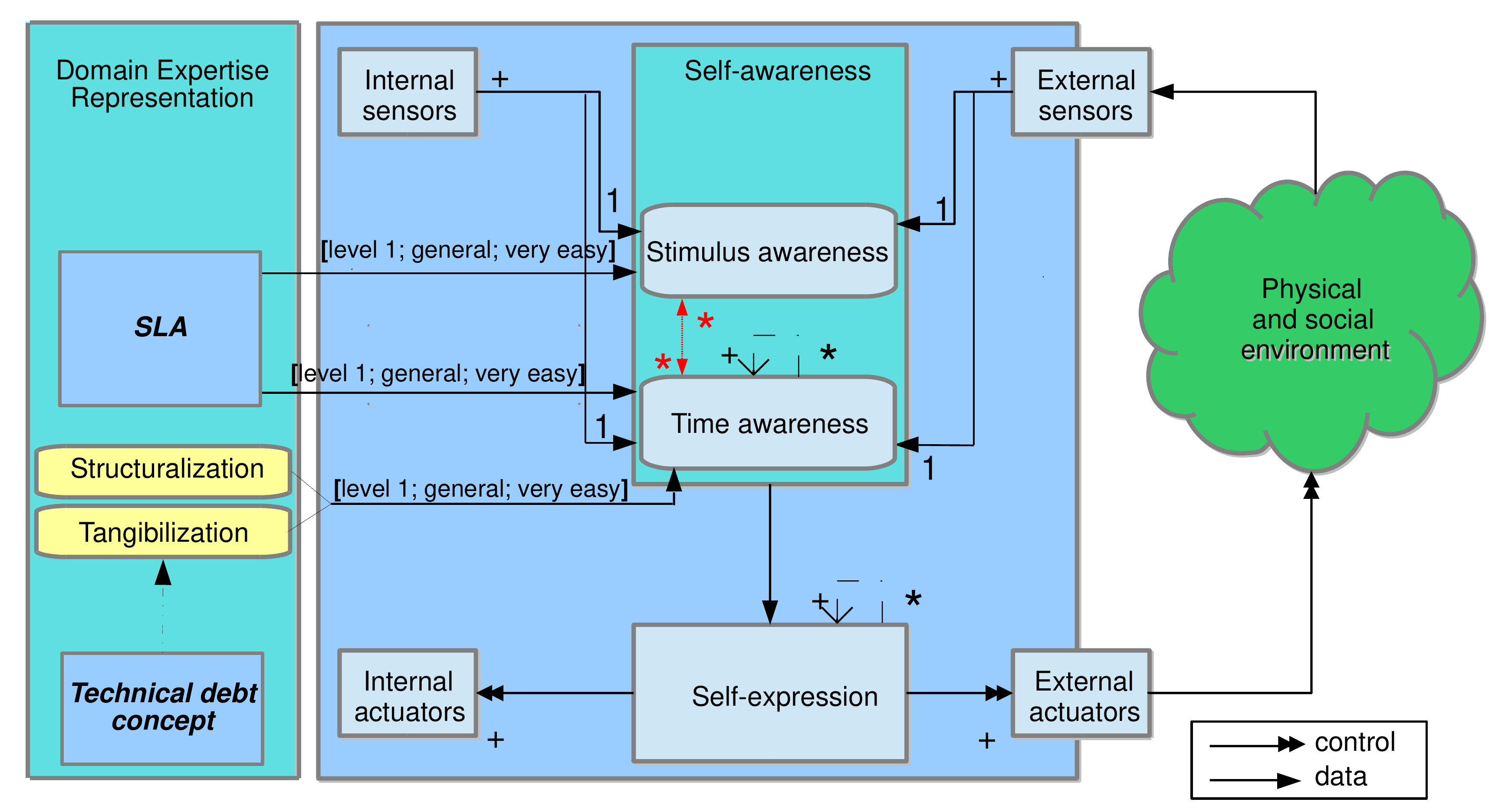}
\subcaption{candidate $C_2$}
 \label{fig:2nd-arch-2}
    \end{subfigure}

    \caption{Possible candidates selected for further investigation in the second case study.}

  \end{figure}

\subsubsection{Further Investigation}

Two candidates, as illustrated in Figure~\ref{fig:2nd-arch}, have been selected due to its superiority on the expected benefit over most other candidates, while causing an acceptable degree of difficulty. In a nutshell, they are discussed as follows:

\begin{itemize}
\item $C_1$ ($\varmathbb{B}_{C_1}=11.865$, $\varmathbb{D}_{C_1}=3.01$): As shown in Figure~\ref{fig:2nd-arch}, the candidate automatically converts the technical debt concept into structural and tangible knowledge, which can be synergized with time-awareness at \emph{level 2}. The SLA has also been parsed to extracted meaningful information to consolidate understanding regarding time as \emph{level 2} of synergy. The stimulus-awareness, however, directly use the information from SLA without additional machine reasoning at \emph{level 1} of synergy.
 
\item $C_2$ ($\varmathbb{B}_{C_2}=11.235$, $\varmathbb{D}_{C_2}=2.86$): For the candidate in Figure~\ref{fig:2nd-arch-2}, the synergy of domain expertise on the SLA and technical debt concept with the time-awareness are realized at \emph{level 1}. In this way, the time awareness merely predicts the occurrence of an event, i.e., violation of performance requirement (the only information from SLA), without further parsing on the SLA and technical debt. The prediction results is then further analyzed by statistical inference; thus only the significant, reliable and persistent violations would trigger adaptation. All the other synergies remain the same as those of $C_1$.
 
\end{itemize}

More technical details on the actual synergy approaches can be found in~\cite{chen-icpe}.

\begin{table}[t!]

\centering
   \caption{The cloud-based software system for the experiments of the second case study.}
\label{table:2nd-exp}

\begin{tabularx}{0.8\columnwidth}{p{2cm}X}
\toprule 
\textbf{Attribute}&\textbf{Setup}\\ \midrule
Latency&threshold is 0.05s; reward and penalty rate are \$3.5 per unit\\ 
Power&threshold is 5 watt; reward and penalty rate are \$0.5 per unit\\ 
CPU time of adaptation&charge rate is \$0.01 per unit\\
Enviroment&workload\\
\bottomrule

\end{tabularx}
\end{table}

\subsubsection{Further Investigation Setup}



The candidates are evaluated using \textsc{RUBiS} (detailed in Table~\ref{table:1st-exp}) as the subject software system deployed and run in the real Cloud environment, with the negotiated SLA shown in Table~\ref{table:2nd-exp}. Two distinct machine learning algorithms are run in parallel, i.e., Naive Bayes~\cite{nb} (NB) and Multilayer Perceptron~\cite{mlp} (MLP), each of which is of different complexities. The goal is to optimize the latency and power of the cloud-based software system, and thus both Multi-Objective Planner (MOP) and Single-Objective Planner (SOP), in which all objectives are combined in an equally weighted aggregation, are applied. The actual objective function to be optimized is trained based on machine learning~\cite{Chen:2015:tse,Chen:2013,Chen:2014:ucc}. However, it is worth noting that optimization is not a concern of the designed software system that is self-aware, as it is not part of selected the pattern. The number of repeated runs is 100 for the experiments, based on which the mean is reported. The results are confirmed by Wilcoxon Signed Rank test ($p<$0.05), following the effect sizes categorization in~\cite{kampenes2007systematic}.

Similar to the previous case study, the quality indicators used to assess both benefit and difficulty are shown in Table~\ref{table:2nd-qi}.

\begin{table}[t!]
\centering
   \caption{The quality indicators for benefit and difficulty for the second case study}
\label{table:2nd-qi}

\begin{tabularx}{0.7\columnwidth}{p{1.5cm}X}
\toprule 
\textbf{Attribute}&\textbf{Quality Indicators}\\ \midrule
Benefit&latency, power, debt, number and cost of adaptation\\ 
Difficulty&Lines-Of-Code (LOC)\\
\bottomrule

\end{tabularx}
\end{table}

\subsubsection{Results}

\begin{figure}
  
  \begin{subfigure}[t]{0.33\columnwidth}
\begin{tikzpicture}
\begin{axis}[
 width=5cm,
 height=6cm,
    ybar=0.15cm,
    bar width=8pt,
 enlarge x limits=0.5,
 enlarge y limits=0.05,
 legend cell align={left},
    legend style={at={(0.55,0.6)},
      anchor=north,legend columns=1},
    ylabel={Total Net Debt (\$)},
          y label style={at={(-0.25,0.5)}},
    symbolic x coords={MOP,SOP},
    xtick=data
    ]

\addplot[fill=black] coordinates {(MOP,8.31) (SOP,-34.43)};
\addplot[fill=gray] coordinates {(MOP,23.69) (SOP,-19.61)};
\addplot[fill=white] coordinates {(MOP,885.48) (SOP,1168.83)};

\legend{$C_1$ (NB),$C_1$ (MLP),$C_2$}

\end{axis}
\end{tikzpicture}
    \end{subfigure}
    ~\hspace{-0.2cm}
      \begin{subfigure}[t]{0.33\columnwidth}
\begin{tikzpicture}
\begin{axis}[
 width=5cm,
 height=6cm,
    ybar=0.15cm,
    bar width=8pt,
 enlarge x limits=0.5,
 enlarge y limits=0.05,
  legend cell align={left},
    legend style={at={(0.62,0.6)},
      anchor=north,legend columns=1},
    ylabel={\# of Adaptations},
          y label style={at={(-0.25,0.5)}},
    symbolic x coords={MOP,SOP},
    xtick=data
    ]

\addplot[fill=black] coordinates {(MOP,12) (SOP,15)};
\addplot[fill=gray] coordinates {(MOP,24) (SOP,14)};
    
\addplot[fill=white] coordinates {(MOP,34) (SOP,43)};

\legend{$C_1$ (NB),$C_1$ (MLP),$C_2$}

\end{axis}
\end{tikzpicture}
     \end{subfigure}
    ~\hspace{-0.2cm}
      \begin{subfigure}[t]{0.33\columnwidth}
\begin{tikzpicture}
\begin{axis}[
 width=5cm,
 height=6cm,
    ybar=0.15cm,
    bar width=8pt,
 enlarge x limits=0.5,
 enlarge y limits=0.05,
  legend cell align={left},
    legend style={at={(0.55,0.93)},
      anchor=north,legend columns=1},
    ylabel={Costs of Adaptations (\$)},
          y label style={at={(-0.25,0.5)}},
    symbolic x coords={MOP,SOP},
    xtick=data
    ]

\addplot[fill=black] coordinates {(MOP,3.81) (SOP,4.60)};
\addplot[fill=gray] coordinates {(MOP,7.60) (SOP,4.30)};
    
\addplot[fill=white] coordinates {(MOP,8.33) (SOP,11.73)};

\legend{$C_1$ (NB),$C_1$ (MLP),$C_2$}

\end{axis}
\end{tikzpicture}

        \end{subfigure}

    \caption{The total debt, number and cost of adaptation under further investigated candidates over all runs.}
  \label{fig:debt-info}
  \end{figure}
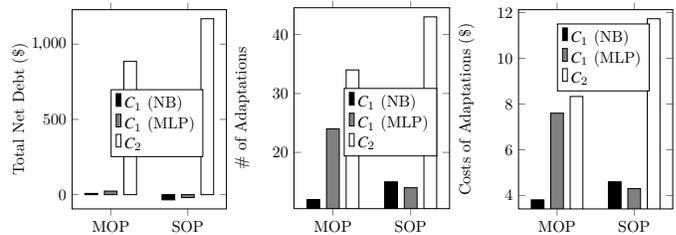

From Table~\ref{table:dlda-t}, we see that for all cases, the $C_1$ under all algorithms outperforms the $C_2$ on both quality attributes, with statistical significance and non-trivial effect size on at least one attribute. In Figure~\ref{fig:debt-info}, we also observe that $C_1$ has led to less debt, meaning that the monetary value generated by the software system, after synergizing the domain expertise with automatic machine reasoning, is higher than the case when the synergy is limited. We can also note that such benefit is achieved by using remarkably smaller amount and cost of adaptation.

\begin{figure}[!t]
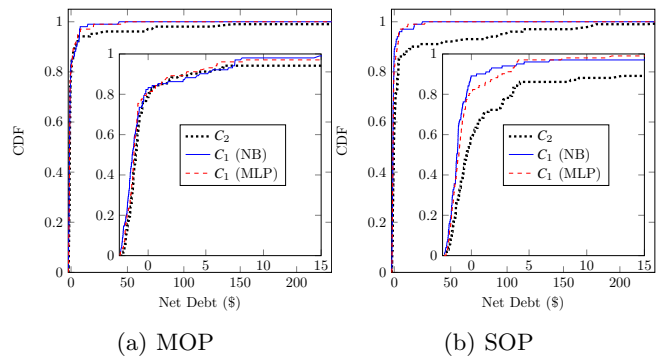

  {
\centering
     \begin{subfigure}[t]{0.5\columnwidth}
  \includestandalone[width=\columnwidth]{tikz/femosaa-cum-debt}
      \subcaption{MOP}
  \label{fig:femosaa-debt}
    \end{subfigure}
 ~\hspace{-0.5cm}
     \begin{subfigure}[t]{0.5\columnwidth}
  \includestandalone[width=\columnwidth]{tikz/plato-cum-debt} 
     \subcaption{SOP}
  \label{fig:plato-debt}
    \end{subfigure}
      
  }

    \caption{The cumulative distribution function of debt under further investigated candidates over all runs.}

  \end{figure}

To gain a better understanding about the total debt, we plot the debt throughout for an entire run. Figure~\ref{fig:femosaa-debt} shows the cumulative distribution of debt for different levels of synergy, when using multi-objective planner. We can see clearly that, in contrast to others, $C_1$ with the two machine learning algorithms reduces the debt quicker as their slopes are much steeper than the $C_2$. Yet, the superiority of  $C_2$ on debt reduction is much more obvious when the debt is greater than about \$9. Figure~\ref{fig:plato-debt} compares the cumulative debt of approaches when using single-objective planner. Here, we see that $C_1$ is again significantly outperforms the case when there is no actual synergy, with faster reduction on the debt.

The difficulty in terms of LOC is shown in Table~\ref{table:2nd-bd}. As can be seen, $C_1$ requires higher LOC than $C_2$, which implies higher difficulty in implementation and maintenance. This is predictable based on the benefit/difficult plot. However, we did not expected that the margin is as little as 3,672 lines, suggesting that the difficulty difference is in fact negligible given the much better benefits brought by $C_1$.

\begin{table}[t!]
\centering
   \caption{Benefits in terms of latency and power, their statistical significance and effect sizes (ES) over all runs.}
\label{table:dlda-t}
\begin{tabularx}{\columnwidth}{P{1.8cm}P{0.6cm}YP{0.6cm}Y}
\toprule 
\multirow{2}{*}{\textbf{{MOP}}}&
\multicolumn{2}{c}{Latency (ms)}&
\multicolumn{2}{c}{Power (watt)}

\\ \cmidrule{2-5}
&Mean&$p$ value (ES)&Mean&$p$ value (ES)
 \\ \midrule

$C_2$&
2.69 &
-&
3.58 &
-
\\

$C_1$ (NB)&
\bfseries0.19 &
\cellcolor{gray!50}.003 (large) &
4.10 & 
.810 (trivial)
\\

$C_1$ (MLP)&
0.32 &
\cellcolor{gray!50}.017 (small)&
3.42& 
.576 (trivial)
\\ 

\midrule

\textbf{{SOP}} &
&
&

 \\ \midrule 

$C_2$&
3.32 &
-&
5.06 &
-
\\

$C_1$ (NB)&
\bfseries0.19 &
\cellcolor{gray!50}$<$.001 (large) &
\bfseries3.22&
\cellcolor{gray!50}$<$.001 (large)
\\

$C_1$ (MLP)&
0.18 &
.108 (small)&
3.64& 
\cellcolor{gray!50}$<$.001 (large)
\\

\bottomrule

\end{tabularx}
\end{table}

\begin{bclogo}[couleur=gray!10,arrondi=0.1,epBord=1.5,couleurBord=black!70,logo=\bctrombone]{{\normalsize Final choice for the second case study:}}According to the verified results from the further investigation, $C_1$ is deemed as more suitable and thus it is chosen for production.
\end{bclogo}

\begin{table}[t!]
\centering
   \caption{The LOC for the second case study}
\label{table:2nd-bd}

\begin{tabularx}{0.4\columnwidth}{YY}
\toprule 
\textbf{Candidate}&\textbf{LOC}\\ \midrule
$C_1$&69,343\\ 
$C_2$&65,671\\
\bottomrule

\end{tabularx}
\end{table}

\subsection{Self-Awareness and Self-Adaptation for Rapidly Composed Software Systems}

\textbf{Context:} Service systems, unlike the others, do not have the actual implementation. Instead, they have a set of abstract services, each of which can be adapted to select different concrete services published in the Internet, according to a given workflow with different predefined connectors (sequential or parallel)~\cite{wada2012e3,DBLP:conf/icws/KumarB0B19,seeding,DBLP:conf/icse/ChenB20}. Such a process, namely service composition, is the key to enable rapid realization and integration of different functionalities that are required by the stakeholders. This is also a benefit of service systems, such that they share some similarities which make the exploitation of past problem instances and experiences possible.

\textbf{Problem:} In the third case study, the aim is to conduct multi-objective optimization for rapidly composing self-adaptive service systems at runtime, leveraging the benefits from the capabilities of self-awareness. 

\textbf{Challenge:} The challenge here is that there is often a large number of services to fulfil the same functional requirement, but come with different levels on some possibly conflicting non-functional Quality-of-Service (QoS) attributes, e.g., latency, throughput and cost. Thereby optimizing and finding the good service composition plans, i.e., a set of selected concrete services, and their trade-offs becomes a complex and challenging problem which is known to be NP-hard~\cite{ramirez2017evolutionary,wada2012e3}. In addition, given the potentially rapid needs of composing the services, the optimization requires fast convergence to ensure the effectiveness of the optimized composition plan.

\subsubsection{Patterns and Algorithms}

Following the procedures from the handbook~\cite{2014epicshandbook}, it has been concluded that the requirements in this case study do not involve interaction awareness, because there is no way to know in advance what are the concrete services available, thus there is often a service broker that act as a centralized point to compose a service system. Further, the environment is not expected to react on the adaptation of the software system, hence no interaction between it and the environment. The meta-self-awareness has been ruled out as the requirements on the required capabilities is clear, and no need to introduce extra overhead. goal-awareness is again essential in the optimization and time-awareness is also crucial for self-adaptive service systems, because the currently available concrete services, as well as their QoS values, could change over time, and thereby requiring a model that cope with such a change. As a result, these have led to the conclusion that the \emph{Temporal Goal Aware Pattern} as the appropriate choice for the design. The pattern has been illustrated in Figure~\ref{fig:p9}.

Given the NP-hard problem with an explosion of the search space and the nature of multi-objectivity for the self-adaptive services systems, the handbook~\cite{2014epicshandbook} has suggested that the metaheuristic algorithms, particularly the evolutionary algorithms, are promising to realize the capability of goal-awareness in the software systems. Yet, given the high diversity of the workflow structures, it is expected that the solution does not tie to a specific evolutionary algorithm, rather, it should support a wide range of evolutionary algorithms. The time-awareness is supported by an analytical model, which tracks the available set of concrete services and their QoS values, and is capable of evaluating the aggregated QoS value for the workflow. The stimulus-awareness can be realized by event driven detection, such that the stimulus is captured through passive detection. A complete list of the algorithms and techniques, with respect to the capabilities of self-awareness involved, are show in Table~\ref{table:3rd-alg}.

\begin{table}[t!]
\centering
   \caption{The algorithms and techniques that realize the capabilities of self-awareness for the third case study.}
\label{table:3rd-alg}

\begin{tabularx}{0.9\columnwidth}{p{2.5cm}X}
\toprule 
\textbf{Self-Awareness}&\textbf{Algorithms and Techniques}\\ \midrule
stimulus awareness&event driven detection \\
time-awareness&analytical model \\
goal-awareness&evolutionary algorithm \\
\bottomrule

\end{tabularx}
\end{table}

\begin{figure}[!t]
  \centering 
   \includegraphics[width=\columnwidth]{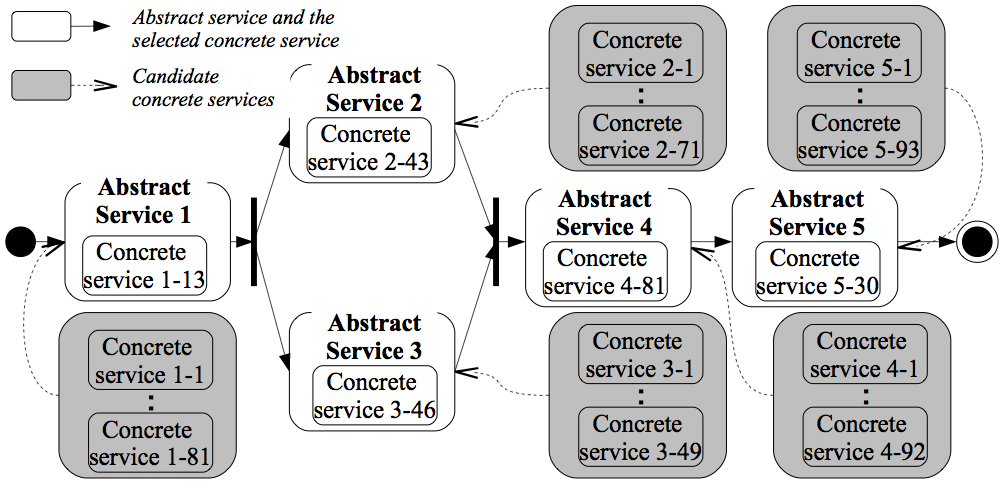}
    \caption{An example model of workflow structure for a service composition.}
 \label{fig:worflow}
  \end{figure}

\subsubsection{Representations of Expertise}

There are two fundamental representations of the expertise in this case: the workflow structure of the service composition and past problem instances/experience about the optimization when composing services.

As shown in Figure~\ref{fig:worflow}, where we can see that the workflow is represented as a graph and each vertex represents an abstract service. The edge denotes the connector between vertices, e.g., they can be either \emph{sequential} where the users' requests are proceed in strict order or \emph{parallel} such that different users' requests are handled by simultaneously. 

Another important representation of expertise is past problem instances and experience about the service composition. In the context of service composition, adaptation is required when change occur, e.g., the QoS of concrete services changes or some concrete services becomes unavailable. These changes, albeit can occur rapidly, often occur in relatively small extents. As a result, past problem instances and experience can still provide useful information for the scenario after changes occur. For example, changes on the QoS for a few concrete services may not affect the search and objective space significantly. Further, composition plans for service composition with similar workflow structure can also be rather useful.

  \subsubsection{Candidates Creation}

At this step, the team considers all the candidates of synergy by instantiating the enriched self-awareness architectural pattern. In particular, they answer the questions as presented in Section~\ref{sec:method} as follows:

\begin{enumerate}

\item Which category does the expertise representation belong to? 
\begin{itemize}
\item[---] \textbf{\emph{Answer:}} Workflow structure belongs to the \texttt{Model} category but past problem instances/experience belongs to the \texttt{Assumption}.
\end{itemize}
\item If such a representation structural? is it tangible?
\begin{itemize}
\item[---] \textbf{\emph{Answer:}} Workflow structure is both structural and tangible while past problem instances/experience is neither structural nor tangible.
\end{itemize}
\item The expertise representation can be synergized with which algorithm/technique that realizes the self-aware capability? What are the possible levels of synergy?
\begin{itemize}
\item[---] \textbf{\emph{Answer:}} The workflow structure needs to be synergized with both stimulus- and time-awareness at \emph{level 1}; its synergy with goal-awareness is also required, but can be at any level except \emph{level 0}. The past problem instances/experience needs to be synergized with goal-awareness only, at all levels, including \emph{level 0}.
\end{itemize}
\item What is the possible form for each synergy?
\begin{itemize}
\item[---] \textbf{\emph{Answer:}} The workflow structure can be synergized with goal awareness in either \emph{specific} or \emph{general} form. All other synergies need to be realized in a \emph{general} form.
\end{itemize}
\item What is the difficulty level for each synergy?
\begin{itemize}
\item[---] \textbf{\emph{Answer:}} According to Figure~\ref{fig:difficulty}, the difficulty level ranges between \emph{very easy} to \emph{challenging}.
\end{itemize}

\end{enumerate}

The above answers produce 24 different candidates of synergizing domain expertise represented as  the enriched \emph{Temporal Goal Aware Pattern}.


   \begin{figure}[!t]
    \centering
   \begin{tikzpicture}
    \begin{axis}[
        grid = major,
        grid style={dashed},
       ylabel=Difficulty Score,
        xlabel=Benefit Score,
           label style={font=\fontsize{12}{12}\selectfont},
                    tick label style={font=\fontsize{12}{12}\selectfont},
      legend style={font=\fontsize{12}{12}\selectfont,at={(0.5,0.8)}, anchor=east,legend columns=1}]

 \addplot[only marks,mark=*,mark options={scale=2.5,fill=green!50}] plot coordinates  {
(13.775,3.26)
(14.685,3.64)
(15.53,4.8)
(14.45,3.43)
(14.36,3.81)
(15.783,4.28)
(16.205,4.97)
(14.125,3.6)
(14.035,3.98)
(16.458,4.45)
(16.88,5.14)
(13.235,3.15)
(14.145,3.53)
(14.568,4)
(14.99,4.69)
(13.82,3.3)
(14.73,3.68)
(15.158,4.15)
(15.575,4.84)
(14.405,3.44)
(14.315,3.82)
(15.738,4.29)
(16.16,4.98)
 };
 
  \addplot[only marks,mark=otimes*,mark options={scale=2.5,fill=red!50}] plot coordinates  {
(15.108,4.11)
(14.125,3.6)
  };

\addlegendentry{Unselected}
\addlegendentry{Selected}
    \end{axis}

    \end{tikzpicture}
    \caption{Difficulty and benefit scores for all candidates in the third case study..}
        \label{fig:exp-d3}
  \end{figure}
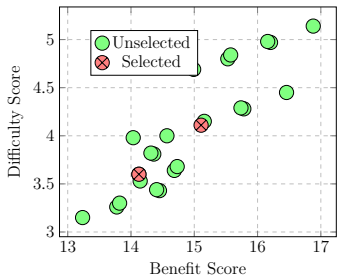

\subsubsection{Difficulty and Benefit Scores}

For all the 24 candidates, their overall scores with respect to both the difficulty and benefit are shown in Figure~\ref{fig:exp-d3}. Here, the $w$ between \emph{specific} and \emph{general} form of synergy is set as 1.3 and 1.5, respectively. For each candidate, the proficiency is set as 1.8 for all synergies related to workflow structure and 1.3 for those related to past problem instances/experience.

\begin{figure}[!t]
  \centering 
    \begin{subfigure}[t]{\columnwidth}
   \includegraphics[width=\columnwidth]{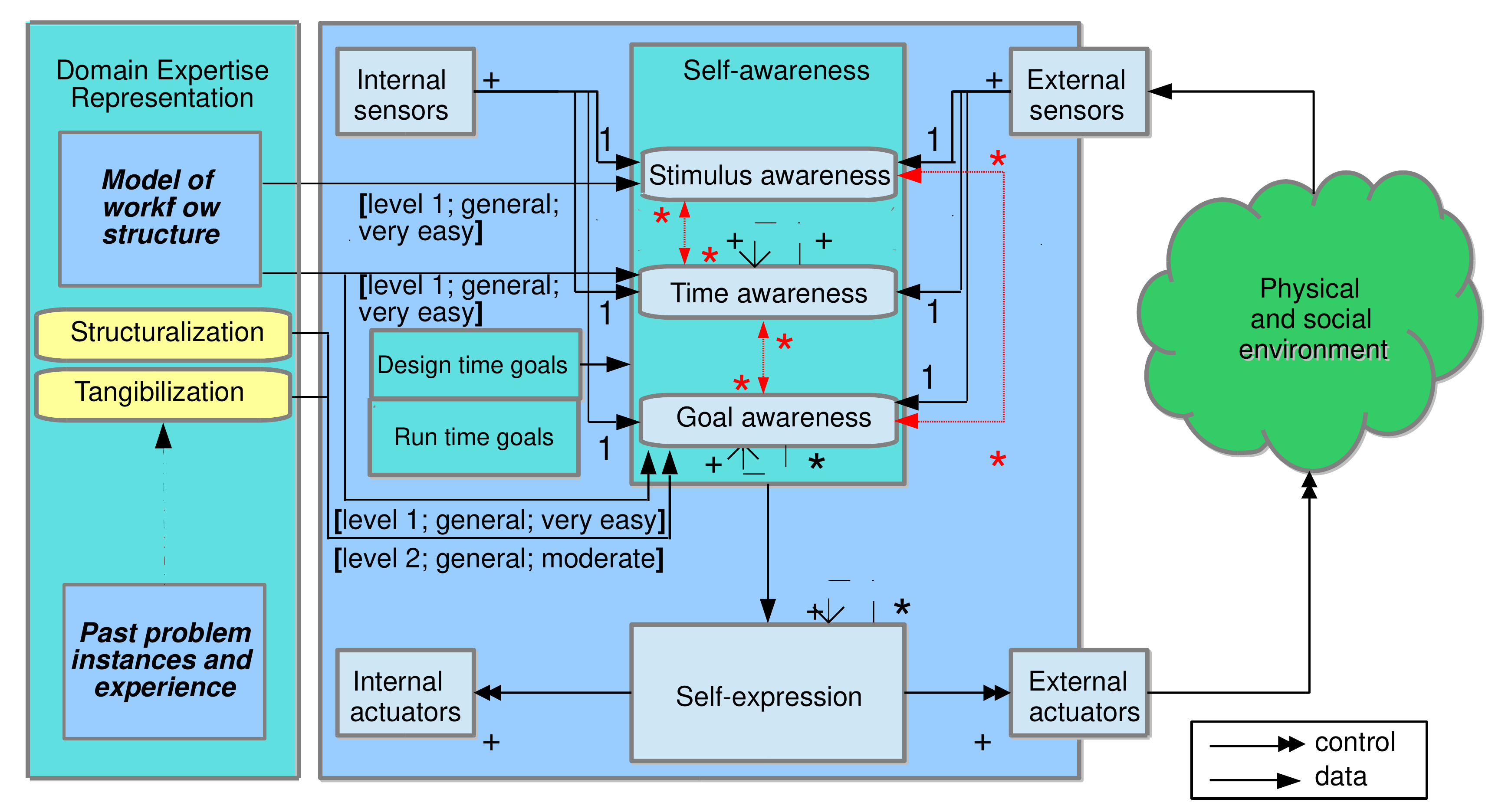}
\subcaption{candidate $C_1$}
 \label{fig:3rd-arch}
    \end{subfigure}
    
        \begin{subfigure}[t]{\columnwidth}
   \includegraphics[width=\columnwidth]{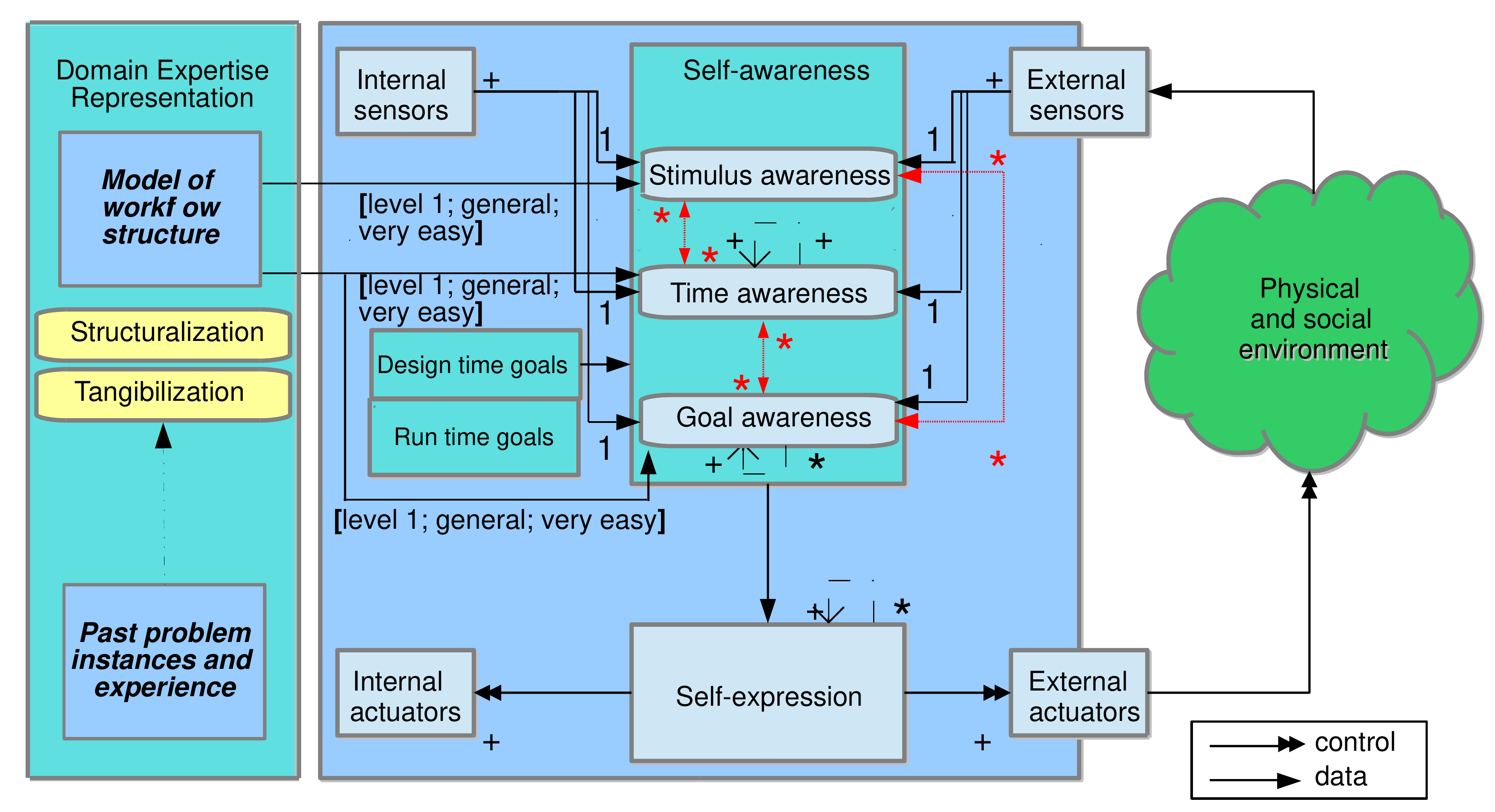}
\subcaption{candidate $C_2$}
 \label{fig:3rd-arch-2}
    \end{subfigure}

    \caption{Possible candidates selected for further investigation in the third case study.}

  \end{figure}

\subsubsection{Further Investigation}

After discussion with the team, two candidates, as shown  in Figure~\ref{fig:3rd-arch}, are selected since they appear to achieve a relatively good balance between the likely difficulty and the expected benefit. In brief, each of them are discussed as follows:

\begin{itemize}
\item $C_1$ ($\varmathbb{B}_{C_1}=15.108$, $\varmathbb{D}_{C_1}=4.11$): The candidate in Figure~\ref{fig:3rd-arch} automatically converts the knowledge of past problem instances and experience into structural and tangible representation. This is then used directly synergized with the goal-awareness to expedite the optimization process at \emph{level 2}. The workflow model, which is a result of human reasoning, is directly utilized by the stimulus-, time- and goal-awareness at \emph{level 1} of synergy.
 
\item $C_2$ ($\varmathbb{B}_{C_2}=14.125$, $\varmathbb{D}_{C_2}=3.6$): As shown in Figure~\ref{fig:3rd-arch-2}, the candidate has no synergy between the past problem instances/experience and goal-awareness. In other words, the evolutionary algorithm realizes goal-awareness, supported by the time aware analytical model, without any additional information on the past problem instances and experiences. All the other synergies remain the same as those of $C_1$.
 
\end{itemize}

More technical details on the actual synergy approaches can be found in~\cite{DBLP:journals/infsof/ChenLY19}.

\begin{table}[t!]
\centering
   \caption{The subject service-based systems for the experiments of the third case study.}
\label{table:3rd-exp}
\begin{threeparttable}
\begin{tabularx}{\columnwidth}{P{0.6cm}P{2cm}P{0.7cm}YP{0.8cm}Y}
\toprule 
\textbf{System}&\textbf{Objective}&\textbf{\#AS}&\textbf{\#CS}&\textbf{Env.}&\textbf{Space}\\ \midrule
\textsc{5AS}&latency; throughput; cost&5&510&services&1.1$\times 10^{10}$\\ \midrule
\textsc{10AS}&latency; throughput; cost&10&1,033&services&2.3$\times 10^{20}$\\\midrule
\textsc{15AS}&latency; throughput; cost&15&1,490&services&3.0$\times 10^{30}$\\\midrule
\textsc{100AS}&latency; throughput; cost&100&12,200&services&3.4$\times 10^{200}$\\
\bottomrule

\end{tabularx}
\begin{tablenotes}
        \item[*] AS denotes abstract services; CS denotes concrete services; Env. denotes environment; Space denotes search space.
    \end{tablenotes}
\end{threeparttable}
\end{table}

\subsubsection{Further Investigation Setup}



\begin{table}[t!]
\centering
   \caption{Benefits under further investigated candidates over all runs.}
\label{table:seed-t}

\begin{tabularx}{\columnwidth}{YYYY}
\toprule
\textbf{System}&\textbf{Metric}&
$C_1$&
$C_2$
\\ \midrule
\multicolumn{4}{c}{NSGA-II}\\
\midrule 
\multirow{4}{*}{5AS}&latency&\bfseries0.113&\bfseries0.113
 \\  
&throughput&\bfseries0.048&\bfseries0.048
\\
&cost&\bfseries7.151&12.476
\\
&hypervolume&\bfseries9.803E-01&9.781E-01
\\ \midrule
\multirow{4}{*}{10AS}&latency&\bfseries0.113&\bfseries0.113
 \\  
&throughput&\bfseries0.050&0.048
\\
&cost&\bfseries16.346&24.832
\\
&hypervolume&\bfseries9.871E-01&9.705E-01
\\ \midrule
\multirow{4}{*}{15AS}&latency&\bfseries0.113&\bfseries0.113
 \\  
&throughput&\bfseries0.065&0.048
\\
&cost&\bfseries43.433&56.330
\\
&hypervolume&\bfseries9.749E-01&9.460E-01
\\ \midrule

\multirow{4}{*}{100AS}&latency&\bfseries0.113&\bfseries0.113
 \\  
&throughput&\bfseries0.098&0.047
\\
&cost&\bfseries259.126&404.584
\\
&hypervolume&\bfseries9.860E-01&9.521E-01
\\ \midrule


\multicolumn{4}{c}{IBEA}\\
\midrule 
\multirow{4}{*}{5AS}&latency&0.276&\bfseries0.115
 \\  
&throughput&\bfseries0.047&\bfseries0.047
\\
&cost&\bfseries7.151&\bfseries7.151
\\
&hypervolume&\bfseries9.572E-01&9.443E-01
\\ \midrule
\multirow{4}{*}{10AS}&latency&0.251&\bfseries0.132
 \\  
&throughput&\bfseries0.051&0.050
\\
&cost&16.148&\bfseries16.088
\\
&hypervolume&\bfseries9.669E-01&9.541E-01
\\ \midrule
\multirow{4}{*}{15AS}&latency&0.170&\bfseries0.118
 \\  
&throughput&\bfseries0.065&0.048
\\
&cost&\bfseries45.304&52.832
\\
&hypervolume&\bfseries9.656E-01&9.351E-01
\\ \midrule

\multirow{4}{*}{100AS}&latency&0.164&\bfseries0.115
 \\  
&throughput&\bfseries0.082&0.053
\\
&cost&\bfseries278.873&447.491
\\
&hypervolume&\bfseries9.755E-01&9.435E-01
\\ \bottomrule

\end{tabularx}
\end{table}

The investigation is conducted by using the real-world WS-DREAM dataset~\cite{zheng2014investigating}, which contains QoS values for 4,500 services. Four distinct workflow structures of the software systems are randomly generated, each with 5, 10, 15 and 100 abstract services, respectively. As shown in Table~\ref{table:3rd-exp}, the number of concrete services and their QoS values on latency, throughput and cost\footnote{The cost values are generated in a way that a concrete service with better latency would also have higher cost.} are randomly selecting form the data set, resulting a range between 510 and 12,200 possible concrete services with a search space over one million. NSGA-II~\cite{nsgaii} and IBEA~\cite{ZitzlerK04} are used as the underlying evolutionary algorithm for goal-awareness, which are set a mutation rate of 0.1 and a crossover rate of 0.9, with 100 population size for 50 generation (300 generations for the case of 100 abstract services). As mentioned, for time-awareness, standard analytical models for service compositions are used~\cite{ramirez2017evolutionary}. All experiments were repeated 30 times and the mean values are reported. Again, the quality indicators used to assess both benefit and difficulty are shown in Table~\ref{table:3rd-qi}.


\begin{table}[t!]
\centering
   \caption{The quality indicators for benefit and difficulty for the third case study}
\label{table:3rd-qi}

\begin{tabularx}{0.8\columnwidth}{p{1.5cm}X}
\toprule 
\textbf{Attribute}&\textbf{Quality Indicators}\\ \midrule
Benefit&latency, throughput, cost and HV\\ 
Difficulty&Lines-Of-Code (LOC)\\
\bottomrule

\end{tabularx}
\end{table}

\subsubsection{Results}

\begin{figure}[!t]
 
   {
\centering
 \begin{subfigure}[t]{0.4\columnwidth}
  \begin{tikzpicture}
    \begin{axis}[
        width=5cm,
    height=5cm,
     xmax=5500,
      xtick={0,2500,...,5000},
    y tick label style={/pgf/number format/.cd,precision=3},
        legend style={at={(0.35,0.4),font=\fontsize{7}{8}\selectfont},
      anchor=north west,legend columns=1}]

\addplot[blue,mark=o,mark size=2pt] plot coordinates {
(0,0.9715731820480545)
(500,1.0129397783979206)
(1000,0.9754437024214044)
(1500,0.977569110246499)
(2000,0.9786689608106371)
(2500,0.9792945282042361)
(3000,0.9795878292847983)
(3500,0.9798134545307554)
(4000,0.9799851965779317)
(4500,0.9800880608098812)
(5000,0.980257290610373)
};

\addplot[red,mark=square,mark size=2pt] plot coordinates {
(0,0.7780836367574666)
(500,0.867368171223884)
(1000,0.925947972362261)
(1500,0.9488033544154267)
(2000,0.9606301946232675)
(2500,0.967136817630638)
(3000,0.9717671451802196)
(3500,0.9745580306626656)
(4000,0.9764897510812329)
(4500,0.9774392737500205)
(5000,0.9781364382830301)
};

\addlegendentry{$C_1$}
\addlegendentry{$C_2$}

    \end{axis}
    \end{tikzpicture}
             \subcaption{5AS}
            \label{fig:wf-5}
    \end{subfigure}
 ~
    \begin{subfigure}[t]{0.4\columnwidth}
  \begin{tikzpicture}
    \begin{axis}[
        width=5cm,
    height=5cm,
     xmax=5500,
      xtick={0,2500,...,5000},
    y tick label style={/pgf/number format/.cd,precision=3},
        legend style={at={(0.35,0.4),font=\fontsize{7}{8}\selectfont},
      anchor=north west,legend columns=1}]

\addplot[blue,mark=o,mark size=2pt] plot coordinates {
(0,0.9751585865010385)
(500,0.977097467950126)
(1000,0.9797177948858254)
(1500,0.9816290457190201)
(2000,0.9831878736518224)
(2500,0.9839770459727882)
(3000,0.9848008825173141)
(3500,0.9855195664836547)
(4000,0.9862310284815385)
(4500,0.9866963100148982)
(5000,0.9870478673760269)
};

\addplot[red,mark=square,mark size=2pt] plot coordinates {
((0,0.6343201857084227)
(500,0.7770164196624034)
(1000,0.8574165115106477)
(1500,0.9006567505216335)
(2000,0.9211975939392821)
(2500,0.9346388145049792)
(3000,0.9459542269936023)
(3500,0.9542897431059622)
(4000,0.9618711796360186)
(4500,0.9665449395343516)
(5000,0.9704758267959236)
};

\addlegendentry{$C_1$}
\addlegendentry{$C_2$}

    \end{axis}
    \end{tikzpicture}
                \subcaption{10AS}
             \label{fig:wf-10}
          \end{subfigure}
 
         \begin{subfigure}[t]{0.4\columnwidth}
  \begin{tikzpicture}
    \begin{axis}[
        width=5cm,
    height=5cm,
     xmax=5500,
      xtick={0,2500,...,5000},
    y tick label style={/pgf/number format/.cd,precision=3},
        legend style={at={(0.35,0.4),font=\fontsize{7}{8}\selectfont},
      anchor=north west,legend columns=1}]

\addplot[blue,mark=o,mark size=2pt] plot coordinates {
(0,0.9494809913379986)
(500,0.951225890738334)
(1000,0.9541580655499344)
(1500,0.9579103108690462)
(2000,0.9612076696999613)
(2500,0.9637083677987065)
(3000,0.9660318655993737)
(3500,0.9689992423347271)
(4000,0.9715163356490746)
(4500,0.9731802029330198)
(5000,0.9749370233242349)
};

\addplot[red,mark=square,mark size=2pt] plot coordinates {
(0,0.6355012043548048)
(500,0.7221185085865697)
(1000,0.8105393647769871)
(1500,0.8523391001086935)
(2000,0.8807689973227574)
(2500,0.8994756548152026)
(3000,0.9128880595616429)
(3500,0.9236856442112382)
(4000,0.9330663502543859)
(4500,0.9407316419046972)
(5000,0.9459549145691647)
};

\addlegendentry{$C_1$}
\addlegendentry{$C_2$}

    \end{axis}
    \end{tikzpicture}
         \subcaption{15AS}
           \label{fig:wf-15}
    \end{subfigure}
 ~    
          \begin{subfigure}[t]{0.4\columnwidth}
  \newsavebox{\tenhseednsgaii}
\savebox{\tenhseednsgaii}{
\begin{tikzpicture}
    \begin{axis}[
    scaled ticks = false,
        width=3.5cm,
    height=3.5cm,
    xmax=30000,
   tick label style={font=\tiny},
    y tick label style={/pgf/number format/.cd,precision=4},
        legend style={at={(0.3,0.4),font=\fontsize{7}{8}\selectfont},
      anchor=north west,legend columns=1}]

\addplot[blue,mark size=2pt] plot coordinates {
(20000,0.9793071112398718)
(20500,0.9796101370138977)
(21000,0.9799353011173059)
(21500,0.9802792236083755)
(22000,0.9807523971054268)
(22500,0.9812525308009661)
(23000,0.9816203505212464)
(23500,0.9820181570705143)
(24000,0.9824055372888476)
(24500,0.9827598450503613)
(25000,0.9830033576437571)
(25500,0.9834664055131143)
(26000,0.9837632652904251)
(26500,0.9841236675855208)
(27000,0.9843795708098317)
(27500,0.9847106884148707)
(28000,0.9848686787590838)
(28500,0.9850938831561724)
(29000,0.9854835342562751)
(29500,0.9857458139782437)
(30000,0.9860280034614374)
};

\addplot[red,dashed,mark size=2pt] plot coordinates {
(20000,0.9283635426801926)
(20500,0.929660355249523)
(21000,0.9311844784512366)
(21500,0.9323646598632199)
(22000,0.9339619122247432)
(22500,0.9352208198664279)
(23000,0.9366954509114346)
(23500,0.9379964841376)
(24000,0.939134826140624)
(24500,0.9405190715862761)
(25000,0.9416582775306631)
(25500,0.9426394290274805)
(26000,0.9439341300126658)
(26500,0.944790884611054)
(27000,0.9460389693238866)
(27500,0.9470691844590551)
(28000,0.9479690467982954)
(28500,0.949016633974748)
(29000,0.9501881628668298)
(29500,0.9511580358213022)
(30000,0.9521121650380693)
};

    \end{axis}
    \end{tikzpicture}
      
      }

\begin{tikzpicture}
    \begin{axis}[
        scaled ticks = false,
        width=5cm,
    height=5cm,
    xmax=35000,
    xtick={0,15000,30000},
   y tick label style={/pgf/number format/.cd,precision=3},
        legend style={at={(0.35,0.4),font=\fontsize{7}{8}\selectfont},
      anchor=north west,legend columns=1}]

\addplot[blue,mark=o,mark size=2pt] plot coordinates {
(0,0.957276366859774)
(500,0.9575073981933498)
(1000,0.9580576755380679)
(1500,0.9587491646580075)
(2000,0.9594817557251398)
(2500,0.9602012211836853)
(3000,0.9608800515161509)
(3500,0.9616397912834721)
(4000,0.9622808490399215)
(4500,0.9629386687441327)
(5000,0.9637598132249223)
(5500,0.9644272116345152)
(6000,0.965079921432007)
(6500,0.9658232468709476)
(7000,0.9663914782072304)
(7500,0.967085734402229)
(8000,0.9677176003027533)
(8500,0.9683232742501491)
(9000,0.9688613682844338)
(9500,0.9694548600571431)
(10000,0.970057603466041)
(10500,0.9706762004896687)
(11000,0.9712002566861014)
(11500,0.9718037107162838)
(12000,0.9722648799695051)
(12500,0.9728191899077793)
(13000,0.9733922344513085)
(13500,0.9739227196921278)
(14000,0.9744113132522545)
(14500,0.9747876355174496)
(15000,0.9752262053516235)
(15500,0.975707973418669)
(16000,0.97632213673134)
(16500,0.9767786010381493)
(17000,0.9771119865943095)
(17500,0.9775385120360642)
(18000,0.9779930497084558)
(18500,0.9782987878920061)
(19000,0.9787169548016583)
(19500,0.9790192943677217)
(20000,0.9793071112398718)
(20500,0.9796101370138977)
(21000,0.9799353011173059)
(21500,0.9802792236083755)
(22000,0.9807523971054268)
(22500,0.9812525308009661)
(23000,0.9816203505212464)
(23500,0.9820181570705143)
(24000,0.9824055372888476)
(24500,0.9827598450503613)
(25000,0.9830033576437571)
(25500,0.9834664055131143)
(26000,0.9837632652904251)
(26500,0.9841236675855208)
(27000,0.9843795708098317)
(27500,0.9847106884148707)
(28000,0.9848686787590838)
(28500,0.9850938831561724)
(29000,0.9854835342562751)
(29500,0.9857458139782437)
(30000,0.9860280034614374)
};

\addplot[red,mark=square,mark size=2pt] plot coordinates {
(0,0.3400897848291552)
(500,0.40781777652058365)
(1000,0.5018197442495915)
(1500,0.579726859488306)
(2000,0.6425300584514538)
(2500,0.6897714205734686)
(3000,0.7265042355501791)
(3500,0.7579632196123698)
(4000,0.7811355114196918)
(4500,0.7981598254153695)
(5000,0.8116056862159008)
(5500,0.8230969899300129)
(6000,0.8332050366640759)
(6500,0.841004189089715)
(7000,0.8487144636804912)
(7500,0.854523713109317)
(8000,0.8598625008922517)
(8500,0.8653645464648994)
(9000,0.8701807856073566)
(9500,0.8742832251126624)
(10000,0.8784070011667601)
(10500,0.8821802373919996)
(11000,0.8858415432926372)
(11500,0.8896023506388139)
(12000,0.8929190867138541)
(12500,0.8964440803274744)
(13000,0.8995735351358672)
(13500,0.9021883360624543)
(14000,0.90479032020835)
(14500,0.9072581542807427)
(15000,0.9102057696377633)
(15500,0.9121481322334055)
(16000,0.9148289236078354)
(16500,0.9167391164616826)
(17000,0.9186488853853586)
(17500,0.9206380629604016)
(18000,0.9222559650748231)
(18500,0.9240593627550001)
(19000,0.9255637854844408)
(19500,0.9270891805044368)
(20000,0.9283635426801926)
(20500,0.929660355249523)
(21000,0.9311844784512366)
(21500,0.9323646598632199)
(22000,0.9339619122247432)
(22500,0.9352208198664279)
(23000,0.9366954509114346)
(23500,0.9379964841376)
(24000,0.939134826140624)
(24500,0.9405190715862761)
(25000,0.9416582775306631)
(25500,0.9426394290274805)
(26000,0.9439341300126658)
(26500,0.944790884611054)
(27000,0.9460389693238866)
(27500,0.9470691844590551)
(28000,0.9479690467982954)
(28500,0.949016633974748)
(29000,0.9501881628668298)
(29500,0.9511580358213022)
(30000,0.9521121650380693)};

\addlegendentry{$C_1$}
\addlegendentry{$C_2$}

    \end{axis}
    \end{tikzpicture}
         \subcaption{100AS}
           \label{fig:wf-15}
    \end{subfigure}
    
  }
  \caption{The changes of mean HV (y-axis) when using NSGA-II on all runs with respect to the number of evaluations (x-axis) under further investigated candidates.}
         \label{fig:hv-change}
  \end{figure}
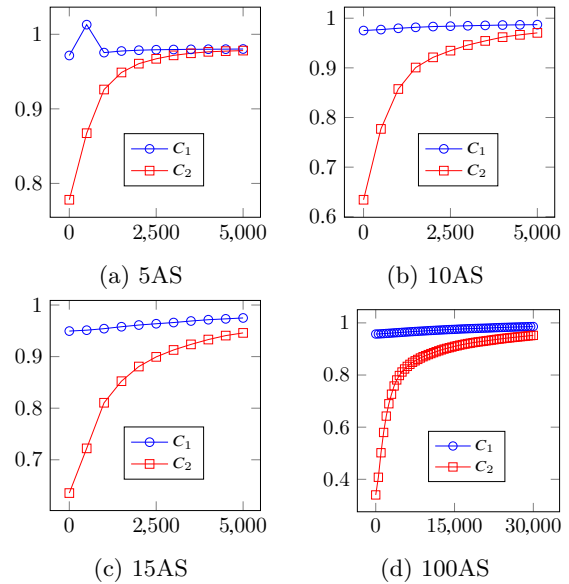

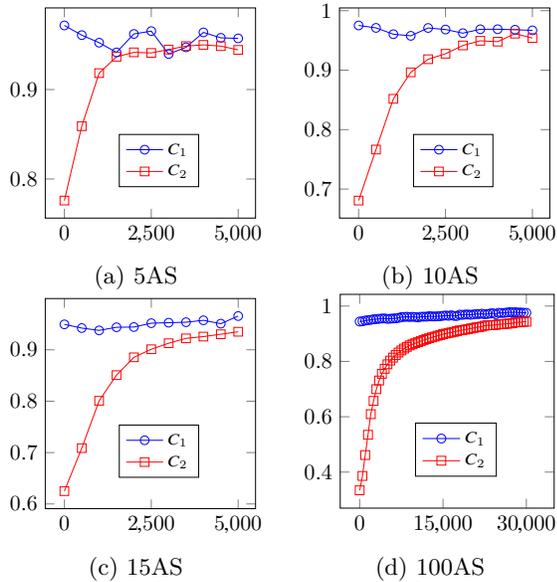
\begin{figure}[!t]
 
   {
\centering
 \begin{subfigure}[t]{0.4\columnwidth}
  \begin{tikzpicture}
    \begin{axis}[
        width=5cm,
    height=5cm,
     xmax=5500,
      xtick={0,2500,...,5000},
    y tick label style={/pgf/number format/.cd,precision=3},
        legend style={at={(0.35,0.4),font=\fontsize{7}{8}\selectfont},
      anchor=north west,legend columns=1}]

\addplot[blue,mark=o,mark size=2pt] plot coordinates {
(0,0.9716376289840044)
(500,0.9608245367718022)
(1000,0.9523697070362458)
(1500,0.9415207042207847)
(2000,0.9621187470101241)
(2500,0.965302969662059)
(3000,0.9396930970511003)
(3500,0.947195693954362)
(4000,0.9638529326795077)
(4500,0.9577979148826546)
(5000,0.9571807633852225)
};

\addplot[red,mark=square,mark size=2pt] plot coordinates {
(0,0.7758941449622054)
(500,0.8589894638361293)
(1000,0.9183411181731943)
(1500,0.9366741716958622)
(2000,0.941569948903349)
(2500,0.940923359616541)
(3000,0.9446926251408096)
(3500,0.9485903237044988)
(4000,0.9499275128476203)
(4500,0.9485750596153861)
(5000,0.9443304410014305)
};

\addlegendentry{$C_1$}
\addlegendentry{$C_2$}

    \end{axis}
    \end{tikzpicture}
             \subcaption{5AS}
            \label{fig:wf-5}
    \end{subfigure}
 ~
    \begin{subfigure}[t]{0.4\columnwidth}
  \begin{tikzpicture}
    \begin{axis}[
        width=5cm,
    height=5cm,
     xmax=5500,
      xtick={0,2500,...,5000},
    y tick label style={/pgf/number format/.cd,precision=3},
        legend style={at={(0.35,0.4),font=\fontsize{7}{8}\selectfont},
      anchor=north west,legend columns=1}]

\addplot[blue,mark=o,mark size=2pt] plot coordinates {
(0,0.9751585865010385)
(500,0.9710711926209868)
(1000,0.9604538291219902)
(1500,0.9577270945318068)
(2000,0.9709524236471443)
(2500,0.9683694079778676)
(3000,0.9621963728388795)
(3500,0.9687497544017547)
(4000,0.9688342991608055)
(4500,0.9679670102319119)
(5000,0.9669302747409997)
};

\addplot[red,mark=square,mark size=2pt] plot coordinates {
(0,0.6806661695714731)
(500,0.7664823082408287)
(1000,0.8517962787576249)
(1500,0.8963244830933593)
(2000,0.9183985569850919)
(2500,0.9275735477513617)
(3000,0.9415774367974726)
(3500,0.9493819947455493)
(4000,0.9479022319247491)
(4500,0.9613562371052797)
(5000,0.9540586092115096)
};

\addlegendentry{$C_1$}
\addlegendentry{$C_2$}

    \end{axis}
    \end{tikzpicture}
                \subcaption{10AS}
             \label{fig:wf-10}
          \end{subfigure}
 
         \begin{subfigure}[t]{0.4\columnwidth}
  \begin{tikzpicture}
    \begin{axis}[
        width=5cm,
    height=5cm,
     xmax=5500,
      xtick={0,2500,...,5000},
    y tick label style={/pgf/number format/.cd,precision=3},
        legend style={at={(0.35,0.4),font=\fontsize{7}{8}\selectfont},
      anchor=north west,legend columns=1}]

\addplot[blue,mark=o,mark size=2pt] plot coordinates {
(0,0.9494809913379986)
(500,0.9422540937373097)
(1000,0.9377761567026321)
(1500,0.9438339046680585)
(2000,0.9442674293064012)
(2500,0.9517574492921591)
(3000,0.9527487112260267)
(3500,0.9536703851199848)
(4000,0.957371780022455)
(4500,0.9508237540441284)
(5000,0.9656456577708599)
};

\addplot[red,mark=square,mark size=2pt] plot coordinates {
(0,0.624889071090223)
(500,0.7086238135817172)
(1000,0.8008124845991506)
(1500,0.85068176769361)
(2000,0.8853408537446399)
(2500,0.9010765048336405)
(3000,0.9127755021276371)
(3500,0.9222744153251982)
(4000,0.9256968393658471)
(4500,0.9302638801173819)
(5000,0.935123340539559)
};

\addlegendentry{$C_1$}
\addlegendentry{$C_2$}

    \end{axis}
    \end{tikzpicture}
         \subcaption{15AS}
           \label{fig:wf-15}
    \end{subfigure}
 ~    
          \begin{subfigure}[t]{0.4\columnwidth}
  \begin{tikzpicture}
    \begin{axis}[
        scaled ticks = false,
        width=5cm,
    height=5cm,
    xmax=35000,
    xtick={0,15000,30000},
   y tick label style={/pgf/number format/.cd,precision=3},
        legend style={at={(0.35,0.4),font=\fontsize{7}{8}\selectfont},
      anchor=north west,legend columns=1}]

\addplot[blue,mark=o,mark size=2pt] plot coordinates {
(0,0.9440728142327305)
(500,0.9450493829387183)
(1000,0.9476097825474209)
(1500,0.9490685958514938)
(2000,0.9500935525003339)
(2500,0.9522671208436507)
(3000,0.9528120928707888)
(3500,0.9538761294096033)
(4000,0.9550832779224457)
(4500,0.95522253188452)
(5000,0.9536801252940391)
(5500,0.9543520852436226)
(6000,0.9551997365752679)
(6500,0.9557616083640499)
(7000,0.9573759176109193)
(7500,0.9600098597913027)
(8000,0.9607335860208952)
(8500,0.961354942819091)
(9000,0.9605249217090325)
(9500,0.9611317482359832)
(10000,0.9604301932638646)
(10500,0.9588833997426018)
(11000,0.9613965424863974)
(11500,0.9617090375387443)
(12000,0.961679497391057)
(12500,0.9634736543775453)
(13000,0.9627319689412122)
(13500,0.9622676163035518)
(14000,0.9630326208307642)
(14500,0.9634955538705918)
(15000,0.9648481754227473)
(15500,0.965952982786991)
(16000,0.9647545254998138)
(16500,0.9666404805652835)
(17000,0.96446726342825)
(17500,0.9641885845040455)
(18000,0.9691203179807263)
(18500,0.9680505426033924)
(19000,0.9694687872395236)
(19500,0.9694567363584903)
(20000,0.968490228111859)
(20500,0.9698784797840855)
(21000,0.9713447683628508)
(21500,0.9712916872271685)
(22000,0.9723912979041232)
(22500,0.9705588056478571)
(23000,0.9713123659490679)
(23500,0.9733045353696528)
(24000,0.9739652294632232)
(24500,0.9703021787432831)
(25000,0.9757493650040099)
(25500,0.975094744161805)
(26000,0.9753108752003138)
(26500,0.9766175156179371)
(27000,0.9778975360562047)
(27500,0.9772682666135986)
(28000,0.9779378325383999)
(28500,0.9765387761064059)
(29000,0.9767115755860852)
(29500,0.9758077010261507)
(30000,0.9755120945303077)
};

\addplot[red,mark=square,mark size=2pt] plot coordinates {
(0,0.334127230974257)
(500,0.3860548501819432)
(1000,0.4615836985798596)
(1500,0.5348000938593945)
(2000,0.6093595337019927)
(2500,0.6574262145061897)
(3000,0.6996127710292881)
(3500,0.731013687550303)
(4000,0.7551360245201201)
(4500,0.77327677490212)
(5000,0.7896211149327826)
(5500,0.8029587604247664)
(6000,0.8136285309034634)
(6500,0.8230723806113983)
(7000,0.832012766343663)
(7500,0.8396560420458504)
(8000,0.8464990185878276)
(8500,0.8524960303278114)
(9000,0.8579495461251191)
(9500,0.8618126481339473)
(10000,0.8657795022761913)
(10500,0.869553064708754)
(11000,0.8733238005503268)
(11500,0.877242243508828)
(12000,0.8804853809989505)
(12500,0.8838062286561186)
(13000,0.8869540996079527)
(13500,0.8896262896084709)
(14000,0.892626338461352)
(14500,0.8951671486079543)
(15000,0.8986361340229708)
(15500,0.9005873280267023)
(16000,0.9030445668244323)
(16500,0.9046926549584959)
(17000,0.9068377826611569)
(17500,0.9093707527625488)
(18000,0.9122570282201892)
(18500,0.9130644747824233)
(19000,0.9151294873305234)
(19500,0.916901608374146)
(20000,0.9192032890760401)
(20500,0.9208906420904038)
(21000,0.9230993762527281)
(21500,0.9241125873182593)
(22000,0.9261066335664465)
(22500,0.9292474007537507)
(23000,0.9305897129118009)
(23500,0.9306308203240851)
(24000,0.9310825465382002)
(24500,0.9320700754405262)
(25000,0.9340619745280726)
(25500,0.9344480101558033)
(26000,0.9352327734147144)
(26500,0.9360637851520687)
(27000,0.937030725874962)
(27500,0.9380003583157985)
(28000,0.941146129479298)
(28500,0.9412852713266154)
(29000,0.9413303437066273)
(29500,0.9433098645987158)
(30000,0.9435066350009583)
};

\addlegendentry{$C_1$}
\addlegendentry{$C_2$}

    \end{axis}
    \end{tikzpicture}
         \subcaption{100AS}
           \label{fig:wf-15}
    \end{subfigure}
    
  }
  \caption{The changes of mean HV (y-axis) when using IBEA on all runs with respect to the number of evaluations (x-axis) under further investigated candidates.}
         \label{fig:hv-change-ibea}
  \end{figure}

From Table~\ref{table:seed-t}, clearly, $C_1$ leads to at least the same results for a quality objective when comparing to the case of $C_2$. In particular, it has also resulted in better Hypervolume (HV) value\footnote{HV measures the region from the non-dominated solutions to a nadir point, which in this case is a vector of the worst possible objective values found. The larger the HV value, the better.}~\cite{DBLP:conf/icse/Li0Y18}. All the comparisons, except those equivalent ones, are statistically significant according to the Wilcoxon Signed Rank test ($p<$0.05), with non-trivial effect sizes. In particular, the improvement tends to be amplified as the number of abstract services increases, implying that the more complex the scenario, the better benefit that the domain expertise on past problem instances can offer when combined with the self-awareness. In Figure~\ref{fig:hv-change} and~\ref{fig:hv-change-ibea}, we see that on all cases, $C_1$ achieves higher HV value than that of the $C_2$ throughout, meaning that it exhibits faster convergence. Again, the improvement is more obvious under more complex scenarios, e.g., when there are 100 abstract services.

\begin{figure}[!t]
    \centering
   \begin{tikzpicture}
\begin{axis}[
    ybar=0.05cm,
    width=5cm,
    height=5cm,
    bar width=0.4cm,
 enlarge x limits=0.5,
 enlarge y limits=0.05,
    legend style={at={(0.3,0.96)},
      anchor=north,legend columns=1},
    ylabel={Mean HV Reduction},
    symbolic x coords={NSGA-II,IBEA},
    xtick=data
    ]

\addplot[fill=black] coordinates {(NSGA-II,0.0069594907) (IBEA,0.01123521009)};
\addplot[fill=white] coordinates {(NSGA-II,0.01369724486) (IBEA,0.04139148237)};



\legend{$C_1$,$C_2$}

\end{axis}
\end{tikzpicture}
    \caption{The mean HV reduction (on all systems and runs) simplified algorithm under further investigated candidates.}
        \label{fig:deg-change}
  \end{figure}
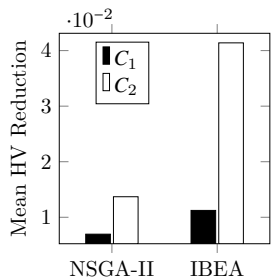

In Figure~\ref{fig:deg-change}, the team examines how the behaviors of the software systems change when the underlying algorithm that realizes self-awareness is simplified. To this end, the crossover operator in NSGA-II is omitted, based on which the results can be compared to the cases when it is present for both approaches. Clearly, we see a considerable reduction on the HV values when the crossover operator is removed, suggesting that a simplified version of the underlying algorithm that realizes self-awareness may negatively affect the performance. Further, the more complex the service system, the greater the reduction. However, we see that $C_1$ is more resilient than $C_2$, which again proves that the domain expertise of past problem instance can be beneficial in guiding the algorithm that achieves self-awareness for even better results.

As shown in Table~\ref{table:3rd-bd}, $C_1$ requires higher LOC than $C_2$, which is as anticipated. Yet, their margin of difficult is remarkably small (635 lines) and hence it is wiser to choose $C_1$ for the actual deployment.

\begin{bclogo}[couleur=gray!10,arrondi=0.1,epBord=1.5,couleurBord=black!70,logo=\bctrombone]{{\normalsize Final choice for the third case study:}}According to the verified results from the further investigation, the team has decided that $C_1$ can better fit the needs.
\end{bclogo}

  \begin{table}[t!]
\centering
   \caption{The LOC for the second case study}
\label{table:3rd-bd}

\begin{tabularx}{0.4\columnwidth}{YY}
\toprule 
\textbf{Candidate}&\textbf{LOC}\\ \midrule
$C_1$&70,429\\
$C_2$&69,794\\
\bottomrule

\end{tabularx}
\end{table}
  
\section{Discussion}
\label{sec:discussion}

\subsection{How DBASES can Help?}

In summary, there are three aspects based on which DBASES can help to engineering synergy between domain expertise and self-awareness in a principled way, as it:

\begin{itemize}

\item Encapsulates various notations and classification that helps to understand, analyze and reason about the information and knowledge that the engineers have for achieving self-awareness.

\item Visualizes the possible synergies, in a form of the enriched self-awareness architectural patterns, to provide intuitive understanding of the candidates.

\item Provides a methodology, building on the above tow points, that offers step-by-step guidance on how to engineering self-awareness with explicit consideration of domain expertise.

\end{itemize}

It is worth noting that, although DBASES aims to help the engineers to finally select a single candidate of synergy, within the engineering process, it is by no mean that we restrict them to select only one nor to choose all of the possible candidates. In this respect, DBASES is similar to CBAM~\cite{DBLP:conf/icse/KazmanAK01}, which is a successful architecture selection methodology that also helps to reveal and quantify the cost, benefit and risk of design options in software development. CBAM also provides some visualisations similar to the way we do, but it is irrelevant to the explicit categories of domain expertise and their synergies to self-awareness. In fact, the precise description of which synergy candidate(s) chosen for implementation is irrelevant to the point of the framework and the message of this work. Our key point is that we have given the engineers a principled, repeatable method for making architectural choices of candidate(s), and understanding the consequences of these choices in terms of difficult and benefit, This method has been successful in that it guided the engineers to consider many ways of synergy that they would have otherwise overlooked, for two reasons: 

\begin{itemize}

\item Taking the difficult/benefit plot in Figure~\ref{fig:exp-dec} as an example, some candidates have extremely high benefits scores whilst relatively easy to realize and hence bear some of the highest desirability.

\item Some candidates have relatively low benefit scores but are still quite
difficult to realize, primarily due to low proficiency. Therefore, they can be ruled out from consideration.
\end{itemize}

It is perfectly normal that more than one candidates are selected for further investigation and profiling, as what we have done in the tutorial case studies. But our framework provides such opportunity to intuitively localize which are the prefer ones and which candidates should be ruled out, thereby saving the valuable human effort in investigating them.

\subsection{Threats to Applicability}

When the number of possible candidate increase, the engineers are likely overwhelmed with identifying and discussing them all. However, the fact that there are many combinations is not uncommon when making architectural design decisions~\cite{paul2002evaluating}; this is in fact a more general problem in Operational Research that how can one makes proper trade-off decision when there is a large number of alternatives. Visualisation and quantification seem to be a promising solution, which is what DBASES provides. In this way, a designer can have more intuitive information on the relative difficulty and benefits on the alternative, and thus making informed decision. 

Of course, when the number of points are too many to conduct analysis visually, it is possible to improve DBASES by incorporating some forms of preferences so that only a particular region of points that is of interest can be focused. This is however subject to future work.

\subsection{Threats to External Validity}

It is known that methodological work is extremely difficult to be evaluated and ensure its generality. The reported case studies, as the name suggested, aim to replicate what would have happen when DBASES is used in a diverse scenarios, in which case it is likely that only some desirable ones can be selected for further investigation while ruling out the others which are of no interests. This is the reason why we have chosen a subset of the candidates in the experiments. Of course, it is indeed possible to evaluate all of the synergy candidate, but this would consume a large amount of time/resource, which we plan to investigate as part of future work.

Another threat is related to whether the industry practitioners will find that the DBASES is practical enough at a real-world industrial scale. Indeed, while this is important, it cannot be achieved without expensive surveying process, which will be extremely time-consuming. Therefore, we see this work as a first step to promote engineering synergy between domain expertise and self-awareness, and a more thorough evaluation with industrial stakeholders is part of our ongoing research.

\section{Conclusion and Future Work}
\label{sec:conclusion}
Architectural patterns and methodology for self-awareness have proven to be effective in guiding the systematic design, knowledge representation and reasoning for software systems that demand self-adaptation. However, when domain expertise needs to be synergized with the capabilities of self-awareness, current patterns and methods lack of guidelines about which domain expertise can be synergized, the extents of synergy and what are the trade-offs involved.

This paper is the first attempt that highlights the importance of synergizing domain expertise with the self-awareness in software systems, relying on well-defined underlying approaches. As part of the contributions, we present a holistic framework, dubbed DBASES, that offers a principled guideline for the engineers to perform difficulty and benefit analysis for synergizing domain expertise and self-awareness,

Using three tutorial case studies from distinct domains, we describe how DBASES can help to assist in making design decision on the synergy of domain expertise with self-aware capabilities, particularly on selecting candidates for further investigation with quantitative profiling.

The notion of synergy in DBASES is a genuine attempt towards keeping domain experts and architects in the loop, a branch of a larger vision that relate to keeping ``engineers-in-the-loop" for self-adaptive software systems, in which human (i.e., software and system engineers for our case) can control the behaviors of the underlying algorithms and techniques that realize the self-awareness at least to certain extents. This will consequently offer greater intuition and transparency into the awareness processes of the self-adaptive software system, improving its interpretable and explainable appeal. 


Drawing on the foundation provided in this work, future research shall investigate how exactly the human can be placed into the loop with DBASES, considering the timeliness and reliability of the their expertise. Those problems will open up a full range of new research directions, drawing on the findings and proposals derived from this the work. This is one of our ongoing research investigation that is evolving into a specialized topic by its own for the discipline of engineering self-aware and self-adaptive software systems.

\section*{Acknowledgment}
This work was supported by the Guangdong Provincial Key Laboratory of Brain-inspired Intelligent Computation, the Program for Guangdong Introducing Innovative and Enterpreneurial Teams (Grant No. 2017ZT07X386), Shenzhen Science and Technology Program (Grant No. KQTD2016112514355531) and the Program for University Key Laboratory of Guangdong Province (Grant No. 2017KSYS008).


\ifCLASSOPTIONcaptionsoff
  \newpage
\fi

\bibliographystyle{myIEEEtrans}
\bibliography{IEEEabrv,references}

\begin{IEEEbiography}[{\includegraphics[width=1in,height=1.25in,clip,keepaspectratio]{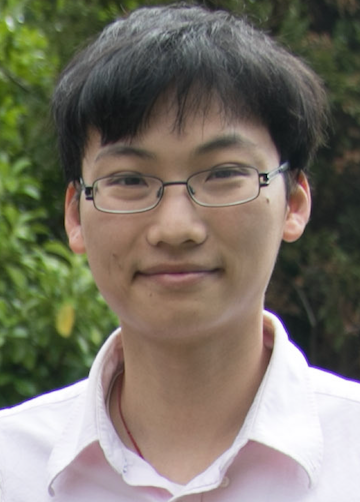}}]{Tao Chen} 
received his Ph.D. from the School of Computer Science, University of Birmingham, United Kingdom, in 2016. He is currently a Lecturer (assistant professor) in Computer Science at the Department of Computer Science, Loughborough University, United Kingdom. He has broad research interests on software engineering, including but not limited to performance engineering, self-adaptive software systems, search-based software engineering, data-driven software engineering and computational intelligence. As the lead author, his work has been published in internationally renowned journals and conference such as \textsc{IEEE Transactions on Software Engineering}, \textsc{ACM Transactions on Software Engineering and Methodology}, \textsc{IEEE Transactions on Services Computing}, \textsc{ACM Computing Surveys} and \textsc{IEEE/ACM International Conference on Software Engineering}. Among other roles, Dr. Chen regularly serves as a PC member for various conferences in his fields and is an associate editor for the \textsc{Services Transactions on Internet of Things}. He is a member of the IEEE.
\end{IEEEbiography}

\begin{IEEEbiography}[{\includegraphics[width=1in,height=1.25in,clip,keepaspectratio]{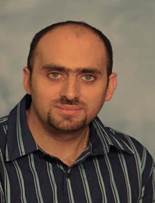}}]{Rami Bahsoon} 
received the Ph.D. degree in software engineering from University College London for his research on evaluating software architecture stability using real options and he attended London Business School for MBA-level studies in technology strategy and dynamics. He is a Senior Lecturer of software engineering (associate professor) and leads the software engineering for/in the Cloud Interest Group, University of Birmingham, Birmingham, United Kingdom. The group's research aims at developing architecture and frameworks to support and reason about dependable complex software systems, where the investigations span cloud computing architectures and their economics. He published extensively in the area of economics driven software engineering, cloud software engineering, and utility computing and co-edited a book on Software Architecture and Software Quality and another on Economics-Driven Software Architecture (published by Elsevier). He is a member of the IEEE.
\end{IEEEbiography}

\begin{IEEEbiography}[{\includegraphics[width=1in,height=1.25in,clip,keepaspectratio]{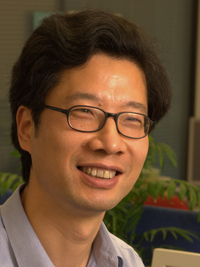}}]{Xin Yao} 
(F'03) received the B.Sc. degree from the University of Science and Technology of China (USTC), Hefei, China, in 1982, the M.Sc. degree from the North China Institute of Computing Technologies, Beijing, China, in 1985, and the Ph.D. degree from USTC in 1990. He is a Chair Professor of Computer Science at the Southern University of Science and Technology (SUSTech), Shenzhen, China, and a part-time Professor of Computer Science at the University of Birmingham,Birmingham, United Kingdom. His current research interests include evolutionary computation, machine learning, and their real world applications, especially to software engineering. He was a recipient of the prestigious Royal Society Wolfson Research Merit Award in 2012, the IEEE Computational Intelligence Society (CIS) Evolutionary Computation Pioneer Award in 2013 and the IEEE Frank Rosenblatt Award in 2020. His work won the 2001 IEEE Donald G. Fink Prize Paper Award, the 2010, 2016, and 2017 \textsc{IEEE Transactions on Evolutionary Computation} Outstanding Paper Awards, the 2011 \textsc{IEEE Transactions on Neural Networks} Outstanding Paper Award, and many other best paper awards. He was the President of IEEE CIS from 2014 to 2015 and the Editor-in-Chief of \textsc{IEEE Transactions on Evolutionary Computation} from 2003 to 2008. He is an IEEE fellow since 2003 and was a Distinguished Lecturer of IEEE CIS.
\end{IEEEbiography}


\end{document}